\documentclass[12pt]{article}

\usepackage{arxiv}
\usepackage{graphicx}

\usepackage{amssymb}
\usepackage{amsmath}
\usepackage{amsthm}
\usepackage{amsfonts}
\usepackage{newtxtext, newtxmath}
\usepackage{enumitem}
\usepackage{titling}
\usepackage[colorlinks=true]{hyperref}
\usepackage{cancel}

\usepackage[utf8]{inputenc} 
\usepackage[T1]{fontenc}    
\usepackage{url}            
\usepackage{booktabs}       
\usepackage{nicefrac}       
\usepackage{microtype}      
\usepackage{lipsum}
\usepackage{subfiles}
\usepackage{comment}
\usepackage{natbib}
\usepackage{color}
\usepackage[dvipsnames,cmyk]{xcolor}
\usepackage{tcolorbox}
\newtcolorbox{mybox}{colback=red!5!white,colframe=red!75!black}

\usepackage[english]{babel}

\usepackage{authblk}

\theoremstyle{definition}

\theoremstyle{remark}

\graphicspath{ {./images/} }

\title{
\normalfont \normalsize 
\textsc{} \\
\LARGE Shaping manifolds in equivariant recurrent neural networks
}

\author[1]{Arianna Di Bernardo}
\author[1]{Adrian Valente}
\author[2]{Francesca Mastrogiuseppe}
\author[1]{Srdjan Ostojic}
\affil[1]{Laboratoire de Neurosciences Cognitives et
Computationnelles, INSERM U960, École Normale Supérieure - PSL Research University, 75005 Paris, France}
\affil[2]{SISSA - Theoretical and Scientific Data Science, Via Bonomea 265, 34136 Trieste, Italy}

\begin{document}
\maketitle

\pagestyle{plain}

\begin{abstract}
Recordings of increasingly large neural populations have revealed that the firing of individual neurons is highly coordinated. When viewed in the space of all possible patterns, the collective activity forms non-linear structures called neural manifolds. Because such structures are observed even at rest or during sleep, an important hypothesis is that activity manifolds may correspond to continuous attractors shaped by recurrent connectivity between neurons. Classical models of recurrent networks have shown that continuous attractors can be generated by specific symmetries in the connectivity.
Although a variety of attractor network models have been studied, general principles linking network connectivity and the geometry of attractors remain to be formulated. Here, we address this question by using group representation theory to formalize the  relationship between the symmetries in recurrent connectivity and  the resulting fixed-point manifolds.
We start by revisiting the classical ring model, a continuous attractor network generating a circular manifold. Interpreting its connectivity  as a circular convolution, we draw a parallel with  feed-forward convolutional neural networks. 
Building on principles of geometric deep learning, we then generalize this architecture to a broad range of symmetries using group representation theory. Specifically, we introduce a new class of \textit{equivariant recurrent neural networks}, where the connectivity is based on group convolution. Using the group Fourier transform, we reduce such networks to low-rank models. This formulation  gives us a low-dimensional non-linear description that can be fully analyzed to determine the symmetry, dimensionality and stability of fixed-point manifolds. Our results underline the importance of stability considerations: for a connectivity with a given symmetry, depending on parameters, several manifolds with different symmetry subgroups can coexist, some stable and others consisting of saddle points.
Our framework unifies a variety of existing models and offers a principled approach to build new types of continuous attractor models of neural systems.
\end{abstract}

\newpage 
\addcontentsline{toc}{section}{Introduction}
\section*{Introduction}

Neural computations rely on the collective dynamics of populations of neurons, which exhibit rich structure in their coordinated activity \citep{Vyas, Chung_2021, so-fusi, Duncker2021}. Large scale recordings in a variety of species and brain areas have identified a prominent type of structure, in which population activity is organized along low-dimensional manifolds in the space of all possible firing patterns \citep{jazayeri-so, Perich2025}. Such neural manifolds have in particular been observed in systems representing and manipulating continuous variables, such as the oculo-motor integrator \citep{Seung-Tank, Seung1996},  the head direction system \citep{Chaudhuri2019,rubin2019}, the entorhinal cortex \citep{Gardner2022, Hermansen2024}, the primary visual cortex \citep{Blumenfeld2006, DiPoppa2025}, motor cortex \citep{Churchland2012, Gallego2017} or working memory \citep{wimmer, Cueva2021}. Because they are often observed even when the animal is at rest or sleeping \citep{Chaudhuri2019, trettel2019}, an important hypothesis posits that these manifolds of activity may correspond to continuous sets of stable states - or attractors  - generated by recurrent connectivity in the underlying networks \citep{fiete_review, engel_review}. To identify the underlying mechanisms, a range of continuous attractor network (CAN) models have been developed for various systems that encode continuous variables \citep{amari, ben1995theory, zhang1996representation, Compte2000,  Burak2009, Stringer2002, CAN, Couey2013, Sagodi2024, Si2014, Xie2002, Romani2010, spalla2021}. In these models, the geometry of the resulting activity manifolds is naturally linked to the symmetries of the continuous variable being encoded, and arises from either explicit or implicit symmetries in the recurrent connectivity \citep{Darshan-Rivkind, Clark2015}. General, unifying principles linking the properties of encoded variables to the symmetry in  network connectivity and the geometry of the resulting attractor manifolds however remain to be formulated.

Remarkably, a set of such unifying principles has been proposed in the adjacent field of artificial neural networks, and assembled within the framework of Geometric Deep Learning \citep{GDL}. Starting from the premise that classical convolutional neural networks (CNNs) extract invariant features from images based on the symmetries of the physical world, GDL developed an abstract formulation of CNNs using the language of group theory. This broad generalization of CNNs then enabled GDL to derive feed-forward network architectures based on the symmetries and invariances present in a vast set of problems beyond classical image classification. These principles were in particular applied to derive spherical CNNs \citep{Cohen2018-jb}, different types of graph neural networks \citep{Scarselli2009-st, monti2017}, or networks operating on arbitrary sets \citep{segol2020}. More recently, these principles have been extended beyond static transformations to recurrent architectures that handle temporal symmetries, addressing scaling equivariance \citep{zuo} and flow equivariance \citep{keller2025} in sequence processing.

Here we build on the theory of Geometric Deep Learning  to formalize the general relationship between the symmetries in connectivity of a recurrent neural network and  the resulting  manifolds of fixed points.
We start by revisiting the classical ring model, a continuous attractor network which generates a circular manifold \citep{ben1995theory}, 
and extract the key underlying ingredients: (i) neurons are indexed by a one-dimensional angular variable;
(ii) the recurrent connectivity has the structure of a circular convolution \citep{zhang}; 
(iii) this confers translational symmetry to the set of attractors of the dynamics; 
(iv)  expanding the convolutional connectivity on Fourier modes leads to a low-dimensional description that allows for an analysis of the stability and dimensionality of fixed-point solutions \citep{mastrogiuseppe2018}.
By formulating the classical ring model in the GDL language, we then generalize it to a broad range of continuous variables and symmetries using the theory of group representations. Specifically, we introduce a new class of \textit{equivariant recurrent neural networks}, where the connectivity is defined via a group convolution \citep{Kondor_Trivedi}, which confers group symmetry to the manifolds of fixed points. Using the concept of group Fourier transform, the network can then be reduced to a low-dimensional model, which preserves the original symmetries and determines the manifold's dimensionality, geometry and stability. As concrete examples, we apply this approach to derive toroidal and spherical RNN models. Analyzing the low-dimensional dynamics, we show that connectivity with
the same  symmetry can generate a variety of  manifolds with different dimensionalities and geometries.
More generally, 
our framework reveals the importance of non-linearity for relating connectivity to emerging manifolds, and offers a new method to generate models of neural systems. \\

\newpage
\section{Recurrent neural networks}

Throughout this manuscript, 
we consider recurrent neural networks (RNNs) consisting of $N$ interacting neurons evolving in discrete time. The activation of neuron $i$ at time $t$ is denoted as $x_{i}^{t} \in \mathbb{R}$, and the dynamics are given by:

\begin{eqnarray}\label{RNN_definition}
{x}_{i}^{t+1}={x}_{i}^{t}+\Delta t (-x_{i}^{t}+F_i^{t}) & i=0,\ldots,N-1,
\end{eqnarray}
where 
\begin{equation} \label{eq:F_i_def}
F_i^{t}=\frac{1}{N}\sum_{j=0}^{N-1} J_{ij} \Phi [x_{j}^{t}]
\end{equation}
is the recurrent input to neuron $i$ at time-step $t$, with $J_{ij} \in \mathbb{R}$  the connection strength from neuron $j$ to neuron $i$, and $\Phi: \mathbb{R} \to \mathbb{R}$ the non-linear function transforming the activation of each unit into its firing rate. 
The general framework is independent of the form of $\Phi$, but for specific examples we use $\Phi[x]=1+tanh[x]$.

\section{Ring models as convolutional RNNs}\label{ring_section}

In this section, we revisit ring models, the most fundamental type of continuous attractor networks \citep{amari,ben1995theory,zhang1996representation}. We formulate this class of models within the language of geometric deep learning \citep{GDL}, and  outline their relation with convolutional neural networks (CNNs). We outline how the symmetry properties of CNNs - in particular their equivariance \citep{Cohen2016-rd} -  induce structure in the sets of fixed points in ring models that we denote as {\em fixed point manifolds}. Expanding the recurrent connectivity on Fourier modes, we then introduce {\em low-rank ring models}, and show how this subset of ring models leads to an explicit parametrization of fixed point manifolds.

\subsection{Convolutional structure of ring models}

\begin{figure}[!ht]
    \centering
    \includegraphics[width=15cm]{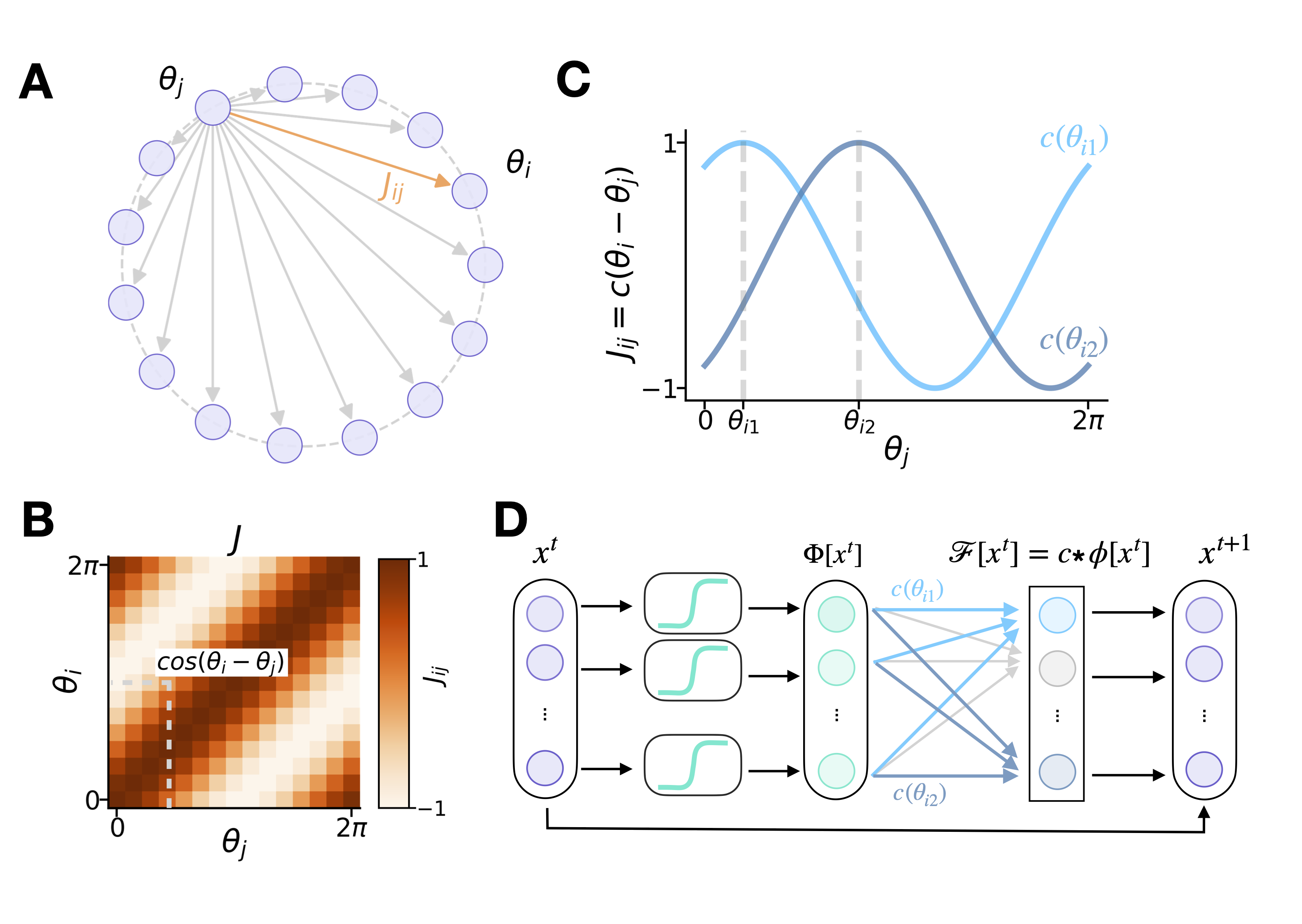} 
    \caption{A) Ring structure of neurons with recurrent connectivity where the weight $J_{ij}$ between neurons $i$ and $j$ only depends on their angular distance.
B) Connectivity matrix showing cosine-shaped weights as a function of neuron indices.
C) Connectivity weights of selected neurons share the same connectivity profile, but shifted according to their preferred angle. D) Feed-forward implementation of recurrent dynamics. The recurrent input is computed by applying point-wise non-linearity $\Phi$ to input $x$, then convolving with shifted connectivity profiles $c(\theta_i)$ corresponding to each neuron's preferred angle.}
    \label{fig1}
\end{figure}

Ring models are a subclass of recurrent neural networks in which each neuron $i$  is  labeled by an angle $\theta_i=\frac{2\pi i}{N}$ for $i=0,\ldots,N-1$, so that the network can be visualised as lying on a ring (Fig.~\ref{fig1} A). Following the language of geometric deep learning \citep{GDL}, here we formalize the collective activation in the network as a function of neural labels.
Writing $\Omega=\{\theta_i=\frac{2\pi i}{N}|i=0,\ldots,N-1\}$, we represent the  network activity  at time $t$  as a function  $x^t: \Omega \to \mathbb{R}$ such that the activation of neuron $i$ is obtained by applying $x^t$ to its angular label $\theta_i$:

\begin{eqnarray}\nonumber
x^t:\theta \mapsto x^t(\theta) &\mathrm{such\, that}&  x^t(\theta_i)=x_{i}^{t}.
\end{eqnarray}
The firing rate and the collective recurrent input in the network  can be similarly expressed as  functions $\Phi^t$ and  $F^t: \Omega \to \mathbb{R}$ such that, for neuron $i$ at time $t$, the firing rate is given by $\Phi^t(\theta_i)=\Phi[x^t(\theta_i)]=\Phi[x_{i}^{t}]$ and the recurrent input is 
 is $F^t(\theta_i)=F_i^{t}$ defined in Eq.~\eqref{eq:F_i_def}. Since $\Phi^t$ and $F^t$ depend intrinsically on the neural activity at time $t$, they can be expressed in terms of time-independent operators $\mathcal{P}$ and $\mathcal{F}$ applied to $x^t$: 
\begin{eqnarray}
\mathcal{P}:x^t\mapsto \Phi^t & \Phi^t=\mathcal{P}[x^t]\\
\mathcal{F}:x^t\mapsto F^t & F^t=\mathcal{F}[x^t].  \label{eq:F-def}
\end{eqnarray}
Denoting as $L_{\mathbb{R}(\Omega)}$ the space of square-integrable  functions $\{f: \Omega \to \mathbb{R}\}$, we have $x^t, F^t \in L_{\mathbb{R}(\Omega)}$ and $\mathcal{P},\mathcal{F}:L_{\mathbb{R}(\Omega)}\to L_{\mathbb{R}(\Omega)}$. In the following, we identify the set of $2\pi-$periodic functions $\mathbb{R}\to \mathbb{R}$ with the set $L_{\mathbb{R}(\Omega)}$ of functions $\Omega\to \mathbb{R}$.  
Because the functional $\mathcal{P}$ is obtained by applying $\Phi$ "pointwise" to $x^t$ (i.e. the value of $\Phi^t$ at $\theta_i$ depends only on the value of $x^t$ at $\theta_i$), in the following we  write  $\Phi^t=\Phi[x^t]$.

With these notations, the network dynamics in Eq.~\eqref{RNN_definition} can therefore be formulated as a temporal evolution within $L_{\mathbb{R}(\Omega)}$:

\begin{equation}\label{RNN_definition_abstract}
{x}^{t+1}={x}^{t}+\Delta t (-x^{t}+\mathcal{F}\left[x^t \right]).
\end{equation}

A key ingredient of  ring models is that the connection strength $J_{ij}$ from neuron $j$ to neuron $i$  depends only on the difference $\theta_i-\theta_j$ between their angular variables:
\begin{equation}\label{eq:ring-connectivity}
J_{ij}=c(\theta_i-\theta_j),
\end{equation}
where $c: \mathbb{R} \to \mathbb{R}$ is a $2\pi-$periodic function, which we call the {\em connectivity kernel}. Classical examples for $c$ are a shifted cosine  \citep{ben1995theory} or a circular Gaussian  \citep{amari}. 
We moreover assume that $c$ is  even, which is equivalent to having a symmetric connectivity matrix and ensures the existence of fixed points.
 
The specific connectivity structure in Eq.~\eqref{eq:ring-connectivity} confers a periodicity and symmetry to ring models: each neuron $i$ in the network receives connections from other neurons through the same set of weights,  translated by an angle $\theta_i$ (Fig.~\ref{fig1} C). Equivalently, the connectivity matrix $J$ is circulant, meaning that each row is obtained by shifting the previous row (Fig.~\ref{fig1} B).

The recurrent input $F^t_i$ to neuron $i$ in Eq.~\eqref{eq:F_i_def} can therefore be written as  

\begin{eqnarray}
    F^t_i &=&\frac{1}{N}\sum_{j=0}^{N-1} J_{ij} \Phi [x_{j}^{t}]\\ \label{circular_conv}
    &=&\frac{1}{N}\sum_{\theta' \in \Omega} c(\theta_i-\theta') \Phi [x^t(\theta')]  \label{eq:F-def-conv}\\ 
    &=&\frac{1}{N}\sum_{\theta' \in \Omega} c(\theta_i-\theta') \Phi^t (\theta')\nonumber\\ 
    &=&F^t(\theta_i)\nonumber\\
    &=&\mathcal{F}[x^t](\theta_i) \nonumber
\end{eqnarray}
so that $\mathcal{F}[x^t]$ is given by the \textit{convolution} between the kernel $c$ and the firing rate $\Phi^t$ of the network:
\begin{eqnarray} \label{ring_convolution}
\mathcal{F}[x^t]=c * \Phi^t.
\end{eqnarray}
The convolution between two functions $f_1,f_2 \in L_{\mathbb{R}(\Omega)}$, is a function $f_1*f_2: \Omega \to \mathbb{R}$  defined as
\begin{equation}
 (f_1*f_2)({\theta})=\frac{1}{N}\sum_{\theta' \in \Omega} f_1({\theta}-\theta') f_2(\theta'). \nonumber
\end{equation} 
In Eq.~\eqref{ring_convolution}, $f_1=c$ and $f_2=\Phi^t=\Phi[x^t]$.

The recurrent input $F^t$ at time-step $t$ is therefore obtained by applying to the activation $x^t$ the point-wise non-linear function $\Phi$, followed by a convolution with the kernel $c$.  This is equivalent to applying to the set of activities $\{x^{t}_i\}_{i=0,\ldots N-1}$ a single-layer feed-forward convolutional neural network, in which the order of the convolution and point-wise non-linearity are reversed with respect to standard CNNs. The output of that CNN is the set of recurrent inputs $\{F^{t}_i\}_{i=0,\ldots N-1}$ (Fig.~\ref{fig1} D). 

In the limit $N\to \infty$, $x^t, \Phi^t, c$ and $F^t$ become  functions on the continuous set $\Omega=[0,2 \pi)$. In that  limit, the discrete sum over $\Omega$ in Eq.~\eqref{circular_conv} is replaced by an integral \citep{gerstner2014neuronal}.

\begin{equation} \label{continuous_convolution}
F^t(\theta)=\frac{1}{2 \pi} \int_{\theta'\in \Omega} c(\theta-{\theta}^{\prime}) \Phi[x^t({\theta}^{\prime})] d {\theta}^{\prime}.
\end{equation}
For $N\to \infty$, the recurrent input is obtained by applying to $x^t$ the point-wise non-linearity $\Phi$, followed by a continuous convolution with the kernel $c$. Following the formalism of geometric deep learning \citep{GDL}, this sequence of operations can be interpreted as a continuous-valued CNN \citep{monti2017}.
Throughout this study we will assume that this limit is well-defined, and we will interchangeably consider domains $\Omega$ corresponding to finite and infinite $N$.

Altogether, the recurrent input in ring models  has the form of a single layer  CNN that can be either discrete or continuous. Ring models are therefore closely related to feed-forward convolutional neural networks and form a class of models that we denote as convolutional RNNs. In particular, ring models directly inherit the symmetry properties of convolutional neural networks.

\subsection{Equivariance under angular translations}
    
A central property of feed-forward convolutional networks is that a translation of their input leads to an equivalent translation of their output. This property is more generally called {\em equivariance} \citep{Cohen2016-rd}. Since the recurrent input $\mathcal{F}$ in ring models has the form of a CNN, it directly inherits the equivariance property. Using the language of geometric deep-learning \citep{GDL}, here we  formulate more formally the equivariance of the recurrent input $\mathcal{F}$ under translations of the activity in the network. 

The operator $\mathcal{F}$ defined in Eqs.~\eqref{eq:F-def} and \eqref{eq:F-def-conv}  takes as input the activation $x^t$ of the network, and outputs the recurrent input $F^t$. As  $x^t$ is a function of an angle $\theta \in \Omega$, an angular translation by an angle $\phi \in \Omega$ is obtained by mapping any  $\theta \in \Omega$ to $\theta + \phi$ modulo $2\pi$. More formally, we represent this angular translation  by the function $g_{\phi}:\Omega \to \Omega$ defined  by
\begin{equation}\nonumber
    g_{\phi}:\theta \mapsto \theta '\quad \quad \theta'=\theta +\phi  \,\,\mathrm{mod}(2\pi).
\end{equation}

For later reference, it is important to note that the set of all angular translations $G=\{g_{\phi}|\phi \in \Omega \}$ has some remarkable properties, which confer to it the structure of a {\em group} \citep{Kosmann-Schwarzbach2009-uj}. In particular, composing any pair of angular translations $g_{\phi_1}, g_{\phi_2}$ yields an angular translations by the angle $\phi_1+\phi_2$:
\begin{equation} \label{composing_translations}
g_{\phi_1} g_{\phi_2}=g_{\phi_1 +\phi_2}.
\end{equation}
Moreover, any translation $g_{\phi}$ can be inverted by a translation $g_{-\phi}$, so that
\begin{equation} \label{inverting_translations}
g_{\phi}^{-1}=g_{-\phi}.
\end{equation}

Angular translations on $\Omega$ naturally induce transformations of functions in $L_{\mathbb{R}(\Omega)}$.
Indeed any translation $g_{\phi}$ maps any function $f \in L_{\mathbb{R}(\Omega)}$ to a translated function $S_{\phi}[f]:\Omega \to \mathbb{R}$ defined by 
\begin{eqnarray} \label{translation_action}
S_{\phi}[ f]:\theta \mapsto f(g_{-\phi}(\theta))=f(\theta')\,\,\,\mathrm{with}\,\,\, \theta'=\theta - \phi  \,\,\mathrm{mod}(2\pi).
\end{eqnarray}
The operator $S_{\phi}:L_{\mathbb{R}(\Omega)}\to L_{\mathbb{R}(\Omega)}$ is more generally called the {\em action} of the translation  $g_{\phi}$ on functions in $L_{\mathbb{R}(\Omega)}$. In particular, applying $S_{\phi}$ to  $x^t$ yields the network activation profile translated by ${\phi}$, while applying $S_{\phi}$ to the kernel $c$ translates the connectivity by ${\phi}$ and therefore yields the profile of incoming connection strengths to the neuron indexed by ${\phi}$ (Eq.~\eqref{eq:ring-connectivity}).

With these definitions, we can formulate more formally
the equivariance property of  $\mathcal{F}$, which states that a translation of the activation profile $x^t$ by $\phi$ leads to a translation of the recurrent input $F^t$ by the same angle $\phi$. Using the operator $S_{\phi}$ to express an angular translation by $\phi$, the equivariance of $\mathcal{F}$ corresponds to the fact that the operators $\mathcal{F}$ and $S_{\phi}$ commute for any $\phi$:

\begin{eqnarray}\label{ring_equivariance}
\mathcal{F} S_{\phi} = S_{\phi} \mathcal{F} \,\,\,\,\,\,\forall \phi \in\Omega.
\end{eqnarray}
This property stems directly from the fact that $\mathcal{F}$ has the form of a convolution, i.e. an angular translation of the kernel $c$ (Eq.~\eqref{continuous_convolution}).
Indeed, $\forall x \in L_{\mathbb{R}(\Omega)}$ and $\forall \theta, \phi \in \Omega$, we have
\begin{eqnarray}
    \mathcal{F} \left[S_{\phi}\left[x\right]\right](\theta) &=& c * \Phi\left[S_{\phi}\left[x\right]\right] (\theta) \nonumber \\
    &=&\frac{1}{N} \sum_{\theta'\in \Omega} c(\theta-\theta') \Phi \left[S_{\phi}\left[x\right](\theta')\right] \nonumber\\
    &=&\frac{1}{N} \sum_{\theta'\in \Omega} c(\theta-\theta') \Phi \left[x(\theta'')\right] \,\,\,\mathrm{with}\,\,\, \theta''=\theta' - \phi  \,\,\mathrm{mod}(2\pi) \nonumber\\
     &=&\frac{1}{N} \sum_{\theta''\in \Omega} c\left((\theta-\theta'')-\phi\right) \Phi \left[x(\theta'')\right] \nonumber\\
     &=&\frac{1}{N} \sum_{\theta''\in \Omega} S_{\phi}\left[c\right]\left(\theta-\theta''\right) \Phi \left[x(\theta'')\right] \nonumber\\
     &=&S_{\phi}\left[\mathcal{F}\left[x\right]\right](\theta).\nonumber
\end{eqnarray}
Here we used the definition of $\mathcal{F}$ (Eq.~\eqref{eq:F-def-conv}), the definition of $S_{\phi}$ (Eq.~\eqref{translation_action}), and the fact that $c$ is $2\pi$-periodic.

\subsection{Structure of fixed points}

The equivariance of the recurrent input has direct consequences on the structure of fixed points in the ring model. Indeed
from Eqs.~\eqref{RNN_definition} and \eqref{RNN_definition_abstract}, an activation profile ${x^*} \in L_{\mathbb{R}(\Omega)}$ is a fixed point of the dynamics if it obeys

\begin{eqnarray}\label{fixed_points_equationF}
{x^*}(\theta) = \mathcal{F}[x^*](\theta) & \forall \theta \in \Omega.
\end{eqnarray}

The equivariance of $\mathcal{F}$ directly implies that, if ${x^*}$ is a fixed point, then the translated activation profile $S_{\phi}{x}^*$ is also a fixed point. Indeed
\begin{eqnarray}
\mathcal{F}(S_{\phi}{x}^*)&=&S_{\phi}\mathcal{F}({x}^*) \nonumber\\
&=&S_{\phi}{x}^*, \nonumber
\end{eqnarray}
where the first equality follows from the equivariance property of $\mathcal{F}$ (Eq.~\eqref{ring_equivariance}), and the second one follows from the fact that ${x}^*$ is a fixed point (Eq.~\eqref{fixed_points_equationF}).  Intuitively, because the interactions depend only on the difference between angular labels, shifting the activation profile ${x^*}$  by an angle $\phi$  leads to  an activation profile $S_{\phi}{x}^*$  that is equivalently a fixed point. Unless the activation profile ${x^*}$ is uniform over $\Omega$, $S_{\phi}{x}^*$ will correspond to a distinct fixed point. This means that if a given bump  of activations is a fixed point of the dynamics, any shift of that bump will also be a fixed point of the dynamics.

The convolutional structure of  ring models therefore induces the existence of sets of fixed points, within which different fixed points are related by angular translations.
Each such set is obtained by starting from a fixed point ${x}^*$ and applying  shift operators $S_{\phi}$ for $\phi \in \Omega$. Denoting ${x}^*$ as the seed, the corresponding set of fixed points $\mathcal{M}_{{x}^*}$ is defined by
\begin{equation} \label{eq:ring_manifold}
    \mathcal{M}_{{x}^*}=\{ S_{\phi}{x}^*|\phi \in \Omega\}.
\end{equation}
If ${x}^*$ is non-uniform over $\Omega$, $\mathcal{M}_{{x}^*}$ consists of more than one fixed point. For finite $N$, $\mathcal{M}_{{x}^*}$ is in general a finite set because $\Omega$ is finite (but see \cite{Noorman2024-dm}). For $N\to \infty$, $\mathcal{M}_{{x}^*}$ becomes a manifold of intrinsic dimension 1, parametrized by the angular parameter $\phi \in [0,2\pi)$. By abuse of language, we will denote $\mathcal{M}_{{x}^*}$ as a {\em fixed-point manifold}  for both finite and infinite networks.

\subsection{Fourier transform of convolutional RNNs}
The equivariance of ring models can give rise to  fixed-point manifolds parametrized by an angle $\phi \in \Omega$. The specific shape of these manifolds  in general depends on the connectivity kernel $c$ and the seed fixed point ${x}^*$. To characterize in more detail the shape of the  manifolds, we will turn to a restricted class of ring models that we denote as {\em low-rank convolutional RNNs}. To define this class of models, we start by introducing the Fourier representation of general ring models.

Any function $f:\Omega \to \mathbb{R}$ can be expressed in terms of its Fourier components $\hat f_r$ for $r=0,1,\ldots$. Distinguishing even and odd indices as $r=2k$ and $r=2k-1$ for $k=0,1,\ldots$, we write the Fourier components as  :
\begin{eqnarray}\label{fourier_components}
\hat f_{0}&=& \frac{1}{2\pi} \int_{\Omega}  f(\theta)d\theta, \\
\hat f_{2k-1}&=&\frac{1}{2\pi} \int_{\Omega}  f(\theta) \cos(k \theta)d\theta, \\
\hat f_{2k}&=&\frac{1}{2\pi} \int_{\Omega}  f(\theta) \sin(k \theta)d\theta.
\end{eqnarray}
Denoting as ${\hat c}_{r}, {\kappa}_{r}, {\hat \Phi}^t_{r}, {\hat F}^t_{r}$ the Fourier components of the 
 connectivity kernel $c$, the activation $x^t$, the firing rate $\Phi(x^t)$ and the recurrent input $F^t$ at time $t$, we next express the dynamics in the ring network in terms of these quantities. 
 
Since we assumed that the connectivity kernel $c$ is even,  only cosine terms have non-zero coefficients, i.e. ${\hat c}_{r}=0$ for $r$ even.
 Since $F^t$ is a convolution between $c$ and $\Phi^t$ (Eq.~\eqref{continuous_convolution}), from the convolution theorem \citep{bracewell} we have

\begin{eqnarray}\label{F_fourier_convolution}
    {\hat F}_{2k-1}^t&=&{\hat c}_{2k-1}{\hat \Phi}_{2k-1}^t \label{f_fourier_cosine} \\
    {\hat F}_{2k}^t&=&{\hat c}_{2k-1}{\hat \Phi}_{2k}^t.\label{f_fourier_sine}
\end{eqnarray}

The RNN dynamics can then be formulated in terms of the evolution of Fourier components $\kappa_{r}^t$ of $x^t$. Taking the Fourier transform of Eq.~\eqref{RNN_definition}, these dynamics read
\begin{eqnarray}\label{kappa_fourier}
\kappa_{r}^{t+1}=\kappa_{r}^{t}+\Delta t (-\kappa_{r}^{t}+{\hat F}_{r}^t)& r=0,1,\ldots
\end{eqnarray}

In particular, the Fourier components $\kappa_{r}$ of any fixed point solution $x^*$ obey

\begin{eqnarray}\label{fp_fourier_ring}
\kappa_{r}&=&{\hat F}_{r} \,\,\,\, r=0,1,\ldots
\end{eqnarray}
where ${\hat F}_{r}$ depends implicitly on $x^*$ (Eqs.~\eqref{f_fourier_cosine},\eqref{f_fourier_sine}) and therefore on all its Fourier components. Eq.~\eqref{fp_fourier_ring} is therefore a system of an infinite number of coupled equations and hence in general intractable. We next introduce a restricted class of models for which this system becomes finite and tractable.

\subsection{Low-rank convolutional RNNs}

A particular sub-class of ring models is obtained by restricting the connectivity kernel $c$ to have only a finite number $K+1$ of non-zero Fourier terms in Eq.~\eqref{fourier_components}. This is the case, for instance, in \cite{ben1995theory} where only the zero-th and first term are non-zero, so that $K=1$.   We call this class of networks {\em low-rank convolutional RNNs} because their connectivity matrices are of rank $R=2K+1$ \citep{mastrogiuseppe2018} (see Appendix \ref{appendix_discretized}). 

For low-rank  convolutional RNNs, ${\hat c}_{r}=0$ for $r>2K$, and therefore from  at steady-state $\kappa_{r}=0$ for $r>2K$. Eq.~\eqref{fp_fourier_ring} then reduces to a system of $R$ equations for the $R$ first Fourier components $\kappa_{r}$ which we group in a $R-$dimensional vector $\underline{\kappa}$. To express Eq.~\eqref{fp_fourier_ring} explicitly as function of $\kappa_{r}$, $r=0. \ldots R-1$, we write the inverse Fourier transform for the fixed-point activation $x^*$:
\begin{eqnarray} \label{fp_inverse_fourier}
    x^*(\theta)=\kappa_0+\sum_{k=1}^K  \left( \kappa_{2k-1} \cos(k \theta)+\kappa_{2k} \sin(k \theta)\right).
\end{eqnarray}

Using Eq.~\eqref{f_fourier_cosine},  for $r=2k-1$ with $k=1\ldots R$ we have

\begin{eqnarray}
    {\hat F}_{2k-1} (\underline{\kappa})&=&{\hat c}_{2k-1}\frac{1}{2\pi} \int  \cos(k\theta) {\Phi}\left[x^*(\theta)\right] d\theta \label{f_kappa_int_1_1} \nonumber \\
&=& \frac{{\hat c}_{2k-1}}{2\pi}  \int  \cos(k\theta) {\Phi}\left[\kappa_0+\sum_{k'=1}^K  \kappa_{2k'-1} \cos(k' \theta)+\kappa_{2k'}  \sin(k' \theta)\right] d\theta. \label{f_kappa_int_1}
\end{eqnarray}
In Eq.~\eqref{f_kappa_int_1_1} we used the definition of the Fourier transform of $\Phi^t$ at the steady state (Eq.~\eqref{f_fourier_cosine}).  In Eq.~\eqref{f_kappa_int_1} we replaced $x^*$ with its inverse Fourier transform given in Eq.~\eqref{fp_inverse_fourier}.

Similarly, from Eq.~\eqref{f_fourier_cosine},  for $r=2k$ with $k=1\ldots R$ we have
\begin{eqnarray}
    {\hat F}_{2k}(\underline{\kappa})&=&{\hat c}_{2k-1}\frac{1}{2\pi} \int \sin(k\theta) {\Phi}\left[x^*(\theta)\right] d\theta \nonumber \\
&=&\frac{{\hat c}_{2k-1}}{2\pi}  \int \sin(k\theta) {\Phi}\left[\kappa_0 +\sum_{k'=1}^K  \kappa_{2k'-1}  \cos(k' \theta)+\kappa_{2k'}  \sin(k' \theta)\right] d\theta .\label{f_kappa_int_2}
\end{eqnarray}

Inserting Eqs.~\eqref{f_kappa_int_1} and ~\eqref{f_kappa_int_2} into Eq.~\eqref{fp_fourier_ring} yields a set of $R$ coupled non-linear equations for $\kappa_{r}$, $r=0.\ldots R-1$.
For special choices of the non-linear function $\Phi$, the integrals in Eqs.~\eqref{f_kappa_int_1} and \eqref{f_kappa_int_2} can be computed analytically, leading to explicit formulas for the functions ${\hat F}_r$ \citep{ben1995theory}.

\subsection{Fixed-point manifolds in low-rank convolutional RNNs}

Restricting the ring model to be of low rank allows us to characterize  more explicitly  the structure of the fixed point manifolds generated by applying the angular translation operator $S_{\phi}$ to a non-trivial fixed point $x^* $ (Eq.~\eqref{eq:ring_manifold}).

The angular translation $S_{\phi}$ of $x^* $  directly induces a linear transformation of its Fourier components.

Directly applying $S_{\phi}$ to the inverse Fourier transform of  $x^* $, we have

\begin{eqnarray} 
    S_{\phi}[x^*] (\theta)&=&x^*(\theta-\phi),\nonumber \\
    &=& {\kappa_0 }+\sum_{k=1}^{K}\left[{\kappa}_{2k-1}  \cos\left(k (\theta-\phi)\right)+{\kappa}_{2k}  \sin \left(k (\theta-\phi)\right)\right],\label{eq:ring-rotation1}\\
    &=& {\kappa_0 }+\sum_{k=1}^{K}\left[ \left(\rho_{k_{1,1}}(\phi)\kappa_{2k-1} +\rho_{k_{1,2}}(\phi)\kappa_{2k}  \right) \cos(k\theta)
    +\left(\rho_{k_{2,1}}(\phi)\kappa_{2k-1} +\rho_{k_{2,2}}(\phi)\kappa_{2k}  \right) \sin(k \theta) \right].\label{eq:ring-rotation2}
\end{eqnarray}
From Eq.~\eqref{eq:ring-rotation1} to Eq.~\eqref{eq:ring-rotation2}, we applied elementary trigonometric identities and introduced $\rho_{k}(\phi) \in \mathbb{R}^{2\times 2}$, the $2\times 2$ rotation matrix with angle $k \phi$

\begin{eqnarray}\label{2d_rotation_matrices}
\rho_{k}(\phi)=\begin{pmatrix}
\cos(k\phi) & -\sin(k\phi)  \\
 \sin(k\phi)& \cos(k\phi)
\end{pmatrix}.
\end{eqnarray}

Grouping the Fourier components of $x^*$ into a vector $\underline{\kappa} \in \mathbb{R}^R$, the transformation $\underline{\kappa}$ under an angular translation is therefore determined by a
 block-diagonal matrix $R^{K}(\phi)$ with $K+1$ blocks:

\begin{eqnarray}\nonumber
R^{K}(\phi)=\begin{pmatrix}
\rho_{0} &  & \\
 &\ddots  &\\
 & & \rho_{K}
\end{pmatrix}
\end{eqnarray}
The first block corresponds to the scalar identity $\rho_{0}=1$. Each subsequent block is a $2\times 2$ rotation matrix with angle $k \phi$ for $k=1\ldots K$.

Any solution ${\underline{\kappa} }$ of Eq.~\eqref{f_kappa_int_1_1}-\eqref{f_kappa_int_2}  therefore leads to a manifold of fixed points in Fourier coordinates parametrized by an angle ${\phi}$ and given by:

\begin{eqnarray}\nonumber
{\underline{\kappa}}(\phi)=R^{K}(\phi){\underline{\kappa} }.
\end{eqnarray}

\subsection{The geometry of the fixed-point manifold}

A geometric description of the manifold of fixed points can be obtained by interpreting the inverse Fourier transform in Eq.~\eqref{fp_inverse_fourier} as a linear parametrization of the manifold, where Fourier components provide coordinates of points on the manifold, in a basis formed by Fourier functions. In finite, $N-$dimensional networks, the Fourier functions correspond to a set of orthogonal directions in the $N-$dimensional activation space. For continuous networks, Fourier functions provide a basis in the infinite-dimensional space of square-integrable functions.
In  low-rank networks, at most $R$ of the Fourier components are non-zero, so that the fixed-point manifold is embedded in a finite-dimensional subspace both for finite-dimensional and continuous networks. 

The most minimal non-trivial manifold is generated by a fixed point $\underline{\kappa} $ with zero values everywhere except on a single Fourier component with $k>0$. If Eq.~\eqref{fp_fourier_ring} admits such a solution, the corresponding manifold is a ring embedded in two dimensions (Fig.~\ref{fig:ring2d} B):

\begin{eqnarray} \nonumber
    x^*(\theta)= \kappa_{2k-1} \cos(k \theta)+\kappa_{2k} \sin(k \theta).
\end{eqnarray}

More generally, if a fixed point $\kappa $ has non-zero entries on $P$ Fourier components, the manifold will be a twisted ring embedded in $2P$ dimensions (Fig.~\ref{fig:ring4d} D).

\newpage 
\section{Equivariant Recurrent Neural Networks}\label{group_section}

The formulation of the ring model in the language of geometric deep learning makes clear that the properties of the fixed-point manifolds stem from the following underlying structure: (i) neurons are indexed by elements of a one-dimensional domain $\Omega$; (ii) any element of $\Omega$ can be mapped onto any other one using an angular translation;  (iii) the recurrent input is equivariant under the set $G$ of all angular translations.

The angular translations in the ring model are a particular type of transformation associated with the ring-like domain $\Omega$. Following the blueprint of geometric deep learning \citep{GDL}, the structure of the ring model can be generalized to a broader class of transformations acting on more general domains $\Omega$. This gives rise to a  family of models which we call {\em equivariant RNNs (eqRNNs)}. This generalization relies on the mathematical formalism of group theory \citep{Kosmann-Schwarzbach2009-uj}. In this section, we introduce  equivariant RNNs using  group theory in an informal manner. A summary of key concepts in group theory is provided in  Appendix \ref{appendix_group_theory}. Specific examples of eqRNNs are developed in the next section.

 We start by defining a class of RNNs in which units are indexed by the elements of domain $\Omega$. Our central assumption is that this domain is associated with  transformations that map any two elements of $\Omega$ onto each other. The set of these transformations forms a
  finite or compact group $G$. Using a generalized concept of  convolution, we define equivariant RNNs as the set of RNNs in which the recurrent input  is equivariant under the action of $G$. We show that as a consequence, eqRNNs give rise to manifolds of fixed points generated by the action of  $G$. We then use the concept of group Fourier transform to introduce low-rank eqRNNs. Similarly to low-rank ring models,  low-rank eqRNNs lead to explicit parametrizations of fixed-point manifolds based on group representation theory.

\subsection{Equivariant recurrent neural networks}

We consider a population of neurons indexed by elements $u$ of a domain $\Omega$, that can be either discrete or compact (i.e. continuous, closed and bounded). We assume that   $\Omega$ is associated with a set of transformations $G$ that form a finite or compact group. Elements of $G$ are  transformations  $g: \Omega \to \Omega$ equipped with a composition operation which obeys group axioms
(Appendix \ref{appendix_group_theory}). The transformations $g$ play a role analogous to angular translations in the ring model, and group axioms correspond to a generalization of properties outlined in Eqs.~\eqref{composing_translations}-\eqref{inverting_translations}.

Our key assumption is that $\Omega$ is a {\em homogeneous space} with respect to $G$. This means that for any fixed $u_0 \in \Omega$, every other element $w \in \Omega$ can be reached by applying a given $g_w \in G$ such that $g_w u_0 = w$. This defines a mapping from  $G$ to $\Omega$ that allows us to index elements of $\Omega$ by elements of $G$.  In the ring model, any angular translation in $G$ corresponds to a unique angle in $\Omega$, so that $\Omega$ and $G$ are isomorphic, which we write as $\Omega=G$. Importantly, in general this needs not be the case, as the mapping from $G$ to $\Omega$ is not necessarily
 bijective. In the following, we will systematically distinguish the case where $\Omega$ and $G$ are isomorphic and the case where they are not, as the second one is technically more complex than the first.

We next define an RNN with units indexed by elements of $\Omega$.
As for the ring model, the  activation  in the network  is defined by a function $x^t:\Omega  \to \mathbb{R}$  that associates to each neuron $u \in \Omega$,  its activation $x^t(u)$ at time $t$. Analogously, the firing rate  and  the recurrent input  at time $t$ are defined by two functions $\Phi^t$ and $F^t:\Omega  \to \mathbb{R}$. In accordance with the structure of interactions in RNNs (Eq.~\eqref{eq:F_i_def}), we assume that the recurrent input has the form of an abstract single-layer feed-forward network applied to the activity in the previous time-step, and is therefore defined as
$F_t=\mathcal{F}\left[x^t \right]$ where $\mathcal{F}:L_{\mathbb{R}(\Omega)}\to L_{\mathbb{R}(\Omega)}$ is a time-independent operator that consists of a point-wise non-linearity followed by a linear operation. The dynamics of $x^t$ obey  Eq.~\eqref{RNN_definition_abstract}.

Since the domain $\Omega$ is associated with the group $G$, any transformation $g_w \in G$  induces a transformation of any function $f \in L_{\mathbb{R}(\Omega)}$. By analogy with Eq.~\eqref{translation_action}, we denote the transformed function as $S_{g_w}[f]$, and  define it as 
\begin{equation} \label{G_action}
    S_{g}[f](u)=f\left({g}^{-1}(u)\right) \quad \forall u \in \Omega.
\end{equation}
 The corresponding operator $S_{g_w}:L_{\mathbb{R}(\Omega)} \to L_{\mathbb{R}(\Omega)}$ is called the action of $g_w$ on $L_{\mathbb{R}(\Omega)}$. Applying $S_{g_w}$ to $x^t$ or $F^t$ is analogous to an angular translation of the activations or the recurrent inputs in the ring model.

A central property of the ring model is that the operator $\mathcal{F}$ defining the recurrent input is equivariant under angular translations, i.e. commutes with their action  (Eq.~\eqref{ring_equivariance}). Our goal here is to define a family of recurrent input operators $\mathcal{F}$ that extends this property to the more general case of a domain $\Omega$ that is homogeneous with respect to the group $G$.

The equivariance of the recurrent inputs in the ring model is a direct consequence of the convolutional structure  of $\mathcal{F}$ with respect to translations on $\Omega=[ 0,  2 \pi )$ (Eq.~\eqref{ring_convolution}). To generalize this property, we rely on the results of \cite{Kondor_Trivedi}, who showed that any single-layer feed-forward network is equivariant under the action of a group $G$ if and only if its linear operation consists of a generalized form of convolution, based on the concept of  {\em group convolution}. 

Introducing a kernel function $c \in L_{\mathbb{R}(\Omega)}$, by analogy with Eq.~\eqref{ring_convolution} we will define the recurrent input operator $\mathcal{F}$  as the generalized convolution on $\Omega$ between $c$ and the firing rate $\Phi[x]$:

\begin{eqnarray}\label{rec_input_G}
\mathcal{F}[x]=c *_{\Omega} \Phi[x].
\end{eqnarray}

The generalized convolution on $\Omega$ in Eq.~\eqref{rec_input_G} is based on the concept of the group convolution on $G$, which is defined for functions acting on the elements of $G$ (rather than $\Omega$). For two functions  $f^1,f^2:G\to \mathbb{R}$ the $G$-convolution is a function $G\to \mathbb{R}$ defined by \citep{Kondor_Trivedi}
\begin{eqnarray}
\label{group_conv0}
    (f_1 *_G f_2)(g)=\sum_{h \in G} f_1(g^{-1}h)  f_2(h)
\end{eqnarray}
for a discrete group and
\begin{eqnarray}
\label{group_conv_cont0}
    (f_1 *_G f_2)(g)=\int_{G} f_1(g^{-1}h)  f_2(h) d\mu(h)
\end{eqnarray}
for a continuous group, where $\mu$ is the Haar measure on $G$ \citep{Kondor_Trivedi}.

In the case where $G$ and $\Omega$ are isomorphic, every element $u$ in $\Omega$ corresponds to a unique element $g_u$ in $G$. The group convolution in Eq.~\eqref{group_conv0} then directly defines a convolution of two functions $f^1,f^2:\Omega\to \mathbb{R}$ that we use in Eq.~\eqref{rec_input_G}. 

The recurrent input $F^t(u)$  to unit $u$ at time $t$ is therefore given by
\begin{eqnarray}
    F^t(u)&=&\left(c *_G \Phi[x]\right)(g_u) \nonumber\\
    &=&\sum_{v \in \Omega} c(g_u^{-1}v) \Phi \left[x^t(v)\right] \label{group_conv}
\end{eqnarray}
for discrete $\Omega$ and $G$, while in the continuous case
\begin{eqnarray}
\label{group_conv_cont}
    F^t(u)&=&\int_{v \in \Omega} c(g_u^{-1}v) \Phi \left[x^t(v)\right]d\mu(v).
\end{eqnarray}

If $G$ and $\Omega$ are not isomorphic, the recurrent input cannot be defined directly in terms of the standard group convolution. Indeed, the group convolution in Eq.~\eqref{group_conv0} is defined on functions on $G$, while the recurrent input acts on functions on $\Omega$. In such a situation, we follow \cite{Kondor_Trivedi}, who define a generalized convolution for functions on
 a domain $\Omega$ that is homogeneous with respect to $G$. Using the fact that $\Omega$  can  be written as a quotient $G/H$ between $G$ and another group $H$, \cite{Kondor_Trivedi} show that their generalized form of convolution inherits the key properties of the standard group convolution.  We provide a summary of this construction in Appendix \ref{app:group_convolution}, and a concrete example in Sec.~\ref{sec:sphere}. We in particular show that defining the recurrent input in terms of this generalized convolution reduces precisely to Eq.~\eqref{group_conv}, the only difference being that $g_u$ is replaced by its representative on  the coset $gH$.

The definition of the recurrent input $\mathcal{F}$ in terms of the group convolution (Eq.~\eqref{rec_input_G})  directly ensures that $\mathcal{F}$ satisfies the equivariance property with respect to the group action defined in Eq.~\eqref{G_action}:

\begin{eqnarray}\label{G_equivariance}
\mathcal{F} S_{g} = S_{g} \mathcal{F} \,\,\,\,\,\,\forall g  \in G.
\end{eqnarray}

Here we provide the proof for the case where $G$ and $\Omega$ are isomorphic. It
follows the same steps as for Eq.~\eqref{ring_equivariance}.
Indeed, $\forall x \in L_{\mathbb{R}(\Omega)}$ and $\forall u,w \in \Omega$, we have

\begin{eqnarray}
\mathcal{F}\left[S_{g_w}\left[ x\right]\right](u)&=& c *_G \Phi\left[S_{g_w}\left[ x \right]\right](g_u) \nonumber \\
&=& \sum_{v \in \Omega} c(g_u^{-1}v) \Phi\left[S_{g_w}[x](v)\right] \nonumber \\
&=& \sum_{v \in \Omega} c(g_u^{-1}v) \Phi\left[x(g_w^{-1}v)\right] \nonumber \\
&& \text{(change of variables: $v' = g_w^{-1}v $, $v = g_w v' $)} \nonumber \\
&=& \sum_{v' \in \Omega} c(g_u^{-1}g_w v') \Phi\left[x(v')\right] \nonumber \\
&=& \sum_{v' \in \Omega} c((g_w^{-1}g_u)^{-1}v') \Phi\left[x(v')\right] \nonumber \\
&=&  \left(c *_G \Phi[x]\right)(g_w^{-1} g_u) \nonumber \\ 
&=& S_{g_w} \mathcal{F}[x](u). \nonumber
\end{eqnarray}

In summary, an equivariant RNN is defined by a triplet $(\Omega,G, c)$ consisting of a domain $\Omega$, a group $G$ acting on it, and a kernel $c$ that specifies a convolutional recurrent input that is equivariant with respect to the action of $G$ on $\Omega$.

\subsection{Structure of fixed points}

Similarly to the ring model, the equivariance of the recurrent input (Eq.~\eqref{G_equivariance}) directly implies the existence of sets of fixed points generated by the action of the group $G$. Indeed, an activation  ${x^*} \in L_{\mathbb{R}(\Omega)}$ is a fixed point of the dynamics (Eq.~\eqref{RNN_definition_abstract}) if it obeys

\begin{equation}\label{fixed_points_eqRNN}
{x^*}(u) = \mathcal{F}[{x^*}](u)\,\,\,\forall u \in \Omega.
\end{equation}
The equivariance of $\mathcal{F}$ directly implies that  $\forall g \in G$, $S_{g}{x}^*$ is then also a fixed point. Indeed
\begin{eqnarray} \nonumber
\mathcal{F}[S_{g}{x}^*]&=&S_{g}\mathcal{F}[{x}^*]\\
&=&S_{g}{x}^*,\nonumber
\end{eqnarray}
where the first equality follows from the equivariance property of $\mathcal{F}$ (Eq.~\eqref{G_equivariance}), and the second one follows from the fact that ${x}^*$ is a fixed point (Eq.~\eqref{fixed_points_eqRNN}). As in the ring model, the generalized convolutional structure of equivariant RNNs therefore induces a fundamental symmetry in the  set of fixed points.

Starting from a fixed point ${x}^*$ and applying  the group action operators $S_{g}$ for $g \in G$ 
therefore yields a set of fixed points 

\begin{equation} \label{eq:manifold_eqrnn}
    \mathcal{M}_{{x}^*}=\{S_{g}[{x}^*]|g \in G\}.
\end{equation}
Here we denote $\mathcal{M}_{{x}^*}$ as the {\em manifold of fixed points} originating from ${x}^*$ even if $G$ is finite or discrete, but strictly speaking $\mathcal{M}_{{x}^*}$ is a manifold only if $G$ is a continuous group.

As for the ring model, different seed fixed points ${x}^*$ can give rise to different manifolds generated by applying elements of $G$. More specifically, defining $H_{x^*}=\{ h\in G | S_h[x^*]=x^* \}$, the subgroup of $G$ whose action keeps  $x^*$ identical, the manifold $\mathcal{M}_{{x}^*}$ is homeomorphic to the quotient $G/H$. 
The intrinsic dimension of $\mathcal{M}_{{x}^*}$ is therefore less or equal to the dimension of $G$, and depends on the dimension of $H_{x^*}$.

Importantly, all fixed points belonging to a given manifold $\mathcal{M}_{{x}^*}$ have identical stability properties. 
This follows from differentiating the equivariance property (Eq.~\eqref{G_equivariance}) with respect to $x$, which yields
\begin{equation}\nonumber
\text{Jac}_{S_g x^*} \mathcal{F} = S_g \, \text{Jac}_{x^*} \mathcal{F} \, S_g^{-1},
\end{equation}
where $\text{Jac}_{x^*} \mathcal{F} = \partial \mathcal{F}/\partial x|_{x^*}$. This similarity transformation ensures that the Jacobians at all points $S_g x^* \in \mathcal{M}_{x^*}$ share identical eigenvalues (and therefore identical stability properties). If $G$ is a continuous, Lie group, and $\mathcal{M}_{{x}^*}$ is non-trivial (i.e. not consisting of  a single fixed point), each fixed point in $\mathcal{M}_{{x}^*}$ will be marginally stable, with zero eigenvalues associated to directions tangent to the manifold.

In summary, the seed fixed point $x^*$ therefore determines both the dimensionality and the stability of the resulting fixed-point manifold $\mathcal{M}_{{x}^*}$.

\subsection{The group Fourier transform of equivariant RNNs}\label{Group_Fourier_Transform}

For classical ring models, expressing the activity on Fourier components and restricting the number of non-zero components in the connectivity kernel led  to an explicit parametrization of the fixed-point manifolds. Here we extend this approach to equivariant RNNs by using the generalization to {\em group Fourier transforms}. We first introduce group Fourier transforms, and then apply them to equivariant RNNs.

Group Fourier transforms are based on the concept of {\em irreducible representations}, which play a role analogous to the classical Fourier basis. A $d$-dimensional group representation $\rho$ is a mapping $G\to \mathbb{R}^{d \times d}$ which represents each group element $g$ by a $d \times d$ invertible matrix, in such a way that the group composition corresponds to matrix multiplication:

\begin{equation}\label{representation_composition}
\rho(g_1 g_2)=\rho(g_1)\rho(g_2).
\end{equation}
A reducible representation is a representation in which all matrices $\rho(g)$ for $g \in G$ can be put in block-diagonal form through a common change of basis. An irreducible representation (irrep) is conversely a representation that cannot be reduced to smaller blocks. A {\em system of irreps} of a group $G$ therefore forms its set of elementary representations. Representation theory  is a prominent branch of mathematics that has in particular extensively mapped and characterized the irreps of finite or compact groups. In particular, for any compact group, any system of irreps is countable and can therefore be indexed by positive integers \citep{Kosmann-Schwarzbach2009-uj}. We therefore denote a given system of irreps as $\{\rho_k\}_{k=0,1\dots}$.

The group Fourier transform of a function  $f:G \to \mathbb{R}$ with respect to the group $G$ is defined by the collection of its \textit{Fourier components} on a system of irreps $\{\rho_k\}_{k=0,1\dots}$ of $G$ \citep{Kondor_Trivedi}. For a discrete group, the $k$-th Fourier component $\hat{f}_k$ is a $d_k \times d_k$ matrix given by
\begin{equation}\label{group_fourier_components}
\hat{f}_k=\sum_{g \in G} f(g) \rho_k(g), \quad k = 0,1,2,\ldots
\end{equation}
where $\rho_k:G \to \mathbb{R}^{d_k \times d_k}$ is the $k-$th irrep of the considered system. For a continuous, compact group, the sum is replaced by the integral over the Haar measure:
\begin{equation}\label{group_fourier_components_cont}
\hat{f}_k=\int_{g \in G} f(g) \rho_k(g) d\mu(g).
\end{equation}

The group Fourier transform obeys properties analogous to the classical Fourier transform. In particular, any function $f \in L_{\mathbb{R}(G)}$ can be expressed in terms of its inverse Fourier transform (Theorem 4.5  in \cite{Kosmann-Schwarzbach2009-uj}) as

\begin{equation}\label{Peter-Weyl}
    f(g)=\frac{1}{|G|} \sum_{k=0}^{\infty} \tilde{d}_k \operatorname{tr}\left[\hat{f}_k \rho_k^{-1}(g)\right] \quad \forall{g} \in \Omega,
\end{equation}
where $\tilde{d}_k$ is the dimension of the $k$-th complex-valued irreducible representation.

Moreover, the group convolution theorem (Proposition 2 in \cite{Kondor_Trivedi}) states that the Fourier components of the convolution $f^1*_Gf^2$ of two functions $f^1, f^2 L_{\mathbb{R}(G)}$ is given by the product of their Fourier components:
\begin{equation}\label{group-convolution-thm}
\widehat{\left(f^1*_Gf^2 \right)}_k=\hat{f^1}_k \cdot \hat{f^2}_k , \quad k = 0,1,2,... 
\end{equation}

In the case where $G$ and $\Omega$ are isomorphic, the group Fourier  transform on $G$ directly defines a Fourier transform for functions $f:\Omega\to\mathbb{R}$.  When $G$ and $\Omega$ are not isomorphic, \cite{Kondor2018-lo} show that a Fourier transform on $\Omega=G/H$ can be naturally defined by restricting the irreducible representations of $G$ to specific columns. We sketch this construction in Appendix \ref{app:group_fourier}, and provide a concrete example in Sec.~\ref{sec:sphere}. For simplicity, in the following we focus on the case where $G=\Omega$, so that every element $u$ in $\Omega$ corresponds to a unique element $g_u$ in $G$. Eq.~\eqref{group_fourier_components} then directly defines a generalized Fourier transform for any $f:G \to \Omega$:

\begin{equation}\label{group_fourier_components_omega}
\hat{f}_k=\sum_{u \in \Omega} f(u) \rho_k(g_u), \quad k = 0,1,2,\ldots.
\end{equation}

or in the continuous case: 
\begin{equation}\label{group_fourier_components_omega_cont}
\hat{f}_k=\int_{u \in \Omega} f(u) \rho_k(g_u) d\mu (\Omega), \quad k = 0,1,2,\ldots
\end{equation}
where $d\mu (\Omega)$ is the invariant measure on $\Omega$.

Analogously to the ring model, we next express in the Fourier space the dynamics of an eqRNN defined on a domain $\Omega=G$. We denote
 as ${\kappa}^t_k, \hat{\Phi}^t_k, \hat{F}^t_k, \hat{c}_k$ the $k$-th Fourier components of the activation function $x^t$, the firing rate $\Phi^t$, the recurrent input $F^t$ and the connectivity kernel $c$,
  with respect to a system of irreps $\{\rho_k\}_{k=0,1\dots}$ of G. In contrast to the classical ring model where the Fourier components were scalar, in the general case the $k$-th component is a $d_k\times d_k$ matrix, where $d_k$ is the dimension of $\rho_k$.

Since $F^t$ is given by the group convolution between $c$ and $\Phi^t$ (Eq.~\eqref{rec_input_G}), applying the group convolution theorem we have:
\begin{equation}\label{f_group_fourier}
    \hat{F}^t_k=\hat{c}_k\hat{\Phi}^t_k , \quad k = 0,1,2,... 
\end{equation}

Applying the group Fourier transform to Eq.~\eqref{RNN_definition_abstract} then allows us to express  the RNN dynamics in terms of the Fourier components ${\kappa}^t_k$ of the activation $x^t$:

\begin{eqnarray}\label{kappa_group_fourier}
\kappa_{k}^{t+1}&=&\kappa_{k}^{t}+\Delta t \left(-\kappa_{k}^{t}+{\hat F}_{k}^t\right)\\
&=&\kappa_{k}^{t}+\Delta t \left(-\kappa_{k}^{t}+\hat{c}_k\hat{\Phi}^t_k\right).\nonumber
\end{eqnarray}

The Fourier components $\kappa_{k}$ of any fixed point solution $x^*$ obey

\begin{eqnarray}\label{fp_group_fourier}
\kappa_{k}&=&{\hat F}_{k} \\
&=&{\hat c}_{k}{\hat \Phi}_{k}.\nonumber
\end{eqnarray}
As for the classical ring model, Eq.~\eqref{fp_group_fourier} forms in general an infinite system of equations coupled through ${\hat \Phi}_{k}$ which depends implicitly on $x^*$ and therefore all the Fourier components. For general equivariant RNNs, the entries of each equation are however $d_k\times d_k$ matrices instead of scalars.

\subsection{Low-rank equivariant RNNs \label{sec:low-rank-eqRNN}}

Following the approach used for ring models, we next define {\em low-rank eqRNNs} as the subset of eqRNNs for which the Fourier components $\hat{c}_k$ of the kernel $c$ are zero for $k>K$.
For low-rank eqRNNs, from Eq.~\eqref{f_group_fourier} we have ${\hat F}_{k}^t=0$ and therefore  $\kappa_{k}=0$ for $k>K$. Any steady state $x^*$ of a low-rank eqRNN is therefore characterized by a set of at most $K+1$ non-zero Fourier components. As the $k$-th Fourier component is a $d_k\times d_k$ matrix, this corresponds to at most $R=\sum_{k=0}^K d_k^2$ scalar values. These $K+1$ Fourier components obey $K+1$ matrix-valued equations given by  Eq.~\eqref{fp_group_fourier}.

To express these equations explicitly, we write the inverse Fourier transform (Eq.~\eqref{Peter-Weyl}) for the fixed-point activation $x^*$:
\begin{eqnarray} \label{FP_group_inverse_Fourier}
x^*(u)=\frac{1}{|G|} \sum_{k=0}^{K} \tilde{d}_k \operatorname{tr}\left[\kappa_k \rho_k^{-1}(g_u)\right].
\end{eqnarray}
Inserting this expression in Eq.~\eqref{fp_group_fourier} for $k=0,\ldots K$, we get a closed set of fixed point equations for the Fourier components $\{{\kappa}_{k}\}_{k=0\ldots K}$ of a fixed point:

\begin{eqnarray}
{\kappa}_{k}&=&{\hat F}_{k}\left(\kappa_0, \ldots,\kappa_K \right) \nonumber\\
&=&\hat{c}_k \sum_{u \in \Omega}  \Phi\left(x^t(u)\right) \rho_k(g_u) \label{F_hat_group_Fourier_1} \\
&=& \hat{c}_k \sum_{u \in \Omega}  \Phi \left( \frac{1}{|G|} \sum_{k'=0}^{K} \tilde{d}_{k'} \operatorname{tr}\left[\kappa_{k'} \rho_{k'}^{-1}(g_u)\right] \right)\rho_{k}(g_u). \label{F_hat_group_Fourier}
\end{eqnarray}
In Eq.~\eqref{F_hat_group_Fourier_1} we used the definition of the group Fourier transform of $\Phi$, while  in Eq.~\eqref{F_hat_group_Fourier} we replaced $x^*$ with its inverse Fourier transform (Eq.~\eqref{FP_group_inverse_Fourier}).

Note that the stability of any fixed point satisfying Eq.~\eqref{F_hat_group_Fourier_1} is determined by the eigenvalues of the Jacobian $\partial {\hat F}_{k}/\partial {\kappa}_{k'}$.

\subsection{Fixed-point manifolds in low-rank equivariant RNNs}

Restricting the ring model to be of low rank allows us to characterize  more explicitly  the structure of the fixed point manifold generated by applying the group action $S_{g}$ to a non-trivial fixed point $x^* $ (Eq.~\eqref{eq:manifold_eqrnn}).

The action $S_{g}$ on $x^* $  directly induces a linear transformation of its Fourier components.
 Indeed, directly applying $S_{g}$ to the inverse Fourier transform of  $x^*$ (Eq.~\eqref{FP_group_inverse_Fourier}), we have

\begin{eqnarray}
    S_{g}[x^*](u)&=&x^*(g^{-1}_u) \nonumber\\
    &=&\frac{1}{|G|} \sum_{k=0}^{K} d_k \operatorname{tr}\left[\kappa_k \rho_k^{-1}(g^{-1}g_u)\right] \nonumber\\
    &=& \frac{1}{|G|} \sum_{k=0}^{K} d_k \operatorname{tr}\left[\kappa_k\rho_k(g) \rho_k^{-1}(g_u)\right] \label{eq:steadystate-transformation}
\end{eqnarray}
where we used Eq.~\eqref{representation_composition} to write:
\begin{eqnarray}\nonumber
    \rho_k^{-1}(g^{-1}g_u)=\rho_k(g) \rho_k^{-1}(g_u). 
\end{eqnarray}
From Eq.~\eqref{eq:steadystate-transformation}, we see that the transformation $x^*\mapsto S_{g}x^*$ induces a transformation of Fourier components given by 
\begin{equation}
    \kappa_k \mapsto \rho_k(g) \kappa_k. \label{eq:fourier_comp-transformation}
\end{equation}
Equivalently, grouping the $K+1$ Fourier components of $x^*$ into a block-diagonal matrix $\underline{\kappa} \in \mathbb{R}^{R\times R}$ defined as
\begin{eqnarray}\label{kappa_matrix}
\underline{\kappa}=\begin{pmatrix}
\kappa_{0} &  & \\
 &\ddots  &\\
 & & \kappa_{R}
\end{pmatrix},
\end{eqnarray}
the transformation of Fourier components is determined by a block-diagonal matrix $R^{K}(g)$: 
\begin{eqnarray}\label{rotation_matrix0}
R^{K}(g)=\begin{pmatrix}
\rho_{0}(g) &  & \\
 &\ddots  &\\
 & & \rho_{K}(g)
\end{pmatrix}.
\end{eqnarray}

The equivariance of the fixed point equation for $x$ with respect to $S_{g}$ directly implies the equivariance of the  fixed point equation for ${\underline{\kappa}}$ with respect to the matrix $R^{(K)}(g)$: 

\begin{equation}\label{f_equiv}
\hat {F}(R^{(K)}(g) \underline{\kappa})=R^{(K)}(g)\hat { F}(\underline{\kappa})   \,\,\, \forall \phi \in \Omega.
\end{equation}

Starting from a seed fixed point $x^*$ with Fourier components $\underline{\kappa}^*$ 
 therefore leads to a manifold of fixed points in Fourier coordinates parametrized by elements of $G$ and given by:
\begin{eqnarray}\label{parametrized_solution0}
{\underline{\kappa}}(g)=R^{K}(g){\underline{\kappa}^*}.
\end{eqnarray}

All fixed points parametrized by Eq.~\eqref{parametrized_solution0} have identical stability properties. This follows from differentiating the equivariance property (Eq.~\eqref{f_equiv}) with respect to $\underline{\kappa}$, which yields
\begin{equation}\label{jac_equiv_latent}
\text{Jac}_{R^{(K)}(g) \underline{\kappa}} \hat{F} = R^{(K)}(g) \, \text{Jac}_{\underline{\kappa}} \hat{F} \, R^{(K)}(g)^{-1},
\end{equation}
where $\text{Jac}_{\underline{\kappa}} \hat{F}$ denotes the Jacobian matrix of $\hat{F}$ with respect to $\underline{\kappa}$. This similarity transformation ensures that the Jacobians at all points $R^{(K)}(g)\underline{\kappa}$ on the manifold share identical eigenvalues.

\subsection{The geometry of the fixed-point manifold}

Similarly to the ring model, in  eqRNNs the inverse Fourier transform in Eq.~\eqref{FP_group_inverse_Fourier},  can be interpreted as a linear parametrization of the fixed-point manifold where the Fourier components provide coordinates in a basis-set provided by the entries of irreducible representations. In  low-rank eqRNNs, at most $R$ of the Fourier components are non-zero, so that the fixed-point manifold is embedded in a finite-dimensional subspace of the full space of square-integrable functions on $\Omega$. The embedding dimensionality of a fixed point manifold $\mathcal{M}_{{x}^*}$, i.e. its number of non-zero Fourier components, is determined by the number of non-zero Fourier components of the seed fixed point ${x}^*$ through Eq.~\eqref{parametrized_solution0}.
The intrinsic dimension of $\mathcal{M}_{{x}^*}$, i.e. the minimal number of parameters needed for a non-linear parametrization is instead  
 determined by the dimensionality of the subgroup $H$ of $G$ that leaves $x^*$ invariant. As we will illustrate in the examples, these two properties lead to variety of possible fixed-point manifolds for a given group $G$.
 
\section{Examples of specific equivariant RNNs}

Having laid out the general framework of equivariant RNNs, in this section we provide three specific instantiations that correspond to three different domains $\Omega$ and associated groups $G$. We start by the ring model, reformulating it this time in the group-theoretic language of eqRNNs. We then consider toroidal and spherical RNNs as two more general examples of eqRNNs on higher-dimensional domains $\Omega$.

\subsection{Ring RNNs}\label{Ring_equiRNNs}

In this section, we revisit the ring model introduced in Section \ref{ring_section} and now formulate it within the  formalism of group theory presented in Section \ref{group_section}. We first explicitly differentiate between the discrete and continuous versions of the model, and then focus on the continuous case.

\subsubsection{Domain structure and symmetry group}

In the discrete ring model, neurons are indexed by angles $u=\theta_i = \frac{2\pi i}{N}$ for $i = 0, \ldots, N - 1$, so that the domain $\Omega$ forms a discretized circle $\Omega=\{\theta_i=\frac{2\pi i}{N}|i=0,\ldots,N-1\}$.
The associated group $G$ consists of the set of angular translations on $\Omega$, and is directly isomorphic to $\Omega$ equipped with addition modulo $2\pi$. $G$ is a finite group known as the cyclic group $G=\mathbb{Z}_N$.

In the continuous limit $N \rightarrow \infty$, neurons in the ring model are labeled with a continuous angle $u=\theta$, so that $\Omega = [0,2\pi)$ is homeomorphic to the unit circle $\mathcal{S}^1$. The associated continuous, compact group of angular translations is the circle group $G=\mathbb{S}^1$, which is isomorphic to the special orthogonal group $SO(2)$, to the group of real numbers modulo the integers $\mathbb{R}/2\pi\mathbb{Z}$, and to the first unitary group $U(1)$ \citep{zee2016group}. 

In both the discrete and continuous ring models, any element of the group $G$ corresponds to a unique element in $\Omega$, which allows us to identify the group  $G$ with the the domain $\Omega$.

\subsubsection{Group convolution}

For the discrete ring model, the recurrent input in Eq.~\eqref{rec_input_G} is defined as the $\mathbb{Z}_N$-convolution between the  kernel function $c: \Omega\to \mathbb{R}$ and the firing rate, $\Phi[x]: \Omega  \to \mathbb{R}$:
\begin{eqnarray}
\label{group_conv_ring}    F^t(\theta_i)&=& \frac{1}{N}\sum_{\theta_j \in \frac{2\pi}{N}\mathbb{Z}_N} c(g_{\theta_i}^{-1}(\theta_j)) \Phi \left[x_t(\theta_j)\right]=\frac{1}{N}\sum_{j =0}^{N-1} c(\theta_j-\theta_i) \Phi \left[x^t(\theta_j)\right].
\end{eqnarray}

In the continuous case, as in  Eq.~\eqref{group_conv_cont}, the recurrent input is the $\mathbb{S}^1$-convolution between $c, \Phi[x]: \Omega=\mathcal{S}^1 \to \mathbb{R}:$
\begin{eqnarray}
\label{group_conv_cont_ring}  F^t(\theta)&=&\int_{0}^{2\pi} c(g_{\theta}^{-1}(\phi)) \Phi [x^t(\phi)]\frac{d \phi}{2 \pi}=\int_{0}^{2\pi} c(\phi-\theta) \Phi [ x^t(\phi)]\frac{d \phi}{2 \pi},
\end{eqnarray}
where we integrate with respect to the normalized Haar measure $d\mu(\mathcal{S}^1)=\frac{d\phi}{2\pi}$.

\subsubsection{Group representations}

The groups $\mathbb{Z}_N$ and $\mathbb{S}^1$ associated with the discrete and continuous ring models have a well-studied set of irreducible representations (irreps) that directly correspond to the classical discrete and continuous Fourier bases.
Working in the real-valued setting, here we provide the standard form of the irreps.

For the discrete ring with $N$ neurons, the group $\mathbb{Z}_N$ has $\lfloor N/2 \rfloor + 1$ distinct irreducible representations, which we denote as $\rho_k$ with $k=0,\ldots,\lfloor N/2 \rfloor + 1$. For $k = 0$ and $k = N/2$ (when $N$ is even), $\rho_k$ are one-dimensional, and given by

\begin{eqnarray}
  \rho_0(\theta_j) &=& 1 \nonumber \\
\rho_{N/2}(\theta_j) &=& (-1)^j  \nonumber .
\end{eqnarray}

For $k = 1, 2, ..., \lfloor (N-1)/2 \rfloor$, the irreps are two-dimensional rotation matrices:
\begin{equation} \nonumber
\rho_k(\theta_j) = 
\begin{pmatrix}
\cos(k \theta_j) & -\sin(k \theta_j) \\
\sin(k \theta_j) & \cos(k \theta_j)
\end{pmatrix}
\end{equation}
for $\theta_j = \frac{2\pi j}{N}$, $j=0\ldots N-1$.

In the continuous case, the group $\mathbb{S}^1$ has a countable set of irreps $\{\rho_k\}_{k \in \mathbb{N}}$. For $k = 0$, 

\begin{equation}
\rho_0(\theta) = 1, \label{eq:ring_rho0}
\end{equation}
while for $k \geq 1$, the irreps are two-dimensional rotation matrices \citep{Feehan2020-dp}:

\begin{equation}
\rho_k(\theta) = 
\begin{pmatrix}
\cos(k\theta) & -\sin(k\theta) \\
\sin(k\theta) & \cos(k\theta)
\end{pmatrix} \label{eq:ring_rhok}
\end{equation}
for $\theta \in [0,2\pi)$.

\subsubsection{Group Fourier transform}

We next use the real-valued irreps
to explicitely write the components $\hat{f}_k$ of the group Fourier transform defined in Eq.~\eqref{group_fourier_components} for any function $f : \Omega \to \mathbb{R}$. From this section, we  focus on the continuous case, where $\Omega=G=\mathcal{S}^1$.

For $k=0$, introducing Eq.~\eqref{eq:ring_rho0} into Eq.~\eqref{group_fourier_components}, $\hat{f}_0$ is a scalar given by

\begin{eqnarray} 
\hat{f}_0 &=& \int_{0}^{2\pi} f(\theta)  \frac{d \theta}{2 \pi} ,
\label{matrix_fourier_ring_0}
\end{eqnarray}

For $k>1$, inserting Eq.~\eqref{eq:ring_rhok} into Eq.~\eqref{group_fourier_components},
\begin{eqnarray}
\hat{f}_k &=& \int_{0}^{2\pi} f(\theta) \begin{pmatrix} \cos(k\theta) & -\sin(k\theta) \nonumber\\ \sin(k\theta) & \cos(k\theta) \end{pmatrix} \frac{d \theta}{2 \pi} \nonumber\\
&=& \int_{0}^{2\pi} \begin{pmatrix} f(\theta)\cos(k\theta) & -f(\theta)\sin(k\theta) \\ f(\theta)\sin(k\theta) & f(\theta)\cos(k\theta) \end{pmatrix} \frac{d \theta}{2 \pi} \\
&=& \begin{pmatrix} \hat{f}_{k,1} & -\hat{f}_{k,2} \\ \hat{f}_{k,2} & \hat{f}_{k,1} \end{pmatrix} \label{matrix_fourier_ring}.
\end{eqnarray}
For $k>1$, $\hat{f}_k$ is therefore a $2\times 2$ matrix, but it 
 contains only two independent elements, which we denote as  $\hat{f}_{k,1}$ and $\hat{f}_{k,2}$, with:

\begin{equation}
\hat{f}_{k,1} = \int_{0}^{2\pi} f(\theta)\cos(k\theta) \frac{d \theta}{2 \pi},\nonumber
\end{equation}
\begin{equation}
\hat{f}_{k,2} = \int_{0}^{2\pi} f(\theta)\sin(k\theta) \frac{d \theta}{2 \pi}.\nonumber
\end{equation}
These elements are the classical Fourier components introduced in Eq.~\eqref{fourier_components}.

\subsubsection{Fixed point equations for low-rank models}

We now focus on low-rank models, i.e. models for which the matrix-valued Fourier components $\hat{c}_k$ of the kernel $c$ are zero for $k>K$. As shown in Sec.~\ref{sec:low-rank-eqRNN}, for such models the steady-state activation is described by a finite number of non-zero Fourier components $\kappa_{k}$ for $k\leq K$. Here we use the expressions for the irreps to write explicitly the fixed-pont equation (Eq.~\eqref{F_hat_group_Fourier}) satisfied by the Fourier components $\kappa_{k}$ of any steady-state activation $x^*$.

Using the Convolution Theorem (Eq.~\eqref{group-convolution-thm}), we get for $k=0$:
\begin{eqnarray}\label{eq:2d_ring_F_0}
\hat{F}_0 &=& \hat{c}_0 \cdot \hat{\Phi}_0
\end{eqnarray}
and, for $k>1$:
\begin{eqnarray}
\hat{F}_k &=& \hat{c}_k \cdot \hat{\Phi}k = \begin{pmatrix} \hat{c}_{k,1} & -\hat{c}_{k,2} \nonumber \\ \hat{c}_{k,2} & \hat{c}_{k,1} \end{pmatrix} \begin{pmatrix} \hat{\Phi}_{k,1} & -\hat{\Phi}_{k,2} \\ \hat{\Phi}_{k,2} & \hat{\Phi}_{k,1} \end{pmatrix}\nonumber \\
&=& \begin{pmatrix} \hat{c}_{k,1}\hat{\Phi}_{k,1}-\hat{c}_{k,2}\hat{\Phi}_{k,2} & -\hat{c}_{k,1}\hat{\Phi}_{k,2}-\hat{c}_{k,2}\hat{\Phi}_{k,1} \\ \hat{c}_{k,2}\hat{\Phi}_{k,1}+\hat{c}_{k,1}\hat{\Phi}_{k,2} & -\hat{c}_{k,2}\hat{\Phi}_{k,2}+\hat{c}_{k,1}\hat{\Phi}_{k,1} \end{pmatrix} \nonumber\\
&=& \begin{pmatrix} \hat{F}_{k,1} & -\hat{F}_{k,2} \\ \hat{F}_{k,2} & \hat{F}_{k,1} \end{pmatrix}\nonumber
\end{eqnarray}

where 
\begin{equation}
\hat{F}_{k,1} = \hat{c}_{k,1}\hat{\Phi}_{k,1}-\hat{c}_{k,2}\hat{\Phi}_{k,2},\nonumber
\end{equation}
\begin{equation}
\hat{F}_{k,2} = \hat{c}_{k,2}\hat{\Phi}_{k,1}+\hat{c}_{k,1}\hat{\Phi}_{k,2}.\nonumber
\end{equation}

To express explicitly  $\hat{F}_{k,1}$ and $\hat{F}_{k,2}$, we  write  $\hat{\Phi}_k$ in terms of the Fourier components $\kappa_{k}$ of the steady-state activation $x^*$.

For $k=0$, we have
\begin{eqnarray}\label{eq:2d_ring_phi_0}
\hat{\Phi}_{0}=\int_{0}^{2\pi} \Phi\left[x^*(\theta)\right] \frac{d \theta}{2 \pi},
\end{eqnarray}
while for $k>1$, the independent matrix elements are given by

\begin{eqnarray}
\hat{\Phi}_{k,1} &=& \int_{0}^{2\pi} \Phi\left[x^*(\theta)\right]\cos(k\theta) \frac{d \theta}{2 \pi}, \label{eq:ring_Phik1_hat}\\
\hat{\Phi}_{k,2} &=& \int_{0}^{2\pi} \Phi\left[x^*(\theta)\right]\sin(k\theta) \frac{d \theta}{2 \pi}. \label{eq:ring_Phik2_hat}
\end{eqnarray}

We next write the steady-state activation $x^*$  using the Inverse Fourier Transform (Eq.~\eqref{Peter-Weyl}):
\begin{eqnarray}
x(\theta) &=&  \sum_{k=0}^K \tilde{d_k} \, \text{tr} (\kappa_k \rho_k^{-1}(\theta)) \nonumber\\
&=&   \kappa_{0}+\sum_{k=1}^K \, \text{tr} \left(\begin{pmatrix} \kappa_{k,1} & -\kappa_{k,2} \\ \kappa_{k,2} & \kappa_{k,1} \end{pmatrix} \begin{pmatrix} \cos (k \theta) & \sin (k \theta) \\ -\sin (k \theta) & \cos (k \theta) \end{pmatrix} \right)  \nonumber\\
&=&   \kappa_{0}+\sum_{k=1}^K \, \text{tr} \begin{pmatrix} \kappa_{k,1}\cos (k \theta) + \kappa_{k,2}\sin (k \theta) & \kappa_{k,1}\sin (k \theta) - \kappa_{k,2}\cos (k \theta) \\ \kappa_{k,2}\cos (k \theta) - \kappa_{k,1}\sin (k \theta) & \kappa_{k,2}\sin (k \theta) + \kappa_{k,1}\cos (k \theta) \end{pmatrix}  \nonumber\\
&=& \kappa_{0}+\sum_{k=1}^K 2\left(\kappa_{k,1}\cos(k \theta) + \kappa_{k,2}\sin(k \theta)\right). \label{eq:ring_x_kappa}
\end{eqnarray}

Notice that we include only the $K+1$ first Fourier components, because at steady-state $\kappa_k=0$ for $k>K$.

Inserting Eq.~\eqref{eq:ring_x_kappa} into Eqs.~\eqref{eq:2d_ring_phi_0}, \eqref{eq:ring_Phik1_hat} and \eqref{eq:ring_Phik2_hat} gives us  explicit expressions for $\hat{\Phi}_{0}, \hat{\Phi}_{k,1}$ and $\hat{\Phi}_{k,2}$ as function of the set of scalars $\kappa_{0}, \kappa_{k,1}, \kappa_{k,2}$ for $k=1\ldots K$.

This leads to a set of $R=2K+1$ fixed point equations for the $R$ scalar variables $\kappa_{0}$, $\kappa_{k,1}, \kappa_{k,2}$:

\begin{eqnarray}  \label{eq:fixed_point_eq0}
\kappa_{0} =  \hat{c}_{0} \int_{0}^{2\pi} \Phi\left[\kappa_{0}+\sum_{k'=1}^K 2\left(\kappa_{k',1}\cos(k' \theta) + \kappa_{k',2}\sin(k' \theta)\right)\right] \frac{d \theta}{2 \pi}, 
\end{eqnarray}
\begin{eqnarray}  \label{eq:fixed_point_eq1}
\kappa_{k,1} = \int_{0}^{2\pi} \Big(\hat{c}_{k,1} \cos(k\theta) -\hat{c}_{k,2}\sin(k\theta)\Big)\Phi\left[\kappa_{0}+\sum_{k'=1}^K 2\left(\kappa_{k',1}\cos(k' \theta) + \kappa_{k',2}\sin(k' \theta)\right)\right]\frac{d \theta}{2 \pi},
\end{eqnarray}
\begin{eqnarray}  \label{eq:fixed_point_eq2}
\kappa_{k,2} = \int_{0}^{2\pi}\Big(\hat{c}_{k,2}\cos(k\theta) +\hat{c}_{k,1}\sin(k\theta)\Big) \Phi\left[\kappa_{0}+\sum_{k'=1}^K 2 \left(\kappa_{k',1}\cos(k' \theta) + \kappa_{k',2}\sin(k' \theta)\right)\right] \frac{d \theta}{2 \pi} . 
\end{eqnarray}

\subsubsection{Fixed point manifold}

Any solution ${\underline{\kappa}}$ of the fixed point equation \eqref{eq:fixed_point_eq0}, \eqref{eq:fixed_point_eq1}, \eqref{eq:fixed_point_eq2}  leads to a manifold of fixed points in Fourier coordinates parametrized by elements of the group $\mathbb{S}^1$:

\begin{eqnarray}\label{parametrized_solution}
{\underline{\kappa}}(\phi)={\underline{\kappa}}R^{K}(\phi).
\end{eqnarray}

where $R^{K}(\phi)$ is block-diagonal matrix containing at most $(K+1)$ rotation matrices that act on the corresponding Fourier components, and given by:

\begin{eqnarray}\label{rotation_matrix_general_ring}
R^{K}(\phi)=\begin{pmatrix}
\rho_{0}(\phi) & & & & \\
& \rho_{1}(\phi) & & & \\
& & \rho_{2}(\phi) & &  \\
& & & \ddots \\
& & & & \rho_{K}(\phi)
\end{pmatrix}.
\end{eqnarray}

\subsubsection{Rank two ring model} 
We now consider the specific instantiation of the  ring model with the connectivity kernel

\begin{equation} \nonumber
    c(\theta) = J_1\cos(\theta),
\end{equation}
where $J_1$ is a scalar parameter that modulates the strength of the recurrent connectivity. In this case, the only non-zero Fourier component of the kernel  is $\hat{c}_{1}$ with
\begin{eqnarray}\nonumber
    \hat{c}_{1,1}&=&\frac{J_1}{2},\\
    \hat{c}_{1,2}&=&0.\nonumber
\end{eqnarray}

The only potentially non-zero Fourier component of the steady state activation $x^*$ is the matrix $\kappa_{1}$, so that 
\begin{eqnarray}\label{eq:x_2d_ring}
x^*(\theta) 
&=& 2 (\kappa_{1,1}\cos( \theta) + \kappa_{1,2}\sin( \theta)).
\end{eqnarray}
The steady state is therefore described by $R=2$ scalar variables, and we denote this instance of the ring model as the {\em rank two model} because the connectivity matrix in the finite-size network is of rank two (Appendix \ref{appendix_discretized}).

Eqs.~\eqref{eq:fixed_point_eq0}-\eqref{eq:fixed_point_eq2} then reduce  to two coupled equations for the scalars $\kappa_{1,1}$ and $\kappa_{1,2}$:
\begin{eqnarray}
\begin{cases}\label{eq:system_eq_ring2d}

{\kappa}_{1,1}  = \frac{J_1}{2} \int_{0}^{2\pi} \cos \theta \Phi \left[ 2 (\kappa_{1,1}\cos \theta + \kappa_{1,2}\sin \theta) \right]  \, \frac{d \theta}{2 \pi} \\
{\kappa}_{1,2}  = \frac{J_1}{2} \int_{0}^{2\pi} \sin \theta \Phi\left[2 (\kappa_{1,1}\cos \theta + \kappa_{1,2}\sin \theta) \right] \, \frac{d \theta}{2 \pi}.
\end{cases}
\end{eqnarray}

Using Eq.~\eqref{parametrized_solution}, it follows that any non-zero solution $\underline{\kappa}=(\kappa_{1,1},\kappa_{1,2})$ of Eq.~\eqref{eq:system_eq_ring2d} leads to a manifold of fixed points in Fourier coordinates parametrized by $\phi \in [0,2\pi)$:
\begin{eqnarray}\label{rotated_kappa}
{\underline{\kappa}}(\phi)=R^{1}(\phi){\underline{\kappa}},
\end{eqnarray}
where 
\begin{eqnarray}
R^1(\phi)=\begin{pmatrix}
\cos(\phi) & -\sin(\phi)  \\
\sin(\phi)& \cos(\phi)
\end{pmatrix}.\nonumber
\end{eqnarray}
Setting $\phi=0$, we obtain a particular solution ${\underline{\kappa}}(0)=(\kappa_{1,1},0)$. The system \eqref{eq:system_eq_ring2d} then reduces to the one dimensional equation:
\begin{eqnarray}
{\kappa}_{1,1}  = \frac{J_1}{2} \int_{0}^{2\pi} \Phi \left( 2 \kappa_{1,1}\cos \theta \right) \cos \theta \frac{d \theta}{2 \pi}. \label{ring2d_1d_eq}
\end{eqnarray}
The stability of the  fixed-point is determined by the sign of the derivative of the r.h.s with respect to $\kappa_{1,1}$, evaluated at the fixed point. This stability is then inherited by all the fixed points on the resulting manifold.

\begin{figure}[!ht]
    \centering
    \includegraphics[width=12cm]{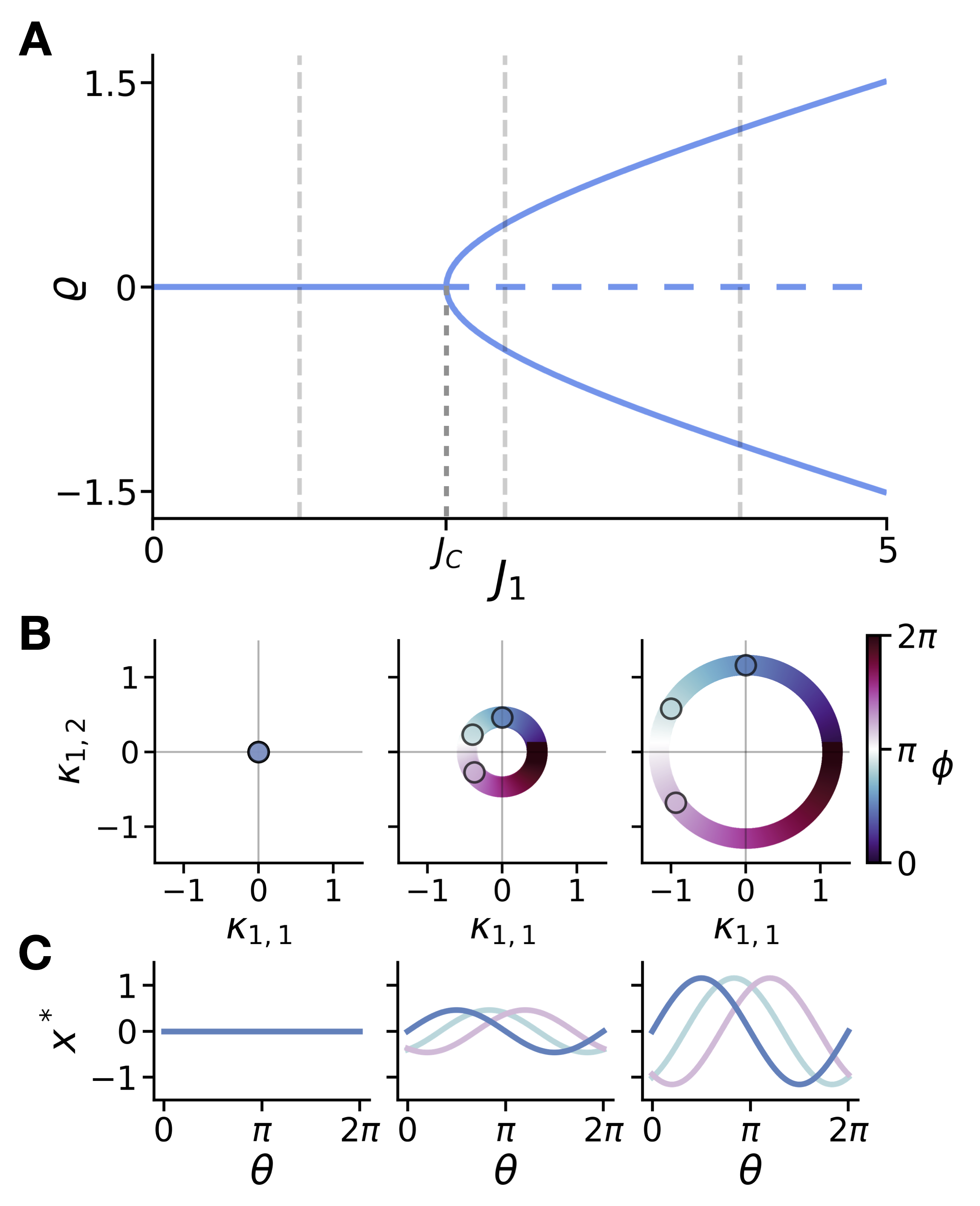} 
    \caption{Bifurcation analysis and manifolds in the rank-two ring model. A) Bifurcation diagram showing the emergence of non-zero fixed points ($\kappa_{1,1}$) as a function of coupling strength $J_1$. Solid lines represent stable solutions while dashed lines indicate unstable solutions. B) Manifolds of stable fixed points in the $\kappa_{1,1}$-$\kappa_{1,2}$ plane at three different values of $J_1$ (increasing from left to right), color-coded by phase $\phi$. Left: At subcritical coupling, only the origin is stable. Middle: Above the bifurcation, a stable ring manifold emerges. Right: At stronger coupling, the ring manifold expands its radius. C) Population activity profiles ($x^*$) corresponding to points on each manifold plotted as function of $\theta$. The profiles transition from uniform/null (left) to sinusoidal patterns (middle) to more pronounced bumps (right) as coupling strength increases. Different colors show $\phi$-shifted versions of the same solution. \textbf{Computational details}: Panel A: $\varrho$ is computed by numerically solving Eq. \eqref{ring2d_1d_eq}. The integral is evaluated numerically using a discretized approximation with N neurons uniformly distributed around the ring. Panel B: Manifolds are generated by applying the rotation \eqref{rotated_kappa} to fixed-point solutions from (A). Panel C: Activity profiles from Eq. \eqref{eq:pop_activity_ring2d_param}.}
    \label{fig:ring2d}
\end{figure}

Eq.~\eqref{ring2d_1d_eq} equation admits a zero solution ${\kappa}_{1,1}=0$ that is stable for $J_1$ smaller than a critical coupling $J_c$, and unstable for $J_1>J_c$. For $J_1>J_c$, Eq.~\eqref{ring2d_1d_eq} admits an additional pair of non-zero solutions with opposite signs: ${\kappa}_{1,1}=\pm \varrho$ (Fig.~\ref{fig:ring2d} A). These solutions are stable, and lead to a manifold of stable fixed points that form a ring embedded in two dimensions (Fig.~\ref{fig:ring2d} B).
 The radius of the ring is given by ${\kappa}_{1,1}$, and increases with $J_c$. Note that the two non-zero solutions of Eq.~\eqref{ring2d_1d_eq} with opposite signs correspond to opposite points $\phi=0$ and $\phi=\pi$ on the ring. Using the inverse Fourier transform (Eq.~\eqref{eq:x_2d_ring}), the  fixed point activation $x_{\phi}$ can be expressed as a cosine function centered at $\phi$, and of amplitude ${\kappa}_{1,1}$ (Fig.~\ref{fig:ring2d} C).

More specifically, applying the relation \eqref{rotated_kappa} to the particular solution ${\underline{\kappa}}(0)=(\varrho,0)$, we get a general solution for $\phi \in [0,2\pi)$:
\begin{eqnarray}\label{eq:kappa_particular_solution_2d}
{\underline{\kappa}}(\phi)=(\varrho \cos(\phi), \varrho \sin(\phi) ).
\end{eqnarray}
Inserting Eq.~\eqref{eq:kappa_particular_solution_2d} into Eq.~\eqref{eq:x_2d_ring} results the classical bump-like population activity for a fixed $\phi$ (Fig.~\ref{fig:ring2d}C):
\begin{eqnarray}
x_{\phi}(\theta) = 2\varrho \cos(\theta-\phi). 
\label{eq:pop_activity_ring2d_param}
\end{eqnarray}

\newpage

\subsubsection{Rank five ring model \label{sec:rank5_ring}} 
To illustrate the effect of including a higher number of Fourier components in the connectivity kernel,
we next consider the rank five ring model, defined by the kernel
\begin{equation}\nonumber
    c(\theta) = J_0 + J_1\cos(\theta)+J_2\cos(2\theta),
\end{equation}
with $J_0, J_1 \text{ and } J_2$ scalar parameters. In this case, the non-zero Fourier components are

\begin{eqnarray}
\hat{c}_0 =   J_0 \nonumber\\
\hat{c}_1 =  \begin{pmatrix} \hat{c}_{1,1} & -\hat{c}_{1,2} \\  \hat{c}_{1,2} & \hat{c}_{1,1} \end{pmatrix} = J_1 \begin{pmatrix} \frac{1}{2} & 0 \\ 0 & \frac{1}{2} \end{pmatrix} \nonumber\\
\hat{c}_2 =  \begin{pmatrix} \hat{c}_{2,1} & -\hat{c}_{2,2} \\  \hat{c}_{2,2} & \hat{c}_{2,1} \end{pmatrix} = J_2 \begin{pmatrix} \frac{1}{2} & 0 \\ 0 & \frac{1}{2} \end{pmatrix}\nonumber
\end{eqnarray}
while for $k>2$
\begin{eqnarray}\nonumber
\hat{c}_k = \begin{pmatrix} 0 & 0 \\ 0 & 0 \end{pmatrix}.
\end{eqnarray}

The steady state activation $x^*(\theta)$ is given by
\begin{eqnarray}\label{eq:x_4d_ring}
x(\theta) 
&=& \kappa_{0} + 2 (\kappa_{1,1}\cos( \theta) + \kappa_{1,2}\sin( \theta)+\kappa_{2,1}\cos( 2\theta) + \kappa_{2,2}\sin( 2\theta)).
\end{eqnarray}

Analogously to the rank two ring model, the fixed point equations \eqref{eq:fixed_point_eq0}-\eqref{eq:fixed_point_eq2} 
reduce to a set of $R=2K+1=5$ coupled equations for the scalars $\kappa_{0},\kappa_{1,1},\kappa_{1,2},\kappa_{1,2},\kappa_{2,2}$: 
\begin{eqnarray}
\begin{cases}\label{eq:system_eq_ring4d}
\kappa_0  =  J_0  \int_{0}^{2\pi} \Phi \left[ \kappa_{0}+2 (\kappa_{1,1}\cos \theta + \kappa_{1,2}\sin \theta+\kappa_{2,1}\cos 2 \theta + \kappa_{2,2}\sin 2 \theta) \right]  \, \frac{d \theta}{2 \pi} \\
{\kappa}_{1,1} = \frac{J_1}{2} \int_{0}^{2\pi} \Phi \left[ \kappa_{0}+2 (\kappa_{1,1}\cos \theta + \kappa_{1,2}\sin \theta+\kappa_{2,1}\cos 2 \theta + \kappa_{2,2}\sin 2 \theta) \right] \cos \theta \, \frac{d \theta}{2 \pi} \\
{\kappa}_{1,2}  = \frac{J_1}{2}  \pi \int_{0}^{2\pi} \Phi\left[ \kappa_{0}+2 (\kappa_{1,1}\cos \theta + \kappa_{1,2}\sin \theta+\kappa_{2,1}\cos 2 \theta + \kappa_{2,2}\sin 2 \theta) \right]\sin \theta \, \frac{d \theta}{2 \pi} \\
{\kappa}_{2,1}  = \frac{J_2}{2}  \pi \int_{0}^{2\pi} \Phi \left[ \kappa_{0}+2 (\kappa_{1,1}\cos \theta + \kappa_{1,2}\sin \theta+\kappa_{2,1}\cos 2 \theta + \kappa_{2,2}\sin 2 \theta) \right] \cos 2\theta \, \frac{d \theta}{2 \pi} \\
{\kappa}_{2,2}  = \frac{J_2}{2}  \pi \int_{0}^{2\pi} \Phi\left[ \kappa_{0}+2 (\kappa_{1,1}\cos 2 \theta + \kappa_{1,2}\sin \theta+\kappa_{2,1}\cos 2 \theta + \kappa_{2,2}\sin 2 \theta) \right]\sin 2 \theta \, \frac{d \theta}{2 \pi}
\end{cases}
\end{eqnarray}

Using Eq.~\eqref{parametrized_solution}, any solution $\underline{\kappa}=(\kappa_0,\kappa_{1,1},\kappa_{1,2},\kappa_{2,1},\kappa_{2,2})$ of the system \eqref{eq:system_eq_ring4d} leads to a manifold of fixed points in Fourier coordinates,  parametrized by element of $SO(2)$, but with a higher dimensional representation:
\begin{eqnarray}\label{eq:ring_rank4_manifold}
\underline{\kappa}(\phi)=R^{2}(\phi)\underline{\kappa}
\end{eqnarray}
where $R^2(\phi) \in \mathbb{R}^{5\times 5}$ is a block diagonal  matrix: 
\begin{eqnarray} \label{eq:rank5_rotation_mtx}
R^2(\phi)=\begin{pmatrix}
1 & 0 & 0 & 0 & 0 \\
0 & \cos(\phi) & -\sin(\phi)  & 0 & 0 \\
0 & \sin(\phi)& \cos(\phi) & 0 & 0 \\
0 & 0 & 0 & \cos(2 \phi) & -\sin(2\phi) \\
0 & 0 & 0 & \sin(2 \phi) & \cos(2\phi)
\end{pmatrix}.
\end{eqnarray}

Setting $\phi=0$, we obtain a particular solution ${\underline{\kappa}}(0)=(\kappa_{0},\kappa_{1,1},0,\kappa_{2,1},0)$. Eq.~\eqref{eq:system_eq_ring4d} then reduces to the three dimensional system:

\begin{eqnarray}
\begin{cases}\label{system_eq_ring4d_reduced}
\kappa_0  =  J_0 \int_{0}^{2\pi} \Phi \left[ \kappa_{0}+2 (\kappa_{1,1}\cos \theta +\kappa_{2,1}\cos 2 \theta)  \right]  \, \frac{d \theta}{2 \pi} \\

{\kappa}_{1,1}  = \frac{J_1}{2}  \pi \int_{0}^{2\pi} \Phi \left[ \kappa_{0}+2 (\kappa_{1,1}\cos \theta +\kappa_{2,1}\cos 2 \theta )  \right] \cos \theta \, \frac{d \theta}{2 \pi} \\

{\kappa}_{2,1} = \frac{J_2}{2}  \pi \int_{0}^{2\pi} \Phi \left[ \kappa_{0}+2 (\kappa_{1,1}\cos \theta +\kappa_{2,1}\cos 2 \theta ) \right] \cos 2\theta \, \frac{d \theta}{2 \pi} .
\end{cases}
\end{eqnarray}
The stability of each  fixed point is determined by the eigenvalues of the Jacobian of  the r.h.s, evaluated at the fixed point. This stability is then inherited by all the fixed points on the resulting manifold.

\begin{figure}[!ht]
    \centering
    \includegraphics[width=14cm]{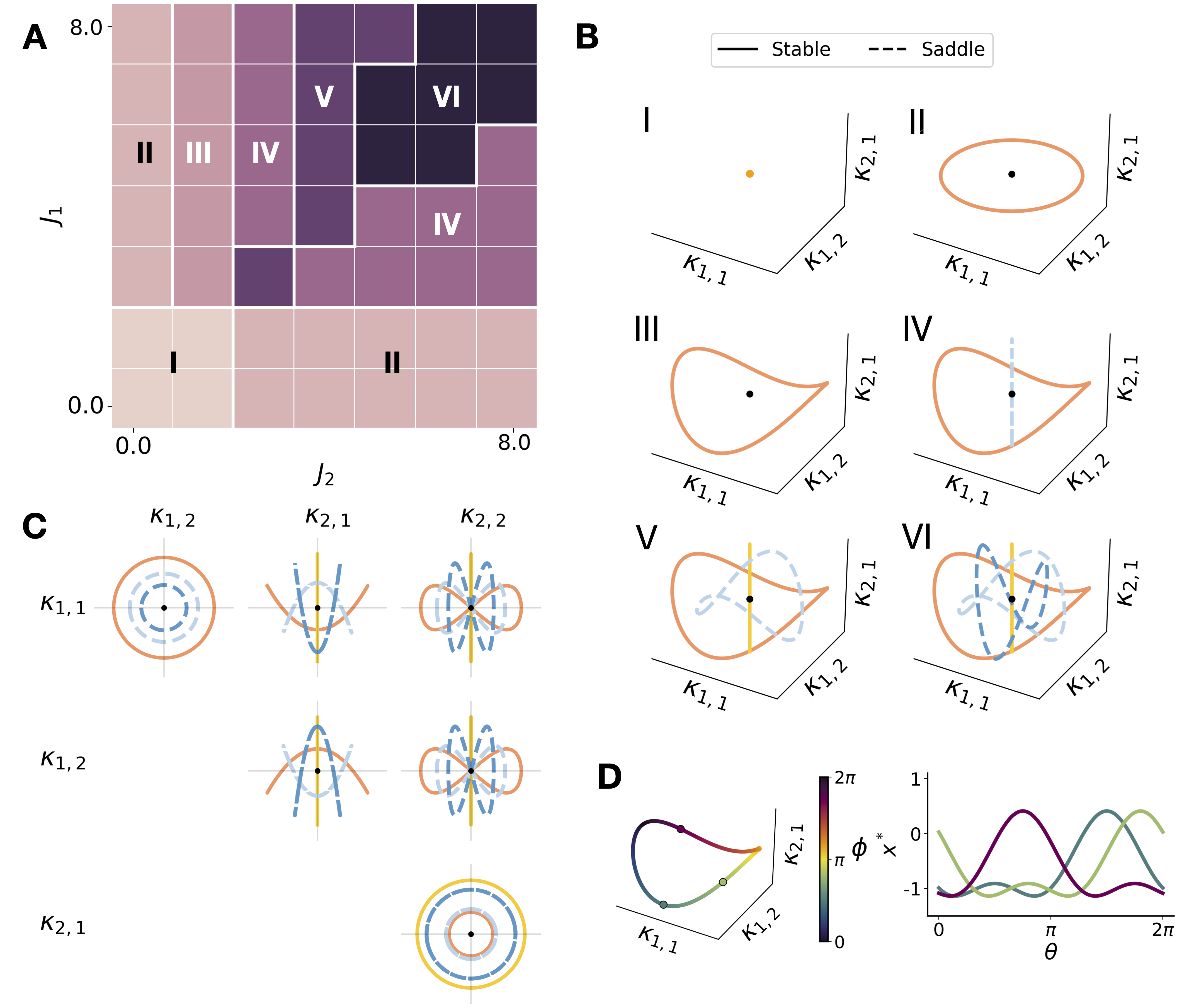} 
    
    \caption{Manifolds of fixed points in the rank five ring model. A) Bifurcation diagram showing six regions of parameters (I-VI) leading to  different combinations of co-existing fixed point solutions. Here we vary $J_1$ and $J_2$, while $J_0=-1$ is fixed.  B) Co-existing manifolds of fixed points within each of the six parameter regions. The manifolds are shown in the three-dimensional space corresponding to the Fourier components $\kappa_{1,1},\kappa_{1,2},\kappa_{2,1}$, while they are actually embedded in four dimensions $\kappa_{1,1},\kappa_{1,2},\kappa_{2,1}, \kappa_{2,2}$. In this projection,  rings  in the $\kappa_{2,1}$-$\kappa_{2,2}$ plane appear as vertical segments. I: single, stable fixed point;  II: unstable single point and stable 2d-ring; III: unstable single point and a stable 4d-ring; IV: unstable single point, a saddle 2d-ring (in the plane $\kappa_{2,1}$-$\kappa_{2,2}$) and  a stable 4d-ring (or vice versa); V: unstable single point, a stable 2d-ring and two 4d-rings (one stable, one saddle); VI: unstable single point and a stable 2d-ring and three 4d-rings (one stable, two saddle). 
     Stable solutions are shown as solid lines,  saddles are shown as dashed lines. C) Two-dimensional projections of the  manifolds of fixed point in  region VI onto planes corresponding to different pairs of Fourier component. D) Left: Three points on a stable solution manifold (type II solution), color-coded by phase $\phi$. Right: The corresponding population activity profiles, showing how shifted positions on the manifold correspond to shifted activity profiles.}
    \label{fig:ring4d}
\end{figure}

Eqs.~\eqref{system_eq_ring4d_reduced} admit three types of solutions that lead to different types of manifolds of fixed points in Eq.~\eqref{eq:x_4d_ring}: (i) solutions with $\kappa_0 = \varrho_o \neq 0$ and ${\kappa}_{1,1}={\kappa}_{2,1}=0$; from Eq.~\eqref{eq:rank5_rotation_mtx} such  solutions lead to single fixed point; (ii) solutions with ${\kappa}_{1,1}=\varrho_1\neq 0$ and ${\kappa}_{2,1}=0$, or ${\kappa}_{1,1}=0$ and ${\kappa}_{2,1} = \varrho_2 \neq0$; from Eq.~\eqref{eq:rank5_rotation_mtx} such  solutions lead to  rings of fixed points embedded in two dimensions, i.e. a manifold of intrinsic dimension $1$ and embedding dimension $2$; (iii) solutions with ${\kappa}_{1,1}=\varrho_1\neq 0$ and ${\kappa}_{2,1}=\varrho_2\neq 0$  lead to a ring of fixed points embedded in four dimensions, i.e. a manifold of intrinsic dimension $1$, but embedding dimension $4$. The existence of these solutions depend on the values of the parameters $J_0,J_1,J_2$. For a fixed set of parameter values, we find that several types of solutions can co-exist, leading to multi-stability between different manifolds of fixed points, some stable and others consisting of saddle points (Fig.~\ref{fig:ring4d}).

Specifically, applying the relation \eqref{eq:ring_rank4_manifold} to the particular solution ${\underline{\kappa}}(0)=(\varrho_0,\varrho_1,0,\varrho_2,0)$,  the solution for general $\phi \in [0,2\pi)$ is:
\begin{eqnarray}\label{eq:kappa_general_solution_2d}
{\underline{\kappa}}(\phi)=(\varrho_0, \varrho_1 \cos(\phi), \varrho_1 \sin(\phi),  \varrho_2 \cos(2\phi), \varrho_2 \sin(2\phi) ).
\end{eqnarray}
Inserting Eq.~\eqref{eq:kappa_general_solution_2d} into Eq.~\eqref{eq:x_4d_ring} results in the population activity for a fixed $\phi$ (Fig.~\ref{fig:ring4d}D, Right):
\begin{eqnarray}
x_{\phi}(\theta) = \varrho_0+2(\varrho_1 \cos(\theta-\phi)+\varrho_2 \cos(2(\theta-\phi))).
\label{eq:pop_activity_ring4d_param}
\end{eqnarray}

\newpage

\subsection{Toroidal RNNs}

A natural extension of the ring models are networks defined on a two-dimensional domain consisting of a product of two one-dimensional ring-like domains. More generally, we can consider domains that are products of $d$ rings, i.e. $d$-dimensional torii. 
The associated transformations then consist of angular translations along the $d$ dimensions, and form so-called {\em toroidal groups}. We denote as {\em toroidal RNNs} equivariant RNNs defined on such  domains and associated toroidal groups.
 While the ring models generate fixed-point manifolds of intrinsic dimension 1, i.e. parametrized by a single variable, 
 a $d$-dimensional toroidal RNN  can additionally generate  manifolds of intrinsic dimension  up to $d$. In the following we focus on the simplest toroidal RNN with $d=2$.

\subsubsection{Domain structure and symmetry group}  

In a two-dimensional toroidal RNN, neurons are  indexed by pairs of angles $(u_1,u_2)=(\theta_1, \theta_2)$ with $\theta_1, \theta_2 \in [0, 2\pi)$. The domain $\Omega$ is therefore the direct product of two circles $\Omega = \mathcal{S}^1 \times \mathcal{S}^1$, forming a two-dimensional torus $\mathcal{T}^2$. The symmetry group $G$ associated  with this domain is the set of two-dimensional angular translations on $\mathcal{T}^2$.  $G$ is called  the 2-dimensional toroidal group $\mathbb{T}^2$, and it is isomorphic the group product $\mathbb{S}^1 \times \mathbb{S}^1$. This group is isomorphic to $SO(2) \times SO(2)$,  to $(\mathbb{R}/2\pi \mathbb{Z})^2$, and to the product of two unitary groups $U(1) \times U(1)$.

We denote the elements of $G$ as $g_{\phi_1,\phi_2}$ for $\phi_1,\phi_2 \in [0, 2\pi)$. Their action on  $\Omega$ is given by
\begin{eqnarray}
g_{\phi_1,\phi_2}: & \Omega \to \Omega \nonumber\\
&(\theta_1, \theta_2) \mapsto (\theta_1 + \phi_1 \mod 2\pi,\,\,\,\, \theta_2 + \phi_2 \mod 2\pi). \nonumber
\end{eqnarray}

\subsubsection{Group convolution}

The recurrent input in Eq.~\eqref{rec_input_G} is defined as the
group convolution between the kernel function $c:\mathcal{S}^1 \times \mathcal{S}^1\to \mathbb{R}$ and the firing rate, $ \Phi[x]$. Following Eq.~\eqref{group_conv_cont0}, it is written as:
\begin{eqnarray}
F^t(\theta_1,\theta_2)&=&\int_{0}^{2\pi} \int_{0}^{2\pi}  c\big(g_{(\theta_1,\theta_2)}^{-1}(\phi_1,\phi_2)\big) \Phi \left[x^t(\phi_1,\phi_2)\right]\frac{d\phi_1 d\phi_2}{4\pi^2}\nonumber\\
&=&\int_{0}^{2\pi} \int_{0}^{2\pi}  c(\phi_1-\theta_1,\phi_2-\theta_2) \Phi \left[x^t(\phi_1,\phi_2)\right]\frac{d\phi_1 d\phi_2}{4\pi^2} \label{group_conv_cont_torus}  
\end{eqnarray}

where we integrate with respect to the normalized Haar measure $d\mu(\mathcal{S}^1 \times \mathcal{S}^1)=\frac{d\phi_1 d\phi_2}{4\pi^2}$.

\subsubsection{Group representations}

The irreps of $\mathbb{T}^2=\mathcal{S}^1 \times \mathcal{S}^1$ are therefore indexed by pairs of integers $(k_1,k_2)$ where $k_1,k_2 \in \mathbb{Z}$, and given by
$\rho_{(k_1,k_2)}(\phi_1, \phi_2)=\rho_{k_1}(\phi_1)\rho_{k_2}(\phi_2)$, with $\rho_{k}(\phi)$  an irrep of $\mathcal{S}^1$ (Eq.~\eqref{eq:ring_rhok}).
In the real-valued setting, these irreps fall into two categories.
For $k_1 = 0$ and $k_2 = 0$, the irrep is one-dimensional:
\begin{equation}\label{eq:torus_rho0}
\rho_{(0,0)}(\phi_1, \phi_2) = 1
\end{equation}

Otherwise, the irreps are two-dimensional rotation matrices, obtained by matrix multiplication of pairs of  $SO(2)$ irreps (Eq. 15.2 in \cite{Bump2013-kd}):
\begin{eqnarray}\label{eq:torus_rhok}
\rho_{(k_1,k_2)}(\phi_1, \phi_2) = \begin{pmatrix} \cos(k_1\phi_1 + k_2\phi_2) & -\sin(k_1\phi_1 + k_2\phi_2) \\ \sin(k_1\phi_1 + k_2\phi_2) & \cos(k_1\phi_1 + k_2\phi_2) \end{pmatrix},
\end{eqnarray}
for $\phi_1, \phi_2 \in [0, 2\pi)$.

While the irreps of $\mathbb{T}^2$ are rotational matrices as in the ring model, they depend on two independent indices and two independent angular variables.

\subsubsection{Group Fourier transform}

We next use the irreps in Eqs.~\eqref{eq:torus_rho0}-\eqref{eq:torus_rhok} to  write the components $\hat{f}_{k_1,k_2}$ of the group Fourier transform  for any function $f: \Omega \rightarrow \mathbb{R}$.

For $k_1 = 0$ and $k_2 = 0$, inserting Eq.~\eqref{eq:torus_rho0} into Eq.~\eqref{group_fourier_components}, $\hat{f}_{(0,0)}$ is a scalar given by

\begin{equation} \nonumber
\hat{f}_{(0,0)} = \int_{0}^{2\pi} \int_{0}^{2\pi}  f(\theta_1, \theta_2) \frac{d\theta_1 d\theta_2}{4\pi^2}.
\end{equation}

When $k_1$ and/or $k_2$ are not zero, inserting Eq.~\eqref{eq:torus_rhok} into Eq.~\eqref{group_fourier_components}, the components $\hat{f}_{k_1,k_2}$ are $2 \times 2$ matrices:

\begin{equation} \nonumber
\hat{f}_{(k_1,k_2)} = \int_{0}^{2\pi} \int_{0}^{2\pi}  f(\theta_1, \theta_2) \begin{pmatrix} \cos(k_1\theta_1 + k_2\theta_2) & -\sin(k_1\theta_1 + k_2\theta_2) \\ \sin(k_1\theta_1 + k_2\theta_2) & \cos(k_1\theta_1 + k_2\theta_2) \end{pmatrix} \frac{d\theta_1 d\theta_2}{4\pi^2}
\end{equation}

\begin{equation}\nonumber
= \begin{pmatrix} \hat{f}_{(k_1,k_2),1} & -\hat{f}_{(k_1,k_2),2} \\ \hat{f}_{(k_1,k_2),2} & \hat{f}_{(k_1,k_2),1} \end{pmatrix}
\end{equation}

where:

\begin{equation}\nonumber
\hat{f}_{(k_1,k_2),1} = \int_{0}^{2\pi} \int_{0}^{2\pi}  f(\theta_1, \theta_2)\cos(k_1\theta_1 + k_2\theta_2) \frac{d\theta_1 d\theta_2}{4\pi^2},
\end{equation}

\begin{equation}\nonumber
\hat{f}_{(k_1,k_2),2} = \int_{0}^{2\pi} \int_{0}^{2\pi}  f(\theta_1, \theta_2)\sin(k_1\theta_1 + k_2\theta_2) \frac{d\theta_1 d\theta_2}{4\pi^2}.
\end{equation}

As for the ring model, the matrix-valued Fourier components $\hat{f}_{(k_1,k_2)}$ have only two intependent scalar entries that we denote as $\hat{f}_{(k_1,k_2),1}$ and $\hat{f}_{(k_1,k_2),2}$.
These elements are the classical two-dimensional Fourier components, corresponding to basis functions on the torus.

\subsubsection{Fixed point equations for low-rank networks}

We now focus on low-rank networks, i.e. models for which the Fourier components $\hat{c}_{k_1,k_2}$ of the kernel $c$ are zero for $k_1>K_1$ or $k_2>K_2$. As shown in Sec.~\ref{sec:low-rank-eqRNN}, for such models the steady-state activation is described by a finite number of non-zero Fourier components $\kappa_{k_1,k_2}$ for $k_1\leq K_1$ and $k_2\leq K_2$. Here we use the expressions for the irreps to write explicitly the fixed-point equation (Eq.~\eqref{F_hat_group_Fourier}) satisfied by the Fourier components $\kappa_{k_1,k_2}$ of any steady-state activation $x^*$.

Using the Convolution Theorem (Eq.~\eqref{group-convolution-thm}), we get for $k_1=0$ and $k_2=0$:
\begin{eqnarray}\label{eq:torus_F_00}
\hat{F}_{(0,0)} &=& \hat{c}_{(0,0)} \cdot \hat{\Phi}_{(0,0)}
\end{eqnarray}
and, for $(k_1,k_2) \neq (0,0)$:
\begin{eqnarray}
\hat{F}_{(k_1,k_2)} &=& \hat{c}_{(k_1,k_2)} \cdot \hat{\Phi}_{(k_1,k_2)} \nonumber\\
&=& \begin{pmatrix} \hat{c}_{(k_1,k_2),1}\hat{\Phi}_{(k_1,k_2),1}-\hat{c}_{(k_1,k_2),2}\hat{\Phi}_{(k_1,k_2),2} & -\hat{c}_{(k_1,k_2),1}\hat{\Phi}_{(k_1,k_2),2}-\hat{c}_{(k_1,k_2),2}\hat{\Phi}_{(k_1,k_2),1} \\ \hat{c}_{(k_1,k_2),2}\hat{\Phi}_{(k_1,k_2),1}+\hat{c}_{(k_1,k_2),1}\hat{\Phi}_{(k_1,k_2),2} & -\hat{c}_{(k_1,k_2),2}\hat{\Phi}_{(k_1,k_2),2}+\hat{c}_{(k_1,k_2),1}\hat{\Phi}_{(k_1,k_2),1} \end{pmatrix} \nonumber\\
&\coloneq& \begin{pmatrix} \hat{F}_{(k_1,k_2),1} & -\hat{F}_{(k_1,k_2),2} \\ \hat{F}_{(k_1,k_2),2} & \hat{F}_{(k_1,k_2),1} \end{pmatrix} \label{eq:torus_F_k1k2}
\end{eqnarray}

where 
\begin{equation} \nonumber
\hat{F}_{(k_1,k_2),1} = \hat{c}_{(k_1,k_2),1}\hat{\Phi}_{(k_1,k_2),1}-\hat{c}_{(k_1,k_2),2}\hat{\Phi}_{(k_1,k_2),2},
\end{equation}
\begin{equation} \nonumber
\hat{F}_{(k_1,k_2),2} = \hat{c}_{(k_1,k_2),2}\hat{\Phi}_{(k_1,k_2),1}+\hat{c}_{(k_1,k_2),1}\hat{\Phi}_{(k_1,k_2),2}.
\end{equation}

To express explicitly $\hat{F}_{(k_1,k_2),1}$ and $\hat{F}_{(k_1,k_2),2}$, we write the expressions for $\hat{\Phi}_{(k_1,k_2)}$ in terms of the Fourier components $\kappa_{(k_1,k_2)}$ of the steady-state activation $x^*$.

For $k_1=0$ and $k_2=0$, we have
\begin{eqnarray}\label{eq:torus_phi_00}
\hat{\Phi}_{(0,0)}=\int_{0}^{2\pi} \int_{0}^{2\pi}  \Phi\left[x^*(\theta_1,\theta_2)\right] \frac{d\theta_1 d\theta_2}{4\pi^2},
\end{eqnarray}
while for $(k_1,k_2) \neq (0,0)$, the independent matrix elements are given by

\begin{eqnarray}
\hat{\Phi}_{(k_1,k_2),1} &=& \int_{0}^{2\pi} \int_{0}^{2\pi}  \Phi\left[x^*(\theta_1,\theta_2)\right]\cos(k_1\theta_1 + k_2\theta_2) \frac{d\theta_1 d\theta_2}{4\pi^2}, \label{eq:torus_Phik1k21_hat}\\
\hat{\Phi}_{(k_1,k_2),2} &=& \int_{0}^{2\pi} \int_{0}^{2\pi}  \Phi\left[x^*(\theta_1,\theta_2)\right]\sin(k_1\theta_1 + k_2\theta_2) \frac{d\theta_1 d\theta_2}{4\pi^2}. \label{eq:torus_Phik1k22_hat}
\end{eqnarray}

We next write the steady-state activation $x^*$  using the Inverse Fourier Transform (Eq.~\eqref{Peter-Weyl}):
\begin{eqnarray}
x^*(\theta_1,\theta_2) &=&  \sum_{k_1=0, \, k_2=0}^{K_1,K_2} \tilde{d}_{(k_1,k_2)} \, \text{tr} \big(\kappa_{(k_1,k_2)} \rho_{(k_1,k_2)}^{-1}(\theta_1,\theta_2)\big)  \nonumber\\ 
&=&   \kappa_{(0,0)}+ \sum_{\substack{k_1=0, \, k_2=0 \\ (k_1,k_2) \neq (0,0)}}^{K_1, K_2} \, \text{tr} \left(\begin{pmatrix} \kappa_{(k_1,k_2),1} & -\kappa_{(k_1,k_2),2} \\ \kappa_{(k_1,k_2),2} & \kappa_{(k_1,k_2),1} \end{pmatrix} \begin{pmatrix} \cos (k_1\theta_1 + k_2\theta_2) & \sin (k_1\theta_1 + k_2\theta_2) \\ -\sin (k_1\theta_1 + k_2\theta_2) & \cos (k_1\theta_1 + k_2\theta_2) \end{pmatrix} \right) \nonumber\\
&=&   \kappa_{(0,0)}+ \sum_{\substack{k_1=0, \, k_2=0 \\ (k_1,k_2) \neq (0,0)}}^{K_1, K_2} 2\left(\kappa_{(k_1,k_2),1}\cos(k_1\theta_1 + k_2\theta_2) + \kappa_{(k_1,k_2),2}\sin(k_1\theta_1 + k_2\theta_2)\right). \label{eq:torus_x_kappa}
\end{eqnarray}

Here we include only the Fourier components up to $K_1$ and $K_2$, because at steady-state $\kappa_{(k_1,k_2)}=0$ for $k_1>K_1$ or $k_2>K_2$.

Inserting Eq.~\eqref{eq:torus_x_kappa} into Eqs.~\eqref{eq:torus_phi_00}, \eqref{eq:torus_Phik1k21_hat} and \eqref{eq:torus_Phik1k22_hat} gives us explicit expressions for $\hat{\Phi}_{(0,0)}, \hat{\Phi}_{(k_1,k_2),1}$ and $\hat{\Phi}_{(k_1,k_2),2}$ as functions of the set of scalars $\kappa_{(0,0)}, \kappa_{(k_1,k_2),1}, \kappa_{(k_1,k_2),2}$ for $k_1=1\ldots K_1$ and $k_2=1\ldots K_2$.

This leads to the following fixed point equations:

\begin{eqnarray}  \label{eq:fixed_point_eq0_torus}
 \kappa_{(0,0)} =  \hat{c}_{0} \int\int_{0}^{2\pi} \Phi\left[\kappa_{(0,0)}+\sum_{\substack{k_1=0, \, k_2=0 \\ (k_1,k_2) \neq (0,0)}}^{K_1, K_2}  2\left(\kappa_{(k_1,k_2),1}\cos(k_1\theta_1 + k_2\theta_2) + \kappa_{(k_1,k_2),2}\sin(k_1\theta_1 + k_2\theta_2)\right)\right] \frac{d\theta_1 d\theta_2}{4\pi^2}, 
\end{eqnarray}
\begin{eqnarray}  \label{eq:fixed_point_eq1_torus}
\kappa_{(k_1,k_2),1} = \int\int_{0}^{2\pi} \Phi\left[ \kappa_{(0,0)}+\sum_{\substack{k'_1=0, \, k'_2=0 \\ (k'_1,k'_2) \neq (0,0)}}^{K_1, K_2}  2\left(\kappa_{(k'_1,k'_2),1}\cos(k'_1\theta_1 + k'_2\theta_2) + \kappa_{(k'_1,k'_2),2}\sin(k'_1\theta_1 + k'_2\theta_2)\right)\right] \\ \times \Big(\hat{c}_{(k_1,k_2),1} \cos(k_1\theta_1 + k_2\theta_2) -\hat{c}_{(k_1,k_2),2}\sin(k_1\theta_1 + k_2\theta_2)\Big)\frac{d\theta_1 d\theta_2}{4\pi^2},
\end{eqnarray}
\begin{eqnarray}  \label{eq:fixed_point_eq2_torus}
\kappa_{(k_1,k_2),2} = \int\int_{0}^{2\pi} \Phi\left[ \kappa_{(0,0)}+\sum_{\substack{k'_1=0, \, k'_2=0 \\ (k'_1,k'_2) \neq (0,0)}}^{K_1, K_2} 2\left(\kappa_{(k'_1,k'_2),1}\cos(k'_1\theta_1 + k'_2\theta_2) + \kappa_{(k'_1,k'_2),2}\sin(k'_1\theta_1 + k'_2\theta_2)\right)\right] \\ \times \Big(\hat{c}_{(k_1,k_2),2}\cos(k_1\theta_1 + k_2\theta_2) +\hat{c}_{(k_1,k_2),1}\sin(k_1\theta_1 + k_2\theta_2)\Big) \frac{d\theta_1 d\theta_2}{4\pi^2} .
\end{eqnarray}

\subsubsection{Fixed point manifold}

Any solution ${\underline{\kappa}}$ of the fixed point equation \eqref{eq:fixed_point_eq0_torus}, \eqref{eq:fixed_point_eq1_torus}, \eqref{eq:fixed_point_eq2_torus} therefore leads to a manifold of fixed points in Fourier coordinates parametrized by elements of the group $\mathbb{S}^1\times \mathbb{S}^1$:

\begin{eqnarray}\label{parametrized_solution_torus}
{\underline{\kappa}}(\phi_1, \phi_2)=R^{K_1,K_2}(\phi_1, \phi_2){\underline{\kappa}}.
\end{eqnarray}

where $R^{K_1,K_2}(\phi_1, \phi_2)$ is block-diagonal containing at most $(K_1+1)(K_2+1)$ rotation matrices that act on the corresponding Fourier components:

\begin{eqnarray}\label{rotation_matrix_torus}
R^{K_1,K_2}(\phi_1, \phi_2)=\begin{pmatrix}
\rho_{(0,0)}(\phi_1, \phi_2) & & & &  &  \\
& \rho_{(1,0)}(\phi_1, \phi_2) & & & & \\
& & \rho_{(0,1)}(\phi_1, \phi_2) & & & \\
&  & & \rho_{(1,1)}(\phi_1, \phi_2)   & \\
& & & & \ddots \\
& & & & & \rho_{(K_1,K_2)}(\phi_1, \phi_2)
\end{pmatrix}.
\end{eqnarray}

\subsubsection{Rank five toroidal RNN \label{sec:rank5-torus}}

As a concrete example, we consider the rank five toroidal model, which we explicitly contrast with the rank five ring model described in Sec.~\ref{sec:rank5_ring}. The rank five toroidal model is
defined by the connectivity kernel
\begin{equation} \nonumber
c(\theta_1,\theta_2)=J_0+J_1 \cos(\theta_1) +J_2 \cos(\theta_2),
\end{equation}
where $J_0, J_1 \text{ and } J_2$ are scalar parameters. In this case, the non-zero Fourier components are

\begin{eqnarray}
\hat{c}_{(0,0)} = J_0 \nonumber\\
\hat{c}_{(1,0)} =  \begin{pmatrix} \hat{c}_{(1,0)1} & -\hat{c}_{(1,0)2} \\  \hat{c}_{(1,0)2} & \hat{c}_{(1,0)1} \end{pmatrix} = J_1 \begin{pmatrix} \frac{1}{2} & 0 \\ 0 & \frac{1}{2} \end{pmatrix} \nonumber\\
\hat{c}_{(0,1)} =  \begin{pmatrix} \hat{c}_{(0,1)1} & -\hat{c}_{(0,1)2} \\  \hat{c}_{(0,1)2} & \hat{c}_{(0,1)1} \end{pmatrix} = J_2 \begin{pmatrix} \frac{1}{2} & 0 \\ 0 & \frac{1}{2} \end{pmatrix} \nonumber
\end{eqnarray}
while for $k_1 \geq 1$ and $k_2 \geq 1$:
\begin{eqnarray}
\hat{c}_{(k_1,k_2)} = \begin{pmatrix} 0 & 0 \\ 0 & 0 \end{pmatrix}.\nonumber
\end{eqnarray}

The Fourier components of the kernel are zero for $k_1 \geq 1 $ and $k_2 \geq 1$, resulting in a rank $5$ model. From Eq.~\eqref{eq:torus_x_kappa}, the steady state activation $x^*(\theta_1,\theta_2)$ is given by
\begin{eqnarray}\label{eq:x_4d_torus}
x(\theta_1,\theta_2) 
&=& \kappa_{(0,0)}+ 2\left(\kappa_{(1,0),1}\cos(\theta_1 ) + \kappa_{(1,0),2}\sin(\theta_1 )+\kappa_{(0,1),1}\cos(\theta_2 ) + \kappa_{(0,1),2}\sin(\theta_2 )\right).
\end{eqnarray}

The fixed point equations \eqref{eq:fixed_point_eq0_torus}-\eqref{eq:fixed_point_eq2_torus} 
reduce to a set of $5$ coupled equations for the scalars $\kappa_{(0,0)}$, $\kappa_{(1,0),1}$, $\kappa_{(1,0),2}$, $\kappa_{(0,1),1}$, $\kappa_{(0,1),2}$:

\begin{eqnarray}
\begin{cases}\label{eq:system_eq_torus}
 \kappa_{(0,0)} =  J_0 \int_{0}^{2\pi} \Phi\left[ \kappa_{(0,0)}+ 2\left(\kappa_{(1,0),1}\cos(\theta_1 ) + \kappa_{(1,0),2}\sin(\theta_1 )+\kappa_{(0,1),1}\cos(\theta_2 ) + \kappa_{(0,1),2}\sin(\theta_2 )\right)\right] \frac{d\theta_1 d\theta_2}{4\pi^2}, \\
\kappa_{(1,0),1} = \frac{J_1}{2} \int_{0}^{2\pi} \Phi\left[  \kappa_{(0,0)}+ 2\left(\kappa_{(1,0),1}\cos(\theta_1 ) + \kappa_{(1,0),2}\sin(\theta_1 )+\kappa_{(0,1),1}\cos(\theta_2 ) + \kappa_{(0,1),2}\sin(\theta_2 )\right)\right] \cos(\theta_1)\frac{d\theta_1 d\theta_2}{4\pi^2},\\
\kappa_{(1,0),2} = \frac{J_1}{2}\int_{0}^{2\pi} \Phi\left[  \kappa_{(0,0)}+ 2\left(\kappa_{(1,0),1}\cos(\theta_1 ) + \kappa_{(1,0),2}\sin(\theta_1 )+\kappa_{(0,1),1}\cos(\theta_2 ) + \kappa_{(0,1),2}\sin(\theta_2 )\right)\right] \sin(\theta_1) \frac{d\theta_1 d\theta_2}{4\pi^2},\\
\kappa_{(0,1),1} = \frac{J_2}{2}\int_{0}^{2\pi} \Phi\left[  \kappa_{(0,0)}+ 2\left(\kappa_{(1,0),1}\cos(\theta_1 ) + \kappa_{(1,0),2}\sin(\theta_1 )+\kappa_{(0,1),1}\cos(\theta_2 ) + \kappa_{(0,1),2}\sin(\theta_2 )\right)\right] \cos(\theta_2)\frac{d\theta_1 d\theta_2}{4\pi^2},\\
\kappa_{(0,1),2} = \frac{J_2}{2}\int_{0}^{2\pi} \Phi\left[  \kappa_{(0,0)}+ 2\left(\kappa_{(1,0),1}\cos(\theta_1 ) + \kappa_{(1,0),2}\sin(\theta_1 )+\kappa_{(0,1),1}\cos(\theta_2 ) + \kappa_{(0,1),2}\sin(\theta_2 )\right)\right] \sin(\theta_2) \frac{d\theta_1 d\theta_2}{4\pi^2}.
\end{cases}
\end{eqnarray}

Using Eq.~\eqref{parametrized_solution_torus}, any solution $\underline{\kappa}$:=$(\kappa_{(0,0)},\kappa_{(1,0),1},\kappa_{(1,0),2},\kappa_{(0,1),1},\kappa_{(0,1),2})$ of the system \eqref{eq:system_eq_torus} leads to a manifold of fixed points in Fourier coordinates, parametrized by elements of $SO(2) \times SO(2)$:
\begin{eqnarray}\label{eq:torus_manifold}
{\underline{\kappa}}(\phi_1,\phi_2)=R^{(1,1)}(\phi_1,\phi_2){\underline{\kappa}}.
\end{eqnarray}

Here, $R^{(1,1)}(\phi_1,\phi_2)$, is a block diagonal 5D matrix containing independent rotation matrices:  
\begin{eqnarray} \label{eq:rank5_rotation_mtx_torus}
R^{(1,1)}(\phi_1,\phi_2)=\begin{pmatrix}
1 & 0 & 0 & 0 & 0 \\
0 & \cos(\phi_1) & -\sin(\phi_1)  & 0 & 0 \\
0 & \sin(\phi_1)& \cos(\phi_1) & 0 & 0 \\
0 & 0 & 0 & \cos(\phi_2) & -\sin(\phi_2) \\
0 & 0 & 0 & \sin(\phi_2) & \cos(\phi_2)
\end{pmatrix}.
\end{eqnarray}

Setting $\phi_1=0 \text{ and } \phi_2=0$, we obtain a particular solution ${\underline{\kappa}}(0,0)=(\kappa_{(0,0)},\kappa_{(1,0),1},0,\kappa_{(0,1),1},0)$. The system \eqref{eq:system_eq_torus} then reduces to the three-dimensional system:

\begin{eqnarray}
\begin{cases}\label{system_eq_torus_reduced}
 \kappa_{(0,0)} =  J_0 \int_{0}^{2\pi} \Phi\left[ \kappa_{(0,0)}+ 2\left(\kappa_{(1,0),1}\cos(\theta_1 ) +\kappa_{(0,1),1}\cos(\theta_2 ) \right)\right] \frac{d\theta_1 d\theta_2}{4\pi^2}, \\
\kappa_{(1,0),1} =\frac{J_1}{2} \int_{0}^{2\pi} \Phi\left[  \kappa_{(0,0)}+ 2\left(\kappa_{(1,0),1}\cos(\theta_1 ) + \kappa_{(0,1),1}\cos(\theta_2 ) \right)\right] \cos(\theta_1)\frac{d\theta_1 d\theta_2}{4\pi^2},\\
\kappa_{(0,1),1} = \frac{J_2}{2} \int_{0}^{2\pi} \Phi\left[  \kappa_{(0,0)}+ 2\left(\kappa_{(1,0),1}\cos(\theta_1 )+\kappa_{(0,1),1}\cos(\theta_2 )\right)\right] \cos(\theta_2)\frac{d\theta_1 d\theta_2}{4\pi^2}.
\end{cases}
\end{eqnarray}
The stability of each  fixed-point is determined by the eigenvalues of the Jacobian of  the r.h.s, evaluated at the fixed point. This stability is then inherited by all the fixed points on the resulting manifold.

\begin{figure}[!ht]
    \centering
    \includegraphics[width=14cm]{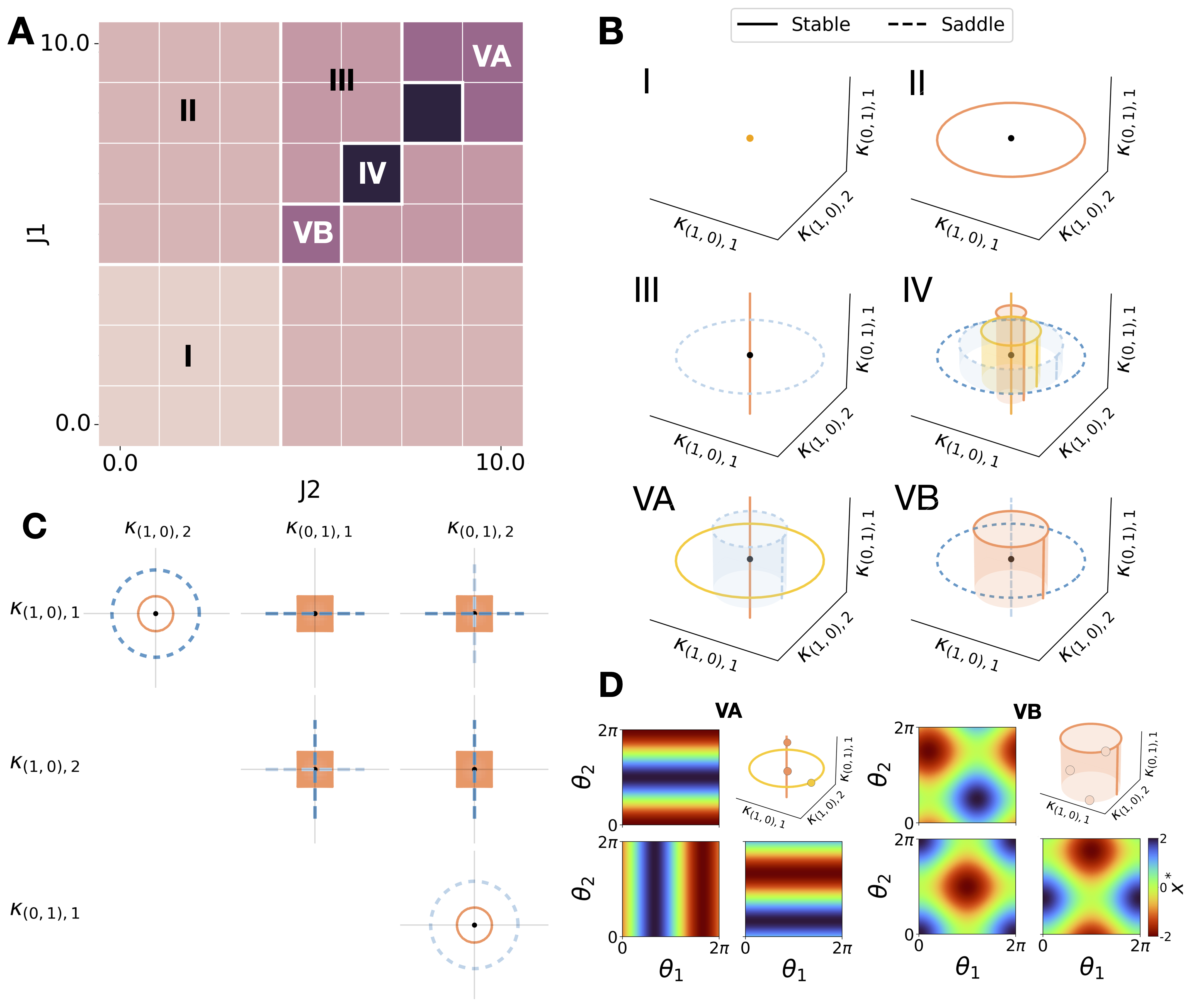} 
    \caption{Manifolds of fixed points in the rank five toroidal RNN. A) Bifurcation diagram showing five regions of parameters (I-V) leading to  different combinations of co-existing fixed point solutions. Here we vary $J_1$ and $J_2$, while $J_0=-3$ is fixed.  B) Illustration of co-existing manifolds of fixed points within each of the five parameter regions.
    The manifolds are shown in the three-dimensional space corresponding to the Fourier components $\kappa_{(1,0),1},\kappa_{(1,0),2},\kappa_{(0,1),1}$, while they are actually embedded in four dimensions $\kappa_{(1,0),1},\kappa_{(1,0),2},\kappa_{(0,1),1},\kappa_{(0,1),2}$. In this projection, torii appear as cylinders, and rings  in the $\kappa_{(0,1),1}-\kappa_{(0,1),2}$ plane as vertical segments.
    I: single, stable fixed point; II: unstable single and a stable 2d-ring; III: unstable single point and two 2d-rings in the planes $\kappa_{(1,0),1}-\kappa_{(1,0),2}$ (saddle) and $\kappa_{(0,1),1}-\kappa_{(0,1),2}$ (stable); IV: unstable single point, two stable 2d-rings (in the planes $\kappa_{(1,0),1}-\kappa_{(1,0),2}$  and $\kappa_{(0,1),1}-\kappa_{(0,1),2}$) and three tori (one stable, two saddles); V: unstable single point, two stable 2d-rings and one saddle torus (or vice versa). 
     Stable solutions are shown as solid lines,  saddles are shown as dashed lines. C) Two-dimensional projections of the  manifolds of fixed point in  region VB onto planes corresponding to different pairs of Fourier component $\kappa_{(1,0),1},\kappa_{(1,0),2},\kappa_{(0,1),1},\kappa_{(0,1),2}$. The projection of a torus (orange) leads to circles in the planes $\kappa_{(1,0),1}-\kappa_{(1,0),2}$ and $\kappa_{(0,1),1}-\kappa_{(0,1),2}$, and squares in the other planes. D) Left: Population activity profiles corresponding to the union of two stable orthogonal 2d-ring manifolds, displaying a band pattern. Right:  Population activity profiles corresponding to the stable torus manifold, displaying a localized bump pattern.}
    \label{fig:torus}
\end{figure}

Eqs.~\eqref{system_eq_torus_reduced} admit three types of solutions that lead to different types of manifolds of fixed points in Eq.~\eqref{eq:x_4d_torus}: (i) solutions with $\kappa_{(0,0)}=\varrho_0\neq 0$ and ${\kappa}_{(1,0)1}={\kappa}_{(1,0)2}=0$; from Eq.~\eqref{eq:rank5_rotation_mtx_torus} such  solutions lead to single fixed point, i.e. a manifold of intrinsic dimension $0$; (ii) solutions with ${\kappa}_{(1,0)1}=\varrho_1\neq 0$ and ${\kappa}_{(1,0)2}=0$, or ${\kappa}_{(1,0)2}=\varrho_2\neq0$ and ${\kappa}_{(1,0)1}=0$; from Eq.~\eqref{eq:rank5_rotation_mtx_torus} each such  solution leads to a ring of fixed points embedded in two dimensions, i.e. a manifold of intrinsic dimension $1$ and embedding dimension $2$; (iii) solutions with ${\kappa}_{(1,0)1}=\varrho_1\neq 0$ and ${\kappa}_{(1,0)2}=\varrho_2\neq 0$  lead to a torus of fixed points embedded in four dimensions, i.e. a manifold of intrinsic dimension $2$, and embedding dimension $4$. The existence of these solutions depends on the values of the parameters $J_0,J_1,J_2$. For a fixed set of parameter values, we find that several types of solutions can co-exist, leading to multi-stability between fixed points manifolds of various intrinsic and embedding dimensions, some stable others consisting of saddle points (Fig.~\ref{fig:torus}).

Specifically, applying the relation \eqref{eq:torus_manifold} to particular solution ${\underline{\kappa}}(0,0)=(\varrho_0,\varrho_1,0,\varrho_2,0)$, the general solution for $(\phi_1,\phi_2) \in [0,2\pi) \times [0,2\pi)$ is:
\begin{eqnarray}\label{kappa_general_solution_torus}
{\underline{\kappa}}(\phi_1,\phi_2)=(\varrho_0, \varrho_1 \cos(\phi_1), \varrho_1 \sin(\phi_1),  \varrho_2\cos(\phi_2), \varrho_2 \sin(\phi_2) ).
\end{eqnarray}
Inserting Eq.~\eqref{kappa_general_solution_torus} into Eq.~\eqref{eq:x_4d_torus} results in the population activity for fixed $(\phi_1,\phi_2)$ (Fig.~\ref{fig:torus}D):
\begin{eqnarray}
x_{(\phi_1,\phi_2)}(\theta_1,\theta_2) = \varrho_0+ 2\left(\varrho_1 \cos(\theta_1-\phi_1)+\varrho_2\cos(\theta_2-\phi_2)\right).
\label{eq:pop_activity_torus_param}
\end{eqnarray}

The rank five toroidal eqRNN and the rank five ring model are both five-dimensional models, in the sense that at most five Fourier components of the steady state can be simultaneously non-zero. The matrices determining the fixed point manifolds have a similar block structure, with the key difference that for the toroidal model the rotation matrix depends on two independent angles (Eq.~\eqref{eq:rank5_rotation_mtx_torus}), while for the ring model the blocks depend on a single angle and its double (Eq.~\eqref{eq:rank5_rotation_mtx}). This leads to an important difference in the resulting fixed-point manifolds.
While both models can generate ring manifolds embedded in two dimensions, toroidal eqRNNs generate torii embedded in four-dimensions, while the ring model generates ring manifolds embedded in four dimensions.

\newpage

\subsection{Spherical RNNs \label{sec:sphere}}

While toroidal RNNs are a generalization of the ring model to a higher-dimensional domain with a product structure, the framework
of eqRNNs allows us to extend ring models to more general higher-dimensional domains. The simplest two-dimensional domain without a product structure is the two-dimensional sphere, which leads to a class of models we call {\em spherical RNNs}. These models are an instance of the situation where the domain $\Omega$ and the associated group $G$ are not isomorphic, and we use this example to illustrate how the eqRNN framework extends to that case.

\subsubsection{Domain structure and symmetry group} 

In spherical RNNs, neurons are indexed by points on the 2-dimensional sphere $\mathcal{S}^2$. The domain is therefore $\Omega=\mathcal{S}^2$, and each unit is parametrized by a pair of angles $(u_1,u_2)=(\theta, \phi)$ with $\theta \in [0, \pi]$ (longitude) and $\phi \in [0, 2\pi)$ (colatitude). Importantly, $\mathcal{S}^2$ does not itself form a group.
The  group associated with $\mathcal{S}^2$ is instead $G = SO(3)$, the group of all rotations in $\mathbb{R}^3$.
Elements of $SO(3)$ are parametrized by three parameters, and we denote them as $g_{\alpha,\beta,\gamma}$, where $\alpha, \beta, \gamma$ are the Euler angles with $\alpha, \gamma \in [0, 2\pi)$ and $\beta \in [0, \pi]$. Any point $(\theta, \phi) \in \mathcal{S}^2$ can be mapped onto any other point $(\theta', \phi') \in \mathcal{S}^2$ by applying an element $g_{\alpha,\beta,\gamma} \in SO(3)$ such that $g_{\alpha,\beta,\gamma}  (\theta, \phi) = (\theta', \phi')$. The sphere $\mathcal{S}^2$ is therefore a homogeneous space under the action of $SO(3)$ \citep{Kondor_Trivedi, Bekkers2019-qy, Cohen2018-lz}. However, in contrast to the two previous examples, $\mathcal{S}^2$ and $SO(3)$ are not isomorphic, because the element of $SO(3)$ mapping $(\theta, \phi)$ onto  $(\theta', \phi')$ is in general not unique. Indeed, 
while $g_{\alpha,\beta,\gamma}$ describes all possible rotations of a rigid body in $\mathbb{R}^3$ using three parameters, on the sphere $\mathcal{S}^2$ only two angles are necessary to describe any point. This is because, the third Euler angle $\gamma$ represents a rotation of a rigid body around the local axis defined by $(\alpha,\beta)$ , which does not change the position of a point on the sphere. 
A bijective mapping between $\mathcal{S}^2$ and $SO(3)$ is therefore obtained by fixing $\gamma=0$. More formally, this amounts to identifying $\mathcal{S}^2$ with the quotient $SO(3)/SO(2)$.

Spherical RNNs are therefore an example of a model in which the domain $\Omega$ is not a group, but can be formulated as a quotient $G/H$ of two groups $G$ and $H$. \cite{Kondor_Trivedi} show how, in such a situation, the concepts of group convolution and group Fourier transform can be extended based on the irreducible representations of $G$ and $H$. Here we follow their construction in the case $G=SO(3)$ and $H=SO(2)$.

\subsubsection{Spherical convolution}
A convolution with respect to a group $G$ is habitually defined on functions $G\to \mathbb{R}$ (Eq.~\eqref{group_conv0}). To construct a convolutional RNN defined on a domain $\Omega$, we instead need to define the recurrent input as a convolution of functions $\Omega \to \mathbb{R}$. In previous examples, we had $\Omega = G$, so this was not an issue, but for spherical RNNs $\Omega$ and $G$ are not isomorphic. Following Kondor and Trivedi, we define the convolution on $\Omega=G/H$ by lifting the convolution from $G/H$ to $G$ (see Appendix \ref{app:group_convolution}).

For $\Omega=SO(3)/SO(2)$, this is equivalent to defining the recurrent input in terms of the {\em spherical convolution} \citep{Bekkers2019-qy, Kondor2018-lo, Kennedy2011-hd, Roddy2021-hp}. For a connectivity kernel $c:\mathcal{S}^2 \to \mathbb{R}$, and the firing rate $\Phi:\mathcal{S}^2 \to \mathbb{R}$, the recurrent input $F^t:\mathcal{S}^2 \to \mathbb{R}$ at time step $t$ is defined as

\begin{eqnarray}
\label{group_conv_cont_sphere}
F^t(\theta,\phi)&=&\int_{0}^{2\pi} \int_{0}^{\pi}c(g_{(\theta,\phi,\gamma=0)}^{-1}(\theta',\phi')) \Phi (x_t(\theta',\phi')) \frac{\sin \theta'  d\theta'  d\phi'}{4\pi},
\end{eqnarray}

where we integrate with respect to the normalized Haar measure $d\mu(\mathcal{S}^2)=\frac{\sin \theta'  d\theta'  d\phi'}{4\pi}$.

\subsubsection{Representations of $SO(3)$ and their restriction to $\mathcal{S}^2$}

The habitual definition of the group Fourier transform for functions $G\to \mathbb{R}$ relies on the irreducible representations of $G$ (Eq.~\eqref{group_fourier_components}). To define low-rank networks on $\Omega$, we instead need a Fourier transform for functions $\Omega\to \mathbb{R}$. For $\Omega=G/H$, Kondor and Trivedi show that a Fourier transform on $\Omega$ is naturally defined by restricting the irreps of $G$ to $G/H$, which amounts to keeping only specific columns of each irrep of $G$. While the group Fourier transform consists of  matrix-valued components (Eq.~\eqref{group_fourier_components}), the Fourier transform on $\Omega=G/H$ has instead  vector-valued components. Here we follow the construction of Kondor and Trivedi for $G=SO(3)$ and $H=SO(2)$. This approach provides a direct relation between the irreps of $SO(3)$ and spherical harmonics that are habitually used to define the Fourier transform on $\mathcal{S}^2$ \citep{Kondor2018-lo, Kennedy2011-hd}.

The irreducible representations of $SO(3)$ are  indexed by a parameter $\ell \in \mathbb{N}$. For a given ${\ell}$, the representation $\rho_{\ell}(g_{\alpha, \beta, \gamma})$ of a group element $g_{\alpha, \beta, \gamma} \in SO(3)$, is a $(2\ell+1)\times(2\ell+1)$ matrix called the $2\ell+1$ dimensional Wigner $D$-matrix \citep{Cohen2014-uw, Kosmann-Schwarzbach2009-uj, Cohen2018-jb}:
\begin{equation}\nonumber
\rho_{\ell}(g_{\alpha, \beta, \gamma}) = D^{\ell}(\alpha, \beta, \gamma).
\end{equation}

Restricting $SO(3)$ to $SO(3)/SO(2)$ corresponds to  setting $\gamma=0$ in $\rho_{\ell}(g_{\alpha, \beta, \gamma})$. This amounts to selecting the central column of the Wigner matrix $D^{\ell}$, which contains a $(2\ell+1)$-dimensional vector called the spherical harmonic $\underline{ Y_{\ell}}$.

For $\ell=0$, $Y_0$ is a scalar given by:
\begin{equation}  
Y_0(\theta,\phi)= 1,\nonumber
\end{equation}
while for $\ell>0$,   the elements $Y_\ell^m$ for $m=-\ell,\ldots,\ell$ of $\underline{ Y_{\ell}}$ are given by

\begin{equation}
Y_{\ell m}(\theta,\phi) =
\begin{cases} 
(-1)^m \sqrt{2} \sqrt{{(2\ell + 1)} \frac{(\ell - |m|)!}{(\ell + |m|)!}} \, P_\ell^{|m|}(\cos\theta) \sin(|m| \phi), & \text{if } m < 0, \\[10pt]
\sqrt{{(2\ell + 1)}} \, P_\ell^m(\cos\theta), & \text{if } m = 0, \\[10pt]
(-1)^m \sqrt{2} \sqrt{{(2\ell + 1)} \frac{(\ell - m)!}{(\ell + m)!}} \, P_\ell^m(\cos\theta) \cos(m \phi), & \text{if } m > 0
\end{cases}\nonumber
\end{equation}
where $P_\ell^{m}: [-1,1] \to \mathbb{R}$ are the associated Legendre polynomials \citep{Kosmann-Schwarzbach2009-uj}.

\subsubsection{Spherical Fourier transform}

Using the restriction of $SO(3)$ irreps, the Fourier transform for a function $f:\mathcal{S}^2\to \mathbb{R}$ consists of vector-valued components indexed by $\ell \in \mathbb{N}$ and obtained by projecting $f$ on spherical harmonics.

For $l = 0$,  the Fourier component $\hat{f}_0$ is a scalar given by
\begin{eqnarray}
\hat{f}_0 &=& \int_{0}^{2\pi} \int_{0}^{\pi}f(\theta, \phi)  \,Y_0(\theta,\phi)\frac{\sin\theta d\theta \, d\phi}{4\pi},
\label{matrix_fourier_sphere_0}
\end{eqnarray}
while for $l > 0$,  $\boldsymbol{\hat{f}_{\ell}} \in \mathbb{R}^{2\ell+1}$ is a vector containing $(2\ell+1)$ elements $\hat{f}_{\ell,m}$, $m=-l,..,l$,  given by:
\begin{equation}
\hat{f}_{\ell,m} = \int_{0}^{2\pi} \int_{0}^{\pi}f(\theta, \phi) Y_{\ell,m}(\theta, \phi) \, \frac{\sin\theta d\theta \, d\phi}{4\pi}.\nonumber
\end{equation}

The function $f$ can be reconstructed from its Fourier components using the real-valued spherical harmonic expansion, which is equivalent to the inverse Fourier transform on $\mathcal{S}^2$ \citep{Kondor2018-lo}:
\begin{equation}
f(\theta, \phi) = \sum_{\ell=0}^\infty \sum_{m=-\ell}^{\ell} \hat{f}_{\ell,m} Y_{\ell,m}(\theta, \phi).\nonumber
\end{equation}

\subsubsection{Fixed point equations for low-rank models}

We now focus on low-rank models on the sphere, i.e., models where the Fourier coefficients $\hat{c}_{\ell,m}$ of the kernel $c(\theta, \phi)$ are zero for $\ell > L$. As shown in Sec.~\ref{sec:low-rank-eqRNN}, for such models the steady-state activation is described by a finite number of non-zero Fourier components $\kappa_{\ell,m}$ for $\ell \leq L$ and $-\ell \leq m \leq \ell$. Here we use the expressions for the irreps to write explicitly the fixed-point equation (Eq.~\eqref{F_hat_group_Fourier}) satisfied by the Fourier coefficients $\kappa_{\ell,m}$ of any steady-state activation $x^*$.

Using the Convolution Theorem (Eq.~\eqref{group-convolution-thm}), we get for $\ell=0$ (and $m=0$):
\begin{eqnarray}\label{eq:sphere_F_0}
\hat{F}_{0} &=& \hat{c}_{0} \cdot \hat{\Phi}_{0}
\end{eqnarray}

For $\ell > 0$, $m \in [-\ell, \ell]$:
\begin{eqnarray}\label{eq:sphere_F_lm}
\hat{F}_{\ell,m} &=& \hat{c}_{\ell,m} \cdot \hat{\Phi}_{\ell,m}.
\end{eqnarray}

To express explicitly $\hat{F}_{0}$ and $\hat{F}_{\ell,m}$, we write the expressions for $\hat{\Phi}_{0}$ and $\hat{\Phi}_{\ell,m}$ in terms of the Fourier components $\kappa_{\ell,m}$ of the steady-state activation $x^*$.

For $\ell = 0$ (and $m = 0$), we have the scalar
\begin{eqnarray}\label{eq:sphere_phi_0}
\hat{\Phi}_{0} = \int_{0}^{2\pi} \int_{0}^{\pi}\Phi\left[x(\theta, \phi)\right] Y_0(\theta,\phi) \frac{\sin\theta d\theta \, d\phi}{4\pi},
\end{eqnarray}
while for $\ell > 0$, the Fourier components are vectors which elements are given by
\begin{eqnarray}
\hat{\Phi}_{\ell,m} &=& \int_{0}^{2\pi} \int_{0}^{\pi}\Phi\left[x(\theta, \phi)\right] Y_{\ell,m}(\theta, \phi) \frac{\sin\theta d\theta \, d\phi}{4\pi}. \label{eq:sphere_Philm}
\end{eqnarray}

We next write the steady-state activation $x(\theta, \phi)$ using the Inverse Fourier Transform:

\begin{eqnarray}
x(\theta, \phi)  &=&  \sum_{\ell=0}^L \sum_{m=-\ell}^\ell \kappa_{\ell,m} Y_{\ell,m}(\theta, \phi) \nonumber\\
&=&  \kappa_{0} Y_{0}(\theta, \phi) +\sum_{\ell=1}^L \sum_{m=-\ell}^\ell \kappa_{\ell,m} Y_{\ell,m}(\theta, \phi)\nonumber\\
&=&  \kappa_{0}
+\sum_{\ell=1}^L \sum_{m=-\ell}^\ell \kappa_{\ell,m} Y_{\ell,m}(\theta, \phi) \label{eq:sphere_x_kappa}.
\end{eqnarray}

Notice that we include only the $L$ first Fourier components, because at steady-state $\kappa_{\ell,m}=0$ for $\ell>L$.

Inserting Eq.~\eqref{eq:sphere_x_kappa} into Eqs.~\eqref{eq:sphere_phi_0} and \eqref{eq:sphere_Philm} gives us explicit expressions for $\hat{\Phi}_{0} $ and $\hat{\Phi}_{\ell,m}$ as function of the set of $2\ell+1$ scalars $\kappa_{0}$ and $ \kappa_{\ell,m}$.

This leads to the explicit form of the fixed point equations:

\begin{eqnarray}  \label{eq:fixed_point_sphere_0}
\kappa_{0} = \hat{c}_{0} \int_{0}^{2\pi} \int_{0}^{\pi}\Phi\left[ \kappa_{0} 
+ \sum_{\ell=1}^L \sum_{m=-\ell}^\ell \kappa_{\ell,m} Y_{\ell,m}(\theta, \phi)\right] \frac{\sin\theta d\theta \, d\phi}{4\pi},
\end{eqnarray}

\begin{eqnarray}  \label{eq:fixed_point_sphere_lm}
\kappa_{\ell,m} = \hat{c}_{\ell,m} \int_{0}^{2\pi} \int_{0}^{\pi}\Phi\left[ \kappa_{0} 
+ \sum_{\ell'=1}^L \sum_{m'=-\ell'}^{\ell'} \kappa_{\ell',m'} Y_{\ell',m'}(\theta, \phi)\right] 
Y_{\ell,m}(\theta, \phi) \frac{\sin\theta d\theta \, d\phi}{4\pi}.
\end{eqnarray}

\subsubsection{Fixed point manifold}

Because of the equivariance of recurrent inputs with respect to rotations in three dimensions, any steady state activation $x^*$ leads to a manifold of fixed points given by $S_g x^*$ for $g\in SO(3)$.

As shown in Eq.~\eqref{eq:fourier_comp-transformation}, applying the group action $S_g$ to $x^*$ leads to a linear transformation of its Fourier components determined by the correponding irreducible representations of $G$. In this case $G=SO(3)$, so that

\begin{equation}
    \kappa_l \mapsto \kappa_l D^\ell(\alpha,\beta,\gamma). \label{eq:fourier_comp-transformation_so3}
\end{equation}
where $D^\ell(\alpha,\beta,\gamma)$ is the $\ell$-th Wigner matrix associated with $g\in SO(3)$ \citep{Cohen2018-jb}, and $\kappa_l$ is a $(2\ell + 1)$-dimensional vector of Fourier components. Since $x^*$ is a function on $S^2$, we can moreover set $\gamma=0$, which amounts to parametrizing the fixed point manifold by elements of $SO(3)/SO(2)$.

In a low-rank spherical RNN, we have at most $K+1$ non-zero vector-valued Fourier components. Grouping them in a single vector $\underline{\kappa}$ of dimension $R=\sum_{\ell=0}^K 2\ell + 1$, the manifold of fixed points in Fourier coordinates is given by

\begin{equation}\label{parametrized_solution_sphere}
\underline{\kappa}(\alpha, \beta) = R^K(\alpha, \beta) \underline{\kappa},
\end{equation}
where $R^K(\alpha, \beta)$ is block-diagonal, with each block corresponding to a Wigner $D$-matrix $D^\ell(\alpha, \beta, 0)$ of increasing dimension $2\ell + 1$, for $\ell = 0, 1, \dots, K$. 

\subsubsection{Rank three spherical RNN}

We now consider the specific instantiation of the sphere model with the connectivity kernel

\begin{eqnarray}\nonumber
c(\theta, \phi) &=&J_1 (\sin\theta \sin\phi  + \cos\theta + \sin\theta \cos\phi ),
\end{eqnarray}

where $J_1$ is a scalar parameter that modulates the strength of the connectivity kernel. In this case, the only non-zero Fourier components of the kernel are for $\ell=1$, with $m=-1,0,1$. 

These components are given by:
\begin{eqnarray}\nonumber
\hat{c}_{1,-1} = \frac{J_1}{\sqrt{3}}, \
\hat{c}_{1,0} = \frac{J_1}{\sqrt{3}}, \
\hat{c}_{1,1} = \frac{J_1}{\sqrt{3}},
\end{eqnarray}

where the $\ell=0$ normalized real spherical harmonics are used:
\begin{equation}\nonumber
Y_{1,-1}(\theta, \phi) = \sqrt{3} \sin\theta \sin\phi, \ \
Y_{1,0}(\theta, \phi) = \sqrt{3} \cos\theta, \ \
Y_{1,1}(\theta, \phi) = \sqrt{3} \sin\theta \cos\phi.
\end{equation}

The only potentially non-zero Fourier components of the steady-state activation $x^*$ are $\kappa_{1,-1},\kappa_{1,0},\kappa_{1,1}$, so that
\begin{eqnarray}
x^*(\theta, \phi) &=& \sum_{m=-1}^1 \kappa_{1,m} Y_{1,m}(\theta, \phi), \nonumber\\
&=& \kappa_{1,-1} Y_{1,-1}(\theta, \phi) + \kappa_{1,0} Y_{1,0}(\theta, \phi) + \kappa_{1,1} Y_{1,1}(\theta, \phi), \nonumber\\
&=& \sqrt{3} \Big( \kappa_{1,-1} \sin\theta \sin\phi + \kappa_{1,0} \cos\theta + \kappa_{1,1} \sin\theta \cos\phi \Big). \label{eq:x_sphere_rank3}
\end{eqnarray}

In this case, Eqs.~\eqref{eq:fixed_point_sphere_0}-\eqref{eq:fixed_point_sphere_lm} reduce to three coupled equations for the scalars $\kappa_{1,-1},\kappa_{1,0},\kappa_{1,1}$

\begin{eqnarray}
\label{eq:system_eq_shpere}
\kappa_{1,-1} &=& \frac{J_1}{\sqrt{3}}\int_{0}^{2\pi} \int_{0}^{\pi}\Phi\left[\sum_{m=-1}^1 \kappa_{1,m} Y_{1,m}(\theta, \phi)\right] Y_{1,-1}(\theta, \phi) \frac{\sin\theta d\theta \, d\phi}{4\pi},\\\nonumber
\kappa_{1,0} &=& \frac{J_1}{\sqrt{3}} \int_{0}^{2\pi} \int_{0}^{\pi}\Phi\left[\sum_{m=-1}^1 \kappa_{1,m} Y_{1,m}(\theta, \phi)\right] Y_{1,0}(\theta, \phi) \frac{\sin\theta d\theta \, d\phi}{4\pi},\\\nonumber
\kappa_{1,1} &=& \frac{J_1}{\sqrt{3}}  \int_{0}^{2\pi} \int_{0}^{\pi}\Phi\left[\sum_{m=-1}^1 \kappa_{1,m} Y_{1,m}(\theta, \phi)\right] Y_{1,1}(\theta, \phi) \frac{\sin\theta d\theta \, d\phi}{4\pi}.
\end{eqnarray}

Any non-zero solution $\underline {\kappa}=(\kappa_{1,-1},\kappa_{1,0},\kappa_{1,1})$ of these equations leads to a manifold of fixed points in Fourier coordinates, parametrized by elements of $SO(3)$ \citep{By-Zhengwei-SuUnknown-yu, Blanco1997-ch, Pinchon2007-gi}:
\begin{eqnarray}\label{eq:rotated_kappa_sphere}
\underline{\kappa}(\alpha, \beta) = R^1(\alpha, \beta) \underline{\kappa},
\end{eqnarray}
where 
\begin{equation}\nonumber
    R^1(\alpha, \beta) = \left(\begin{array}{ccc}
\cos \alpha 
 & \sin \alpha \sin \beta & 
\sin \alpha  \cos \beta
 \\
0 & \cos \beta & - \sin \beta \\
-\sin \alpha 
& \cos \alpha \sin \beta & 
\cos \alpha \cos \beta 
\end{array}\right).
\end{equation}

Setting $\alpha, \beta=0,0$, we obtain a particular solution ${\underline{\kappa}}(0,0)=(0,\kappa_{1,0},0)$. The system \eqref{eq:system_eq_shpere} then reduces to the one dimensional equation:
\begin{equation}\label{eq_sphere_reduced}
    \kappa_{1,0} = \frac{J_1}{\sqrt{3}}\int_{0}^{2\pi} \int_{0}^{\pi}\Phi\left[ \kappa_{1,0} Y_{1,0}(\theta, \phi)\right] Y_{1,0}(\theta, \phi) \frac{\sin\theta d\theta \, d\phi}{4\pi}
\end{equation}
The stability of each  fixed-point is determined by the sign of the derivative of the r.h.s with respect to $\kappa_{1,0}$, evaluated at the fixed point. This stability is then inherited by all the fixed points on the resulting manifold.

\begin{figure}[!ht]
    \centering
    \includegraphics[width=14cm]{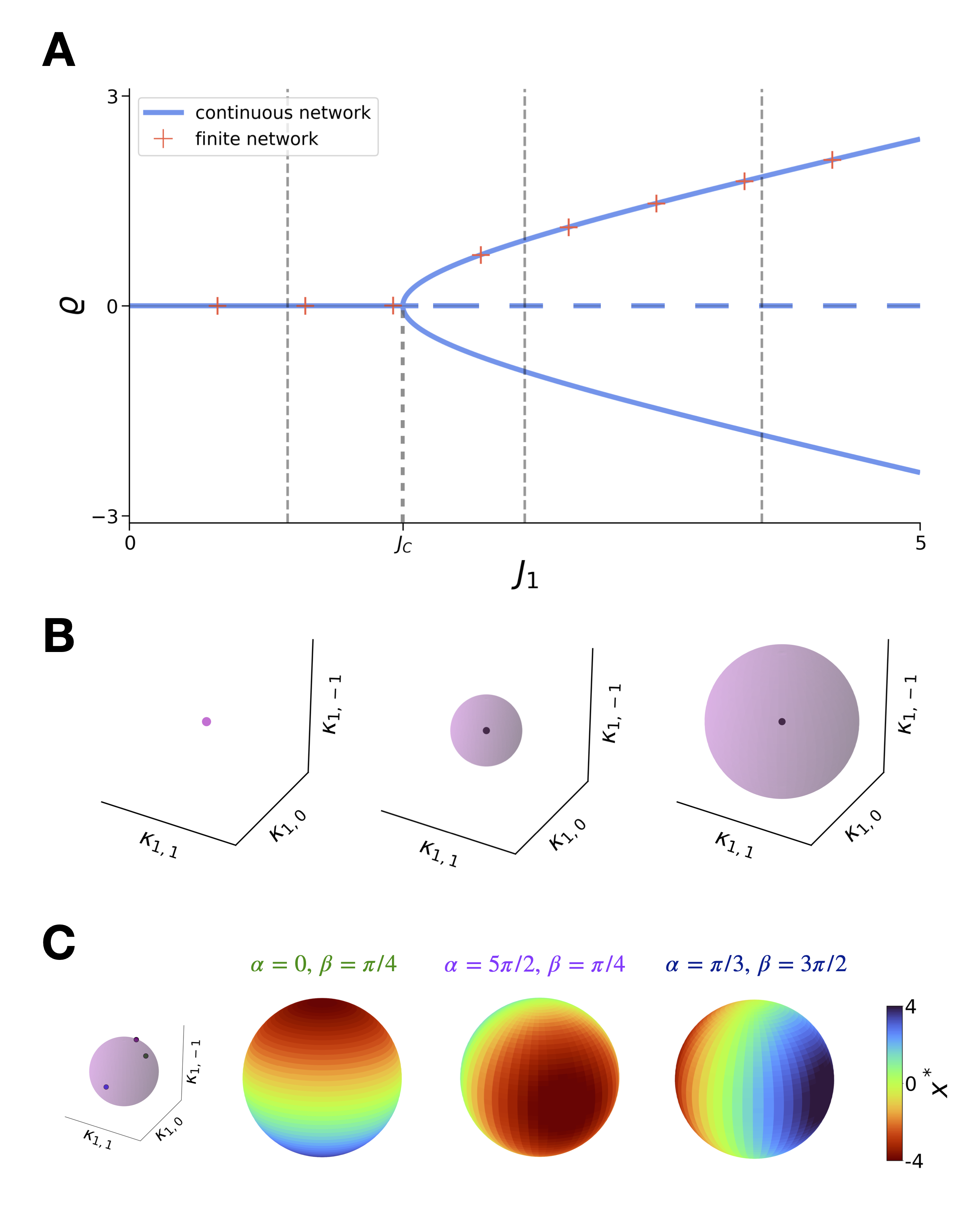} 
    \caption{Bifurcation analysis and manifolds in the rank three spherical model. A) Bifurcation diagram showing the emergence of non-zero fixed points $\kappa_{0,1}$ as a function of coupling strength $J_1$. Solid blue lines represent stable solutions while the dashed blue line indicates an unstable solution. Crosses show simulations of finite size network (N=1000, see appendix \ref{appendix_spherical}). B) Fixed-point manifolds for three different coupling strengths indicated by vertical dashed lines in panel A. Left: At subcritical coupling, only a single fixed point exists. Middle: for $J_1>J_c$, a spherical manifold of stable fixed points emerges. Right: At stronger coupling, the radius of the spherical manifold increases. C) Population activity patterns for three different fixed points on the spherical manifold for $J=1.1$, corresponding to three different sets of coordinates $(\alpha,\beta)$ indicated in the leftmost panel. 
     Each color plot shows the fixed point neural activity $x^*_{\alpha, \beta}(\theta,\phi)$, as function of the coordinates $(\theta,\phi)$ of neurons on $\Omega=S^2$, for a fixed set of $(\alpha,\beta)$ values.
    }
    \label{fig:sphere}
\end{figure}

Eq.~\eqref{eq_sphere_reduced} admits a zero solution $\kappa_{1,0} = 0$ that is stable for $J_1 < J_c$, and unstable for $J_1 > J_c$. For $J_1 > J_c$, Eq.~(\ref{eq_sphere_reduced}) admits an additional pair of non-zero solutions with opposite signs $\kappa_{1,0} = \pm \rho$ (Fig.~\ref{fig:sphere} A). These solutions are stable, and lead to a manifold of stable fixed points that form a sphere embedded in three dimensions (Fig.~\ref{fig:sphere} B). The radius of the sphere increases with $J_1$. Note that the two non-zero solutions of Eq.~(\ref{eq_sphere_reduced}) with opposite signs correspond to antipodal points on the sphere. 

Specifically, applying the relation \eqref{eq:rotated_kappa_sphere} to the particular solution ${\underline{\kappa}}(0,0)=(0,\varrho,0)$, leads to the general solution for $(\alpha,\beta) \in [0,2\pi) \times [0,\pi)$:
\begin{eqnarray}\label{kappa_general_solution_sphere}
{\underline{\kappa}}(\alpha,\beta)=(\varrho \sin \alpha \sin \beta, \varrho \cos \beta, \varrho \cos \alpha \sin \beta).
\end{eqnarray}
Inserting \eqref{kappa_general_solution_sphere} into Eq.~\eqref{eq:x_sphere_rank3} results in the population activity for fixed values of $(\alpha,\beta)$ (Fig.~\ref{fig:sphere}C):
\begin{eqnarray}
x_{\alpha,\beta}(\theta,\phi) = \sqrt{3} \varrho \Big( \sin \beta \sin \theta \cos (\phi - \alpha) + \cos \beta \cos \theta \Big).
\label{eq:pop_activity_sphere_param}
\end{eqnarray}

Note that, unlike the ring and the torus, the sphere does not admit a regular discretization that preserves (discretized) three-dimensional rotations. We provide an example of a discretization in the Appendix \ref{appendix_spherical}. In general, a discretized version of the spherical RNN  leads to at most two exact fixed points, but the rest of spherical manifold forms slow points, on which the speed of dynamics depends on the exact discretization (Fig.\ref{fig7}).

\newpage 
\addcontentsline{toc}{section}{Discussion}
\section*{Discussion}

In this study, we adopted an abstract, mathematical approach  to formulate a broad set of models that we call equivariant recurrent neural networks. This class of models provides a unified framework that relates the symmetries in the connectivity matrix of a recurrent network to the symmetries of its set of fixed points. Various particular instances of equivariant 
RNNs have been previously used in computational neuroscience as models of specific brain phenomena. The most central example is of course the classical ring model \citep{amari}, which is a standard model of orientation selectivity in the visual cortex \citep{ben1995theory}, head direction signals \citep{zhang1996representation}, place cells \citep{tsodyks-sejnowski1995,samsonovich1997,tsodyks1999}  or working memory \citep{Compte2000}. Several recent works have applied the formalism of group theory to ring models \citep{zhang,zuo,zuo2024}.
The rank two version of the ring model was originally introduced in \cite{ben1995theory} in networks of threshold-linear units, and analyzed using a decomposition on Fourier components. The discrete version, consisting of a small number of neurons,  has attracted renewed attention recently \citep{Noorman2024-dm}, with the discovery of the head-direction system in the fly \citep{kim2017} and the zebrafish \citep{petrucco2023}. The class of models we call toroidal RNNs have been previously studied in several contexts. The equivalent of our rank five toroidal RNN with multi-stable ring attractors has been used as a model of storing multiple spatial maps in the navigation system \citep{Romani2010, Si2014}. More general toroidal RNNs leading to two-dimensional toroidal manifolds correspond to classical models of grid cells in the enthorhinal cortex \citep{Burak2009, Gardner2022}. The original model of \cite{Burak2009} was based on a Gaussian connectivity kernel, while the rank five model used here keeps only the first two non-uniform Fourier components of the kernel. Note that, while the rank five toroidal RNN studied in Sec.~\ref{sec:rank5-torus} gives rise to toroidal manifolds, the individual neurons exhibit square rather than hexagonal tuning patterns \citep{sorscher2023, khona2025}. One way to generate hexagonal tuning patterns is to modify the underlying domain $\Omega$ by introducing a twist, which effectively gives rise to a twisted torus \citep{Guanella2007, Spalla2019}. In absence of such a twist, identifying the conditions under which low-rank toroidal RNNs generate hexagonal grid cells is an interesting question for a future study.

Spherical RNNs analyzed in Sec.~\ref{sec:sphere} provide a potential model for neural systems in charge of navigation in three dimensions \citep{stella2013,stella2020}, such as the head orientation system in bats \citep{finkelstein2014}. Such models have been comparatively less studied  as they present two complications \citep{stella2020}. First, the relation between spherical variables and the corresponding set of symmetries, rotations in three dimensions, is less direct than for two-dimensional variables. More specifically, the set of points on a sphere does not form a group, and the associated group of three-dimensional rotations is three-dimensional rather than two-dimensional. As a consequence, the manifold of fixed points in the spherical RNN is not directly indexed by group elements, and therefore does not form a Lie group in contrast to the ring and toroidal models.
Second, unlike the ring and the torus, the sphere and the associated symmetry group  do not admit a regular discretization that preserves a set of discretized three-dimensional rotational symmetries \citep{Cohen2018-jb, Kondor2018-lo}. Because of that, a discretized version of the spherical RNN leads in general to at most two exact fixed points (Appendix \ref{appendix_spherical}) rather than an extensive number as in the ring and toroidal RNN. While the symmetry, and the resulting fixed-point manifold, are recovered in the continuum limit, finite-size models instead exhibit a spherical manifold of slow points (Fig.\ref{fig7}).

Our work is closely related to recent studies that have proposed practical recipes for constructing continuous attractor models with arbitrary topologies \citep{Machens2008,Darshan-Rivkind,Pollock2020,Claudi2022,Claudi2025,romani-barak}. In particular, studies in \cite{Claudi2022,Claudi2025} have highlighted the challenges of modelling memory \citep{Claudi2022} and integration \citep{Claudi2025} on complex topologies that extend beyond the ring, introducing methods from differential geometry to overcome these limitations. Our mathematical approach, based on group theory and Fourier transforms, offers a complementary perspective on this problem. The formalism of group theory, applied within the framework of geometric deep learning, allows us to precisely identify and control the symmetries embedded in the connectivity and activity of these networks. Group Fourier transforms, together with related tools from the framework of low-rank recurrent neural networks, enable us to derive explicit parameterizations and equations for the emergent manifold, from which both dynamical (e.g., stability) and geometric (e.g., embedding dimensionality) properties can be directly controlled.
A key insight obtained from our approach is the importance of stability considerations: for a connectivity with a given symmetry, depending on parameters, several manifolds with different symmetry subgroups and different stability properties can coexist.

The mathematical formalism in this study is directly inspired by Geometric Deep Learning, a broad approach that applies group theory tools to various types of deep networks \citep{GDL}. A distinct line of research has used group theory to study the implications of symmetries on dynamical systems \citep{golubitsky,chossat}. Since recurrent networks are a particular sub-class of dynamical systems, the present manuscript naturally bears similarities to that line of work, but adopts a distinct point of view. The general theory of symmetry in dynamical systems considers  flow fields defined on $\mathbb{R}^N$ that are symmetric, e.g. equivariant, with respect to a set of transformations of $\mathbb{R}^N$ that form a subgroup of $GL(N)$. That theory then seeks to derive general results for the symmetry of individual solutions, and  the change of those symmetries  at bifurcation points, i.e. symmetry breaking. In the present manuscript, we consider recurrent networks where units are indexed by an underlying domain $\Omega$ that is associated with a symmetry group $G$. While an $N$-dimensional RNN is a dynamical system on $\mathbb{R}^N$, we formally define the state of the system as a function $\Omega \to \mathbb{R}$, so that the dynamics in the RNN correspond to dynamics within the space of such functions. Models of this type are usually referred to as neural mass or neural field models \citep{coombes2014}. Although group theoretical approaches to neural field models have been previously developed \citep{bressloff2001}, we are unaware of previous works in that literature focusing on connectivity based on group convolutions and the resulting manifolds of fixed points. The relationship between the present study and the broader field of symmetrical dynamical systems  certainly merits further investigations.

The connectivity in equivariant RNNs is based on the generalized notion of group convolution. Using the concept of group Fourier transforms, we introduced low-rank RNNs as the subset of models  in which the connectivity kernel has a finite number of non-zero Fourier components. For finite networks, this leads to a connectivity matrix that is of a rank determined by the number of non-zero Fourier components (Appendix \ref{appendix_discretized}). Low-rank equivariant RNNs are therefore a subset of the broader class of low-rank RNNs \citep{mastrogiuseppe2018, schuessler2020, beiran2021, dubreuil2022, Logiaco2021, Pezon2024, Schmutz2025}, and correspond to a specific case of the situation where the low-rank connectivity components are indexed by an underlying domain $\Omega$ \citep{Pezon2024}. For both finite and continuous low-rank equivariant RNNs, the dynamics can be exactly reduced to a finite number of variables, similarly to more general low-rank RNNs. An important ingredient in the definition of the group Fourier transform is the notion of irreducible representations. While group representation theory has extensively mapped systems of irreducible representations for compact groups \citep{Kosmann-Schwarzbach2009-uj}, this theory has been generally developed for representations on complex numbers. To avoid defining complex-valued connectivity, here we instead used representations on real numbers, but strictly speaking this might lead to technical difficulties in certain cases because the set of real numbers is not algebraically closed in contrast to the set of complex numbers.

A central assumption in the class of equivariant RNNs is that the connectivity is defined as an exact convolution with a fixed connectivity kernel. While this ingredient induces the equivariance property of the fixed point equations, and therefore the existence of fixed point manifolds, such an exact symmetry is a very strong, and arguably unrealistic constraint for a biological network. Recent works have shown that this strong constraint is not a necessary condition for the appearance of fixed point manifolds. Statistical symmetries in the distribution of connectivity parameters can lead to continuous attractors in the limit of infinitely large networks \citep{mastrogiuseppe2018,beiran2021,beiran2023,Clark2015}. In finite networks, fluctuations in connectivity (or in discretization, see Appendix \ref{appendix_spherical}) typically lead to a breakdown of the manifold into a finite number of stable fixed points. Nevertheless, the dynamics remain very slow on the entirety of the original manifold, so that the fixed point manifold is effectively replaced by a {\em slow manifold} - a region in state space to which the dynamics are quickly attracted, but then slowly evolve on it. This empirical observation can be made mathematically more precise, by showing that finite networks remain in the vicinity of continuous attractors when considered in the parameter space \citep{Sagodi2024}. On the timescales of experimental recordings, such slow manifolds may in practice behave in a manner very similar to continuous attractors. Alternatively, it is often possible to fine-tune connectivity parameters to recover exact continuous attractors even in finite networks without full continuous symmetry \citep{Darshan-Rivkind, Noorman2024-dm}. Extending the theory of equivariant RNNs to statistically symmetric connectivity, or finite, disordered networks, is an important direction for future work.

\section*{Acknowledgments}
This work was supported by the CRCNS project PIND (ANR-19-NEUC-0001-01), the Simons Collaboration on the Global Brain (AN-NC-GB-Culmination-00003154-05), the program “Ecoles Universitaires de Recherche” launched by the French Government and implemented by the ANR, with the reference ANR-17-EURE-0017.











\newpage
\bibliographystyle{agsm}
\bibliography{main}
\appendix
\section{Group Theory} \label{appendix_group_theory}

Here we summarize the main elements of group theory based on \cite{Kosmann-Schwarzbach2009-uj, Blundell2017-gr}.
The approach for extending group convolutions and group Fourier transforms to homogeneous spaces is based on \cite{Kondor_Trivedi}.

\subsection{Groups}

A \textbf{group} $G$ is a set equipped with a binary operation $\cdot : G \times G \rightarrow G$  that satisfies the following axioms:
\begin{enumerate}
    \item \textbf{Closure:} For all $g_1, g_2 \in G$, $g_1 \cdot g_2 \in G$.
    \item \textbf{Associativity:} For all $g_1, g_2, g_3 \in G$, we have $(g_1 \cdot g_2) \cdot g_3 = g_1 \cdot (g_2 \cdot g_3)$.
    \item \textbf{Identity:} There exists an element $e \in G$ such that $e \cdot g = g \cdot e = g$ for all $g \in G$.
    \item \textbf{Inverse:} For each $g \in G$, there exists an element $g^{-1} \in G$ such that $g \cdot g^{-1} = g^{-1} \cdot g = e$.
\end{enumerate}

When $G$ is a finite set of cardinality $n$, we say that $G$ is a \textbf{finite group} and that its \textbf{order} is $n$. If $G$ is an infinite set, then it is an \textbf{infinite group}, with infinite order.

Furthermore, the elements of a group can be a numerable set, in which case they form a \textbf{discrete group}. If the set is not numerable, instead, the group is said to be \textbf{continuous} and the elements depend on a number of continuous parameters, referred to as \textbf{dimension of the group}. If the group is discrete, we say that its dimension is 0. 

Throughout this paper, we focus primarily on compact groups, which are groups that are also compact topological spaces with continuous group operations.

\subsection{Group Actions}

An \textbf{action} of a group $G$ on a set $\Omega$ is a map $\cdot : G \times \Omega \rightarrow \Omega$ satisfying:
\begin{enumerate}
    \item $e \cdot \omega = \omega$ for all $\omega \in \Omega$, where $e$ is the identity element of $G$.
    \item $(g_1 \cdot g_2) \cdot \omega = g_1 \cdot (g_2 \cdot \omega)$ for all $g_1, g_2 \in G$ and $\omega \in \Omega$.
\end{enumerate}

Any action of $G$ on $\Omega$ induces an action of $G$ on the set of functions $\Omega \rightarrow \mathbb{R}$.
For any function $f: \Omega \rightarrow \mathbb{R}$, a group element $g \in G$ induces a transformation  denoted $S_g[f]$ and defined by:
\begin{equation}
    S_g[f](\omega) = f(g^{-1} \cdot \omega) \,\,\,\,\,\,\, \forall \omega \in \Omega.
\end{equation}
We call the resulting operator $S_g$  the \textbf{action} of $g$ on the function space $L_\mathbb{R}(\Omega)$.

\subsection{Cosets and Quotient Spaces}

Given a group $G$, a subgroup $H \subseteq G$, and an element $g \in G$, the {\em left and right cosets} are defined as 

\begin{eqnarray}
gH &=& \{gh | h \in H\} \\
Hg &=& \{hg | h \in H\}.
\end{eqnarray}
The cosets partition $G$ in such a way that each element of $G$ belongs to only one left or right coset.

The set of all left (resp. right) cosets forms the left quotient space $G/H$ (resp. right quotient space $H/G$). If the left and right quotient spaces are equal, $H$ is called a normal subgroup of $G$ and the quotient space $G/H=H/G$ forms a group.

\subsection{Homogeneous Spaces}

A set $\Omega$ is called a \textbf{homogeneous space} with respect to a group $G$ if for any two points $\omega_1, \omega_2 \in \Omega$, there exists a group element $g \in G$ such that $g \cdot \omega_1 = \omega_2$. Fixing an origin $\omega_0 \in \Omega$, any $\omega \in \Omega$ can be reached as $\omega=g \cdot \omega_0$ for some $g\in G$. This element $g$ is however not necessarily unique, in particular when  $\Omega$ is not itself a group \citep{Kondor_Trivedi}. In that case, the set of elements of $G$ that fix the origin forms a subgroup $H=\{h\in G|h\omega_0 =\omega_0\}$, and $\Omega$ can be identified with the left quotient space $G/H$.
This identification then defines a map $g \mapsto [g]_{G/H}$ that maps each element of $G$ to a unique coset representative within $G/H=\Omega$.

\subsection{Haar Measure and Integration on Groups}

For a locally compact topological group $G$, the \textbf{Haar measure} $d\mu(g)$ is a measure that is invariant under the group action. Specifically, for any $h \in G$, and for any integrable function $f$, the Haar measure satisfies
$$\int_G f(hg) \, d\mu(g) = \int_G f \, d\mu(g).$$
We moreover require that $\mu$ satisfies the normalization condition  $$\int_G d\mu(g) = 1.$$ Every locally compact group admits a unique  invariant Haar measure.

The concept of Haar measure extends beyond groups to homogeneous spaces $\Omega = G/H$. When both $G$ and $H$ are compact, the group action on $\Omega$ induces a unique $G$-invariant measure $d\mu(\Omega)$, and normalized so that $\int_{\Omega} d\mu(\Omega) = 1$.

\subsection{Group Representations}

A \textbf{representation} of a group $G$ is a homomorphism $\rho: G \rightarrow GL(V)$, where $GL(V)$ is the general linear group of a vector space $V$. In concrete terms, a $d$-dimensional representation assigns to each group element $g \in G$ a $d \times d$ invertible matrix $\rho(g)$ such that:
\begin{equation}
    \rho(g_1 \cdot g_2) = \rho(g_1) \cdot \rho(g_2)
\end{equation}
for all $g_1, g_2 \in G$.

\subsection{Reducible and Irreducible Representations}

A representation $\rho$ is \textbf{reducible} if the vector space $V$ can be decomposed into proper subspaces that are invariant under the action of all $\rho(g)$ for $g \in G$. Otherwise, the representation is \textbf{irreducible} (or an \textbf{irrep}).

Mathematically, a representation $\rho$ is reducible if there exists a change of basis such that all matrices $\rho(g)$ for $g \in G$ can be simultaneously expressed in block-diagonal form:
\begin{equation}
    P^{-1}\rho(g)P = \begin{pmatrix}
    \rho_1(g) & 0 & \cdots & 0 \\
    0 & \rho_2(g) & \cdots & 0 \\
    \vdots & \vdots & \ddots & \vdots \\
    0 & 0 & \cdots & \rho_k(g)
    \end{pmatrix}
\end{equation}
where $P$ is an invertible matrix representing the change of basis, and each $\rho_i(g)$ is a lower-dimensional representation of $G$. This block-diagonal structure corresponds to a decomposition of the vector space $V$ into proper invariant subspaces $V = V_1 \oplus V_2 \oplus \cdots \oplus V_k$, where each subspace $V_i$ is preserved under the action of $G$, meaning that for any $v \in V_i$ and any $g \in G$, we have $\rho(g)v \in V_i$.

For compact groups, \textbf{Maschke's theorem} ensures that any finite-dimensional representation can be decomposed as a direct sum of irreducible representations, i.e. it is completely reducible. The set of all irreducible representations of a compact group $G$ is countable.

\subsection{Group convolutions and their extension to homogeneous spaces \label{app:group_convolution}}

For two functions  $f^1,f^2:G\to \mathbb{R}$ the $G$-convolution is a function $G\to \mathbb{R}$ defined by
\begin{eqnarray}
    (f_1 *_G f_2)(g)=\sum_{h \in G} f_1(g^{-1}h)  f_2(h).
\end{eqnarray}

To extend this concept of convolutions to functions $f^1,f^2:\Omega\to \mathbb{R}$, where $\Omega$ is a homogeneous space with respect to $G$, we follow \cite{Kondor_Trivedi}. For $f:\Omega\to \mathbb{R}$, we first introduce its {\em lifting} to $G$ as the function $f^\uparrow :\Omega\to \mathbb{R}$ defined as
\begin{eqnarray}\label{eq:lifting}
    f^ \uparrow  (g)=f([g]_{\Omega})  \,\,\,\,\,\,\,  \forall g \in G,
\end{eqnarray}
where $[g]_{\Omega}=[g]_{G/H}$ is the representative of $g$ on $\Omega$, or equivalently on $G/H$. The convolution between $f^1,f^2:\Omega\to \mathbb{R}$ can then be defined as a function $G\to \mathbb{R}$ such that

\begin{eqnarray}
    (f_1 *_G f_2)(g)&=&\sum_{h \in G} f_1^\uparrow (g^{-1}h)  f_2^\uparrow(h) \\
    &=&\sum_{h \in G} f_1 ([g^{-1}h]_{\Omega})  f_2([h]_{\Omega}).
\end{eqnarray}
Since the arguments of $f_1$ and $f_2$ are representatives of elements of $G$ on $\Omega$, and therefore elements of $\Omega$, the sum effectively runs over $\Omega$, and the convolution itself is effectively a function on $\Omega$.

\subsection{Group Fourier Transforms and their extension to homogeneous spaces\label{app:group_fourier}}

For a compact group $G$ with a system of irreducible representations $\{\rho_k\}_{k=0,1,...}$, the \textbf{group Fourier transform} of a function $f: G \rightarrow \mathbb{R}$ is defined as the collection of Fourier components:
\begin{equation}
    \hat{f}_k = \int_G f(g) \rho_k(g) \, d\mu(g)
\end{equation}
where $d\mu(g)$ is the normalized Haar measure on $G$. For finite groups, the integral is replaced by a sum.

The \textbf{Peter-Weyl theorem} states that any function $f: G \rightarrow \mathbb{R}$ can be reconstructed from its \textbf{inverse Fourier transform}:
\begin{equation}
    f(g) = \sum_{k=0}^{\infty} \tilde{d}_k \text{tr}\left(\hat{f}_k \rho_k^{-1}(g)\right)
\end{equation}
where $\tilde{d}_k$ is the dimension of the $k$-th complex-valued irreducible representation.

A fundamental property is the \textbf{convolution theorem} for group Fourier transforms:
\begin{equation}
    \widehat{f^1 *_G f^2}_k = \widehat{f^1}_k \cdot \widehat{f^2}_k
\end{equation}
where $*_G$ denotes the group convolution.

Following \cite{Kondor_Trivedi}, the Group Fourier transform can be directly extended to functions functions $f:\Omega\to \mathbb{R}$, where $\Omega$ is a homogeneous space with respect to $G$ by using the lifting $f^\uparrow$ of  $f$ to $G$ (Eq.~\eqref{eq:lifting}):

\begin{eqnarray}
    \hat{f}_k &=& \int_G f^\uparrow(g) \rho_k(g) \, d\mu(g) \\
           &=& \int_G f([g]_{\Omega=G/H}) \rho_k(g) \, d\mu(g).
\end{eqnarray}
\cite{Kondor_Trivedi} show that this is in general equivalent to integrating over $G/H$ while restricting the irreps $\rho_k$ to specific columns.

\section{Finite networks \label{appendix_discretized}}

In this appendix, we provide finite-size, discretized expressions of the recurrent input, the connectivity matrix and the connectivity vectors for the ring model, the toroidal RNN and the spherical RNN.

\subsection{Ring model}

The ring $\Omega = \mathcal{S}^1$ is parametrized by the angle $\theta \in [0, 2\pi)$. We discretize the ring using $N$ equally-spaced points with spacing $\Delta\theta = \frac{2\pi}{N}$:
\begin{equation}\theta_i = i \cdot \Delta\theta = \frac{2\pi i}{N}, \quad i = 0, 1, \ldots, N-1.
\end{equation}
The discrete measure at each point is $\Delta\mu_i = \frac{\Delta\theta}{2\pi} = \frac{1}{N}$, ensuring that $\sum_{i=0}^{N-1} \Delta\mu_i = 1$.

The continuous convolution in Eq.~\eqref{group_conv_cont_ring} is then discretized as follows:
\begin{equation}F^t(\theta_i) = \frac{1}{N} \sum_{j=0}^{N-1} c(g_{\theta_i}^{-1}(\theta_j)) \Phi[x^t(\theta_j)] = \frac{1}{N} \sum_{j=0}^{N-1} c(\theta_j - \theta_i) \Phi[x^t(\theta_j)]
\end{equation}
where the group action is described by angular translations: $g_{\theta_i}^{-1}(\theta_j) = \theta_j - \theta_i$.

The connectivity matrix $J \in \mathbb{R}^{N \times N}$ has entries
\begin{equation}J_{ij} = \frac{1}{N} c(\theta_j - \theta_i).
\end{equation}

The kernel $c(\theta)$ can be expressed via the Inverse group Fourier transform Eq.~\eqref{Peter-Weyl}:
\begin{eqnarray}
c(\theta) &=& \hat{c}_{0}+\sum_{k=1}^K 2\left(\hat{c}_{k,1}\cos(k \theta) + \hat{c}_{k,2}\sin(k \theta)\right), \label{eq:IFT_c_ring}
\end{eqnarray}
where $\hat{c}_0,\hat{c}_{k,1},\hat{c}_{k,2}$ are the Fourier components of the kernel, and in the hypotesis of an even kernel, the terms $\hat{c}_{k,2}=0$  for each $k$.
The elements of the connectivity matrix are then given by:
\begin{eqnarray}
J_{ij} &=& \frac{1}{N} \big[\hat{c}_{0} + \sum_{k=0}^{K} 2\hat{c}_{k,1}\cos(k (\theta_i-\theta_j)) \big]\\
&=& \frac{1}{N} \big[\hat{c}_{0} + \sum_{k=0}^{K} 2\hat{c}_{k,1}\big(\cos(k\theta_i)\cos(k\theta_j)+\sin(k\theta_i)\sin(k\theta_j) \big) \big]\\
&=& \frac{1}{N} \sum_{k=0}^{2K} z_i^{(k)} z_j^{(k)} 
\end{eqnarray}

where we define the \textbf{connectivity vectors} $z_i^{(k)}$ as \citep{mastrogiuseppe2018}:
\begin{align}
z_i^{(0)} &= \sqrt{\hat{c}_{0}} \\
z_i^{(2k-1)} &= \sqrt{2\hat{c}_{k,1}} \cos(k\theta_i), \quad k = 1, \ldots, K \\
z_i^{(2k)} &= \sqrt{2\hat{c}_{k,1}} \sin(k\theta_i), \quad k = 1, \ldots, K.
\end{align}
The rank of the connectivity matrix is therefore  at most $2K+1$.

In the 2D ring model,  the kernel is $c(\theta) = J_1 \cos(\theta)$ (Figure \ref{fig6}A):
\begin{equation}J_{ij} = \frac{1}{N} J_1\cos(\theta_i - \theta_j) 
\end{equation}
which can be decomposed as $ J_{ij} = \frac{1}{N}\sum_{k=1}^2 z_i^k z_j^k$, where the connectivity vectors are 
\begin{equation}z_i^1 = \sqrt{J_1}\cos(\theta_i), \quad z_i^2 = \sqrt{J_1}\sin(\theta_i).
\end{equation}

In the 5D ring model, the kernel is $c(\theta) = J_0 + J_1 \cos(\theta) + J_2 \cos(2\theta)$ (Figure \ref{fig6}B):
\begin{equation}J_{ij} = \frac{1}{N}\left[J_0 + J_1\cos(\theta_i - \theta_j) + J_2\cos(2(\theta_i - \theta_j))\right]
\end{equation}
which can be decomposed as $J_{ij} = \frac{1}{N}\sum_{k=0}^4 z_i^k z_j^k$, with connectivity vectors
\begin{equation}z_i^0 = \sqrt{J_0}, \quad z_i^1 = \sqrt{J_1}\cos(\theta_i), \quad z_i^2 = \sqrt{J_1}\sin(\theta_i), \quad z_i^3 = \sqrt{J_2}\cos(2\theta_i), \quad z_i^4 = \sqrt{J_2}\sin(2\theta_i).
\end{equation}

\subsection{Toroidal RNNs}

The torus $\Omega = \mathcal{T}^2 = \mathcal{S}^1 \times \mathcal{S}^1$ is the product of two rings, parametrized by $(\theta_1, \theta_2) \in [0, 2\pi) \times [0, 2\pi)$. The discretization follows directly from the ring case by applying it independently to each angular coordinate. We use a grid of $N_1 \times N_2$ equally-spaced points, with spacings:
\begin{eqnarray}
\Delta\theta_1 &=& \frac{2\pi}{N_1} \\
\Delta\theta_2 &=& \frac{2\pi}{N_2}.
\end{eqnarray}

For a total of $N$ neurons, we typically set $N_1 = N_2 = \sqrt{N}$.

Each neuron $i\in {0\ldots N-1}$ is indexed by a pair of indices $(i_1, i_2)$ such that
\begin{equation}
i = i_1 \cdot N_2 + i_2.
\end{equation}

The index pair $(i_1, i_2)$ corresponds to the position of the neuron on the two-dimensional grid:
\begin{equation}
\theta_{i_1} = i_1 \cdot \Delta\theta_1, \quad i_1 = 0, 1, \ldots, N_1-1
\end{equation}
\begin{equation}
\theta_{i_2} = i_2 \cdot \Delta\theta_2, \quad i_2 = 0, 1, \ldots, N_2-1.
\end{equation}

The discrete measure at each grid point is  given by $\Delta\mu_{i_1,i_2} = \frac{\Delta\theta_1}{2\pi} \cdot \frac{\Delta\theta_2}{2\pi} = \frac{1}{N}$, ensuring that $\sum_{i_1=0}^{N_1-1}\sum_{i_2=0}^{N_2-1} \Delta\mu_{i_1,i_2} = 1$.

The continuous convolution in Eq.~\eqref{group_conv_cont_torus} is then discretized as a double sum:
\begin{equation}
F^t(\theta_{i_1}, \theta_{i_2}) = \frac{1}{N} \sum_{j_1=0}^{N_1-1}\sum_{j_2=0}^{N_2-1} c(\theta_{i_1} - \theta_{j_1}, \theta_{i_2} - \theta_{j_2})\Phi[x^t(\theta_{j_1}, \theta_{j_2})].
\end{equation}

The connectivity matrix $J \in \mathbb{R}^{N \times N}$ has entries 
\begin{eqnarray}
J_{i,j}&=&J_{(i_1,i_2)(j_1,j_2)} \\
&=& \frac{1}{N}c(\theta_{j_1} - \theta_{i_1}, \theta_{j_2} - \theta_{i_2}).
\end{eqnarray}

Following the procedure outlined for the ring model, an even kernel $c(\theta_1, \theta_2)$,  is expressed via the Inverse group Fourier transform as:
\begin{equation}
c(\theta_{i1}, \theta_{i2}) = \hat{c}_{(0,0)}+ \sum_{\substack{k'_1=0, \, k'_2=0 \\ (k'_1,k'_2) \neq (0,0)}}^{K_1, K_2} 2\left(\hat{c}_{(k_1,k_2),1}\cos(k_1\theta_{i1} + k_2\theta_{i2})\right)
\label{eq:IFT_c_torus}
\end{equation}
where $\hat{c}_{(k_1,k_2)},\hat{c}_{(k_1,k_2),1},\hat{c}_{(k_1,k_2),2}$ are the Fourier componentsm and $K_1,K_2$ are the numbers of non-zero components along each dimension.

The elements of the connectivity matrix are then given by:
\begin{eqnarray}
J_{i,j}&=&J_{(i_1,i_2)(j_1,j_2)}\\
&=& \frac{1}{N} c(\theta_{i_1} - \theta_{j_1}, \theta_{i_2} - \theta_{j_2}) \\
&=& \frac{1}{N} \bigg[\hat{c}_{(0,0)} + \sum_{\substack{k_1=0, \, k_2=0 \\ (k_1,k_2) \neq (0,0)}}^{K_1, K_2} 2\hat{c}_{(k_1,k_2),1}\cos(k_1(\theta_{i_1}-\theta_{j_1}) + k_2(\theta_{i_2}-\theta_{j_2}))\bigg] \\
&=& \frac{1}{N} \bigg[\hat{c}_{(0,0)} + \sum_{\substack{k_1=0, \, k_2=0 \\ (k_1,k_2) \neq (0,0)}}^{K_1, K_2} 2\hat{c}_{(k_1,k_2),1}\big(\cos(k_1\theta_{i_1} + k_2\theta_{i_2})\cos(k_1\theta_{j_1} + k_2\theta_{j_2}) \nonumber \\
&& \qquad\qquad\qquad\qquad\qquad\qquad + \sin(k_1\theta_{i_1} + k_2\theta_{i_2})\sin(k_1\theta_{j_1} + k_2\theta_{j_2})\big)\bigg] \\
&=& \frac{1}{N} \sum_{r=0}^{R} z_{(i_1,i_2)}^{(r)} z_{(j_1,j_2)}^{(r)}
\end{eqnarray}
where $R = 2(K_1+1)\cdot (K_2+1)-1$ and the \textbf{connectivity vectors}  are defined as:
\begin{align}
z_{(i_1,i_2)}^{(0)} &= \sqrt{\hat{c}_{(0,0)}} \\
z_{(i_1,i_2)}^{(2k-1)} &= \sqrt{2\hat{c}_{(k_1,k_2),1}} \cos(k_1\theta_{i_1} + k_2\theta_{i_2}), \quad k = 1, \ldots, \tilde{K} \\
z_{(i_1,i_2)}^{(2k)} &= \sqrt{2\hat{c}_{(k_1,k_2),1}} \sin(k_1\theta_{i_1} + k_2\theta_{i_2}), \quad k = 1, \ldots, \tilde{K}
\end{align}
where to each $k$ corresponds to a pair $(k_1, k_2)$:

\begin{equation}
    k=k_1+k_2*K_1.
\end{equation}

In the rank-five toroidal model, the kernel is $c(\theta_1, \theta_2) = J_0 + J_1\cos(\theta_1) + J_2\cos(\theta_2)$:
\begin{equation}
J_{(i_1,i_2)(j_1,j_2)} = \frac{1}{N} \big[J_0 + J_1\cos(\theta_{i_1} - \theta_{j_1}) + J_2\cos(\theta_{i_2} - \theta_{j_2})\big]
\end{equation}
which can be decomposed as $J_{(i_1,i_2)(j_1,j_2)} = \frac{1}{N}\sum_{k=0}^{4} z_{(i_1,i_2)}^{(k)} z_{(j_1,j_2)}^{(k)}$, where the connectivity vectors are:

\begin{equation}
    z_{(i_1,i_2)}^{(0)} = \sqrt{J_0}, \quad 
z_{(i_1,i_2)}^{(1)} = \sqrt{J_1}\cos(\theta_{i_1}), \quad z_{(i_1,i_2)}^{(2)} = \sqrt{J_1}\sin(\theta_{i_1}), \quad
z_{(i_1,i_2)}^{(3)} = \sqrt{J_2}\cos(\theta_{i_2}), \quad z_{(i_1,i_2)}^{(4)} = \sqrt{J_2}\sin(\theta_{i_2})
\end{equation}

resulting in a rank-5 connectivity matrix (Figure \ref{fig6}C).

\subsection{Spherical RNNs \label{appendix_spherical}}

The sphere $\Omega = \mathcal{S}^2$ is parametrized by spherical coordinates $(\theta, \phi)$ where $\theta \in [0, \pi]$ is the polar angle and $\phi \in [0, 2\pi)$ is the azimuthal angle.
In contrast to $\mathcal{S}^1$, the sphere and the related rotation group do not admit a regular discretization that preserves exact symmetries. Here, we employ Fibonacci lattice sampling, which generates $N$ points that are approximately uniformly distributed by surface area \citep{fibonacci}. 
The Fibonacci lattice uses the golden angle $\theta_{\text{golden}} = \pi(\sqrt{5} - 1)$ and constructs points via:
\begin{equation}y_i = 1 - \frac{2i}{N-1}, \quad \theta_i = \arccos(y_i), \quad \phi_i = i \cdot \theta_{\text{golden}} \bmod 2\pi, \quad i = 0, \ldots, N-1.
\end{equation}
The key property of Fibonacci sampling is that each point represents approximately equal surface area. Since the total unnormalized surface area is $4\pi$, partitioning into $N$ approximately equal patches gives each point an unnormalized area $\sin\theta_i \Delta\theta_i \Delta\phi_i \approx \frac{4\pi}{N}$. Therefore, the discrete normalized measure for each point is:
\begin{equation}\Delta\mu_i = \frac{\sin\theta_i \Delta\theta_i \Delta\phi_i}{4\pi} \approx \frac{1}{N}.\end{equation}

The convolution Eq.~\eqref{group_conv_cont_sphere} is discretized as:
\begin{equation}F^t(\theta_i, \phi_i) = \frac{1}{N} \sum_{j=0}^{N-1} c(g_{(\theta_i,\phi_i,0)}^{-1}(\theta_j, \phi_j)) \Phi(x_t(\theta_j, \phi_j)).\end{equation}
The connectivity matrix $J \in \mathbb{R}^{N \times N}$ has entries:
\begin{equation}J_{ij} = \frac{1}{N} c(g_{(\theta_i,\phi_i,0)}^{-1}(\theta_j, \phi_j)).\label{eq:spherical_connectivity}
\end{equation}

The kernel $c(\theta, \phi)$ can be expressed via the Inverse group Fourier transform Eq.~\eqref{Peter-Weyl}:
\begin{eqnarray}
c(\theta, \phi) &=&  \hat{c}_{0}
+\sum_{\ell=1}^L \sum_{m=-\ell}^\ell \hat{c}_{\ell,m} Y_{\ell,m}(\theta, \phi), \label{eq:IFT_c_sphere}
\end{eqnarray}

where $\hat{c}_0,\hat{c}_{\ell,m}$ are the Fourier components of the kernel.

Combining Eqs.~\eqref{eq:spherical_connectivity} and \eqref{eq:IFT_c_sphere}, the connectivity matrix can be written as:
\begin{eqnarray}
J_{ij} &=& \frac{1}{N} \left[\hat{c}_0 + \sum_{\ell=1}^L \sum_{m=-\ell}^{\ell} \hat{c}_{\ell,m} Y_{\ell,m}(g_{(\theta_i,\phi_i,0)}^{-1}(\theta_j, \phi_j))\right] \\
&=& \frac{1}{N} \left[\hat{c}_0 + \sum_{\ell=1}^L \sum_{m=-\ell}^{\ell} \hat{c}_{\ell,m} Y_{\ell,m}((\theta_j, \phi_j))Y_{\ell,m}((\theta_i, \phi_i))\right] \\
&=& \frac{1}{N} \sum_{k=0}^{K} z_i^{(k)} z_j^{(k)}
\end{eqnarray}
where  $K = \sum_{\ell=0}^L (2 \ell + 1) $ is the total number of spherical harmonic modes up to degree $L$, and the 
 \textbf{connectivity vectors} $z_i^{(k)}$ are given by:
\begin{align}
z_i^{(0)} &= \sqrt{\hat{c}_0} \\
z_i^{(k)} &= \sqrt{\hat{c}_{\ell,m}} Y_{\ell,m}(\theta_i, \phi_i), \quad \text{for } k \text{ indexing } (\ell,m) \text{ with } \ell = 1, \ldots, L, \, m = -\ell, \ldots, \ell
\end{align}
resulting in a connectivity matrix $J$ with rank at most $\sum_{\ell=0}^L (2 \ell + 1) $.

\begin{figure}[!ht]
    \centering
    \includegraphics[width=16cm]{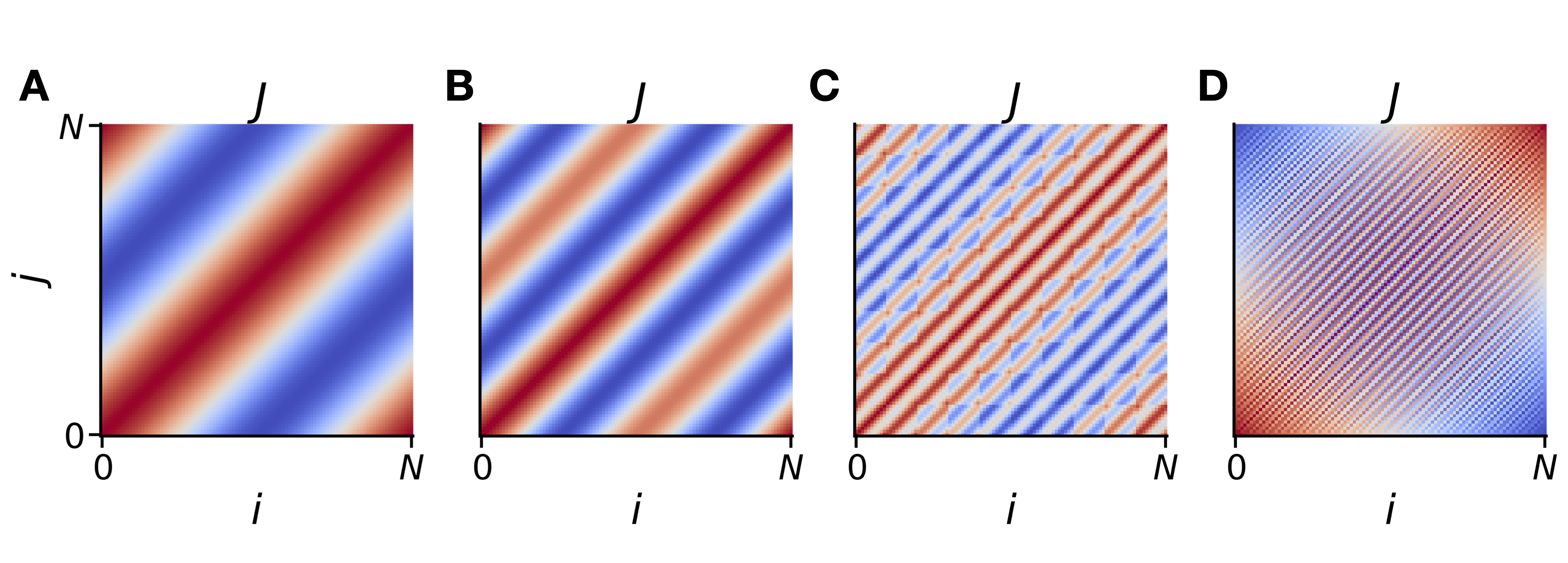} 
    \caption{Illustration of connectivity matrices in finite-size networks. A) Rank two ring model; B) Rank five ring model; C) Rank five toroidal model; D) Rank three sphere model.}
    \label{fig6}
\end{figure}

For the rank-3 spherical model, the kernel is 
\begin{equation}
c(\theta,\phi)=J_1 (\sin\theta \sin\phi  + \cos\theta + \sin\theta \cos\phi).
\end{equation}
The resulting connectivity matrix is rank 3 (Figure \ref{fig6}D): 

\begin{eqnarray}
J_{ij} 
&=&\frac{J_1}{N} \sqrt{3}(\sin\theta_i \cos\phi_i \sin\theta_j \cos \phi_j +\sin\theta_i \sin\phi_i \sin\theta_j \sin\phi_j
 + \cos\theta_i \cos\theta_j),\\
 &=&\frac{1}{N}\sum_{k=1}^3  z_i^{(k)} z_j^{(k)}
\end{eqnarray}
where the connectivity vectors correspond to 

\begin{equation}
z_i^{(1)} = \sqrt{J_1} Y_{1,-1}(\theta_i, \phi_i), \quad
z_i^{(2)} = \sqrt{J_1}Y_{1,0}(\theta_i, \phi_i), \quad z_i^{(3)} = \sqrt{J_1}Y_{1,1}(\theta_i, \phi_i).
\end{equation}

\begin{figure}[!ht]
    \centering
    \includegraphics[width=16cm]{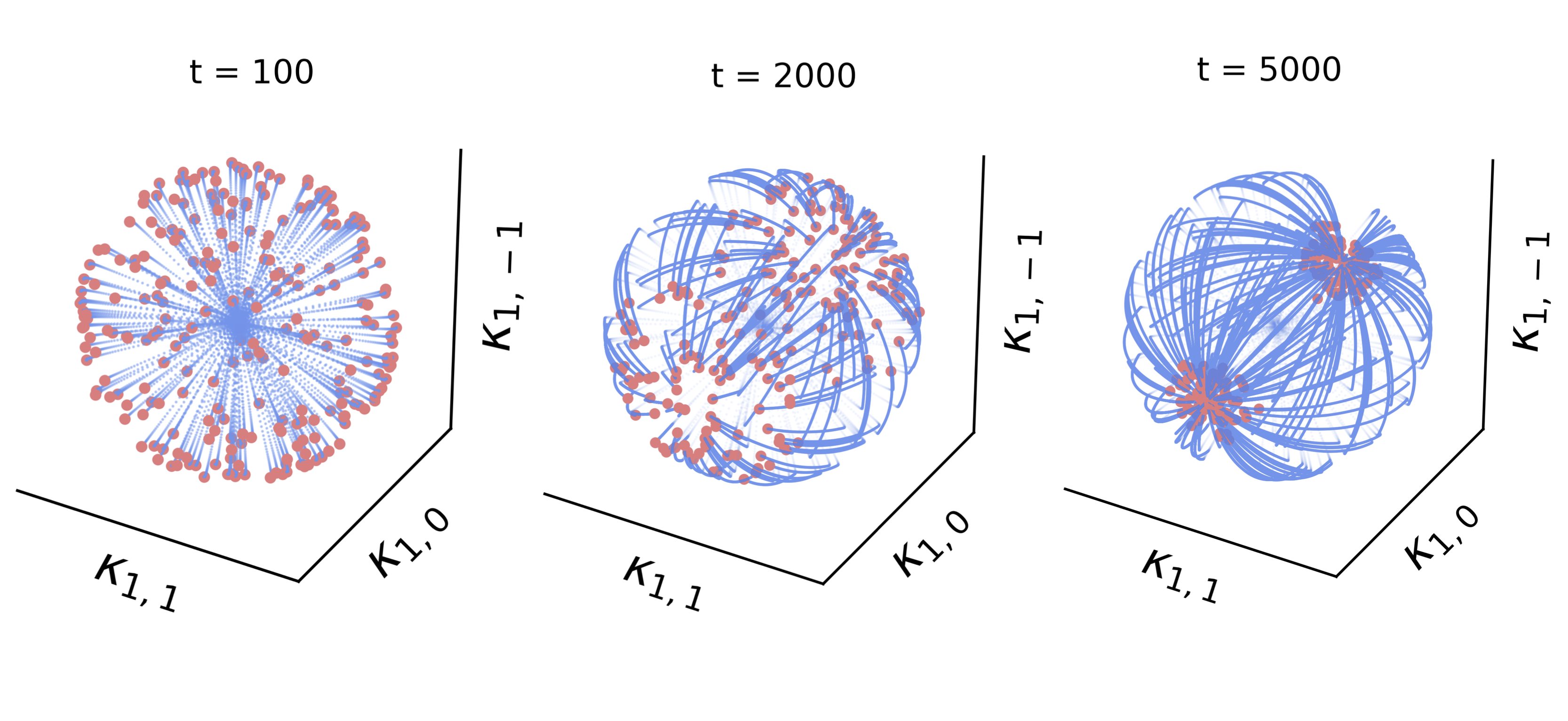} 
    \caption{Dynamics in a finite-size spherical RNN with N=500 neurons. Trajectories (light blue) starting from 200 random initial conditions and their position (light red) at $t$ = 100, 2000, and 5000, showing convergence from random initial conditions to a slow spherical manifold in the $(\kappa_{1,-1},\kappa_{1,0},\kappa_{1,1})$ latent space, and then two fixed points on the slow manifold.}
    \label{fig7}
\end{figure}

\end{document}